\documentclass[12pt,a4paper]{article}

\usepackage{amsmath}
\usepackage{amssymb}
\usepackage{graphicx}
\usepackage{dsfont}
\usepackage{overpic}
\usepackage{cite}

\let\oldappendix=\appendix
\let\oldsection=\section
\renewcommand{\appendix}{\oldappendix%
\def\theequation{\Alph{section}.\arabic{equation}}%
\renewcommand{\section}{\setcounter{equation}{0}\oldsection}}

\newcommand{\beq}{\begin{equation}}
\newcommand{\eeq}{\end{equation}}
\newcommand{\beqa}{\begin{eqnarray}}
\newcommand{\eeqa}{\end{eqnarray}}
\newcommand{\no}{\nonumber}

\newcommand{\tr}{\mbox{tr}}
\newcommand{\sfrac}[2]{{\textstyle\frac{#1}{#2}}}

\def\bra#1{\left\langle #1\right|}
\def\ket#1{\left| #1\right\rangle}

\newcommand{\newop}[2]{\def#1{\mathop{\mathrm{#2}}\nolimits}}
\newop{\artanh}{artanh}
\newop{\det}{det}
\newop{\tr}{tr}
\newop{\diag}{diag}
\newop{\Re}{Re}
\newop{\Im}{Im}

\newcommand{\Lagr}{\mathcal{L}}
\newcommand{\sst}{\scriptstyle}

\begin{document}

\hfill 

\hfill 

\bigskip\bigskip

\begin{center}

{{\Large\bf  Chiral dynamics of kaon-nucleon interactions,\\revisited}}

\end{center}

\vspace{.4in}

\begin{center}
{\large B.~Borasoy\footnote{email: borasoy@itkp.uni-bonn.de}$^{a,b}$,
        R.~Ni{\ss}ler\footnote{email: rnissler@itkp.uni-bonn.de}$^{a,b}$,
        W.~Weise\footnote{email: weise@ph.tum.de}$^{a}$}

\bigskip

\bigskip

$^{a}$Physik Department\\
Technische Universit{\"a}t M{\"u}nchen\\
D-85747 Garching, Germany \\[2ex]

$^{b}$Helmholtz-Institut f\"ur Strahlen- und Kernphysik (Theorie) \\
Universit\"at Bonn \\ 
Nu{\ss}allee 14-16, D-53115 Bonn, Germany \\

\vspace{.2in}

\end{center}

\vspace{.7in}

\thispagestyle{empty} 

\begin{abstract}
The $\bar{K} N$ system close to threshold is analyzed in view of the new accurate DEAR 
kaonic hydrogen data. The calculations are performed using chiral SU(3) effective field theory 
in combination with non-perturbative schemes based on
coupled channels. Several variants of such approaches are compared with experimental data
and the differences in the results are discussed. 
Coulomb and isospin breaking effects turn out to be important
and are both taken into account. The pole structure of the $\Lambda(1405)$ resonance
close to the $\bar{K}N$ threshold is critically examined.
\end{abstract}\bigskip

\begin{center}
\begin{tabular}{ll}
\textbf{PACS:}& 11.80.-m, 12.39.Fe, 13.75.Jz, 36.10.Gv \\[6pt]
\textbf{Keywords:}& Chiral Lagrangians, coupled channels, unitarity.
\end{tabular}
\end{center}


\vfill

\section{Introduction}\label{sec:Intro}

Chiral perturbation theory is an appropriate framework to investigate the dynamics of 
hadrons at low energies, whereby symmetries and symmetry breaking patterns
of QCD are incorporated. A systematic loop expansion can be carried out, but its perturbative
application is often limited to a small range of energies and breaks down in the vicinity of resonances.
In the $K^- p$ channel, for example, the existence of the $\Lambda(1405)$ resonance 
just below the $K^- p$ threshold renders SU(3) chiral perturbation theory (ChPT)
inapplicable. However, the combination with non-perturbative coupled-channels
techniques has proved useful by generating the $\Lambda(1405)$ dynamically
as an $I=0$ $\bar{K} N$ quasibound state and as a resonance in the $\pi \Sigma$
channel \cite{KSW1,OR}.

Such approaches have been applied to a variety of meson-baryon scattering
processes with quite some success \cite{OR, KSW2, Kr, OM, LK, BMW, JOORM, BFMS}. 
All those calculations appear to describe the available scattering data similarly well,
whereas the details of the chosen framework, e.g. the
driving terms in the Bethe-Salpeter equation, differ in most cases.
To our knowledge no attempt has so far been made to compare the different approaches
systematically. To this end, we study in the present work several
variants of the coupled-channels approaches to the $\bar{K} N$ system
with different interaction kernels, hence
providing an estimate for the model dependence of such analyses.

The $\bar{K} N$ channel is of particular interest as a testing ground for 
chiral SU(3) symmetry in QCD and for the role of explicit chiral symmetry breaking
due to the relatively large strange quark mass. High-precision $K^- p$ threshold 
data set important constraints for theoretical approaches and have recently 
been supplemented by the new accurate results for the strong interaction shift and width
of kaonic hydrogen from the DEAR experiment \cite{DEAR} which reduced the mean values
and error ranges of the previous KEK experiment \cite{Ito}.
There is thus renewed interest in an improved analysis of these 
data along with existing information on $K^- p$ scattering, the $\pi\Sigma$ mass 
spectrum and $K^- p$ threshold decay ratios. Some results have already been presented
in \cite{BNW}.

The electromagnetic interaction is responsible for the binding of kaonic hydrogen 
and also contributes significantly in elastic $K^- p$ scattering close to threshold.
It must therefore be included in the investigation of the $K^- p$ system
and we study the importance of Coulomb and isospin breaking effects in the $\bar{K} N$ system.

Another topic of interest is the pole structure of the $\Lambda(1405)$ resonance.
In the context of coupled-channels approaches it has been argued that the $\Lambda(1405)$ 
is a superposition of {\it two} nearby poles \cite{OM, JOORM,GLN, GNRV}.
Both poles couple with different strengths to the $\pi \Sigma$ and $\bar{K} N$
channels, so that by performing experiments with different initial 
states the observed peak structure should change.
Photoproduction of $\Lambda(1405)$ has been studied at ELSA with the
SAPHIR detector at 2.6 GeV and in the charged decay channels,
$\pi^+ \Sigma^-$ and $\pi^- \Sigma^+$, at SPring-8/LEPS with incident photon energies
in the range 1.5 $< E_\gamma^{lab} <$  2.4 GeV \cite{Ahn}. 
Another upcoming experiment at ELSA with the Crystal
Barrel detector is the decay of the $\Lambda(1405)$ into the $\pi^0 \Sigma^0$ 
channel which provides a unique signature of $\Lambda(1405)$, since the
$\pi^0 \Sigma^0$ channel does not have an isospin $I=1$ component and hence does
not couple to the $\Sigma(1385)$ resonance.

On the theoretical side, the positions of the relevant poles in the
complex $\sqrt{s}$-plane have been studied in \cite{JOORM, GLN} by using only the lowest
order effective Lagrangian. We critically examine the changes
in the pole positions by including the contributions from the next-to-leading
order Lagrangian.

The present work is organized as follows. The effective Lagrangian and a short
description of the coupled-channels method is outlined in the next section.
In Sec.~3 the results of different coupled-channels approaches are compared
with experiment and the differences between these frameworks are highlighted.
Moreover, the additional tight constraints set by the new DEAR experiment are
emphasized. Coulomb and isospin breaking effects are discussed.
Sec.~4 summarizes our findings, while some technicalities are relegated to the appendix.

\section{Formalism}\label{sec:Form}

\subsection{Effective Lagrangian}

In this section, we briefly outline the coupled-channel formalism  
of meson-baryon scattering. It is based on the SU(3) chiral 
effective Lagrangian which incorporates the same symmetries and symmetry breaking 
patterns as QCD and describes the coupling of the pseudoscalar octet
$(\pi, K, \eta)$ to the ground state baryon octet $(N, \Lambda, \Sigma, \Xi)$.
The Lagrangian
\beq \label{Lagr}
\Lagr = \Lagr_\phi + \Lagr_{\phi B}
\eeq
includes the mesonic term $\Lagr_\phi$ up to second chiral order \cite{GL},
\beq \label{Lagrphi} 
\Lagr_\phi =  \frac{f^2}{4} \langle u_{\mu} u^{\mu} \rangle 
            + \frac{f^2}{4} \langle \chi_+ \rangle , 
\eeq
where $\langle \dots \rangle$ denotes the trace in flavor space.
The pseudoscalar meson octet $\phi$ is arranged in a matrix valued field
\beq \label{Uphi}
U(\phi) = u^2(\phi) = \exp{\left( \sqrt{2} i \frac{\phi}{f} \right)} 
\eeq
and $f$ is the pseudoscalar decay constant in the chiral limit. The quantity $U$ enters the 
Lagrangian in the combinations $u_\mu = i u^\dagger \partial_\mu U u^\dagger$
and $\chi_+ = 2 B_0 (u^\dagger \mathcal{M} u^\dagger + u \mathcal{M} u)$, 
the latter one involving explicit chiral symmetry breaking via the quark mass
matrix $\mathcal{M} = \diag{(m_u, m_d, m_s)}$, and 
$B_0 = - \bra{0} \bar{q} q \ket{0} / f^2$ relates to the order parameter of 
spontaneously broken chiral symmetry.

The second piece of the Lagrangian in Eq.~(\ref{Lagr}), $\Lagr_{\phi B}$,  describes the 
meson-baryon interactions and reads at lowest order \cite{K}
\beq \label{LagrphiB1}
\Lagr_{\phi B}^{(1)}  =  i \langle \bar{B} \gamma_{\mu} [D^{\mu},B] \rangle
                           - M_0 \langle \bar{B}B \rangle 
                         - \frac{1}{2} D \langle \bar{B} \gamma_{\mu}
                             \gamma_5 \{u^{\mu},B\} \rangle
                           - \frac{1}{2} F \langle \bar{B} \gamma_{\mu} 
                             \gamma_5 [u^{\mu},B] \rangle .
\eeq
The $3 \times 3$ matrix $B$ collects the ground state baryon octet, 
$M_0$ is the common baryon octet mass in the chiral limit and $D$, $F$
denote the axial vector couplings of the baryons to the mesons. Their numerical 
values can be extracted from semileptonic hyperon decays and
we employ the central values determined in \cite{CR}: $D = 0.80$, $F = 0.46$.
The covariant derivative of the baryon field is given by
\beq \label{CoDer}
[D_\mu, B] = \partial_\mu B + [ \Gamma_\mu, B]
\eeq
with the chiral connection
\beq
\label{Conn}
\Gamma_\mu = \sfrac{1}{2} [ u^\dagger,  \partial_\mu u] . 
\eeq
We also need
the next-to-leading order contribution to $\Lagr_{\phi B}$ which is given by
\beqa \label{LagrphiB2}
\Lagr_{\phi B}^{(2)} & = &   b_D \langle \bar{B} \{\chi_+,B\} \rangle
                           + b_F \langle \bar{B} [\chi_+,B] \rangle
                           + b_0 \langle \bar{B} B \rangle \langle \chi_+ \rangle \no \\ 
                     &   & + d_1 \langle \bar{B} \{u_{\mu},[u^{\mu},B]\} \rangle
                           + d_2 \langle \bar{B} [u_{\mu},[u^{\mu},B]] \rangle 
                           + d_3 \langle \bar{B} u_{\mu} \rangle \langle u^{\mu} B \rangle
                           + d_4 \langle \bar{B} B \rangle \langle u^{\mu} u_{\mu} \rangle ,
\eeqa
where only the pieces relevant for our analysis are displayed.
The values of the low-energy constants $b_i$ and $d_i$ utilized in this work
have been constrained also by the 
recent coupled-channel analysis for $\eta$ photoproduction \cite{BMW}. 
We will come back to this point later in Section \ref{sec:Res} and leave their values
undetermined for the moment.

\subsection{Coupled channels} \label{sec:cc}

For $\bar{K} N$ scattering chiral perturbation theory based on the Lagrangian from the 
preceding section
fails due to the presence of the nearby $\Lambda(1405)$ subthreshold resonance.
Unitarity effects from final state interactions are important and must be included
in a non-perturbative fashion.
To this end, one computes 
from the Lagrangian the relativistic tree
level amplitude $V_{j b, i a}(s, \Omega; \sigma, \sigma')$ of the meson-baryon 
scattering processes $\phi_i B_{a}^{\sigma} \to  \phi_j B_{b}^{\sigma'}$ (with spin 
indices $\sigma$, $\sigma'$). This amplitude is the driving term in the coupled-channels 
integral equation determining the meson-baryon T-matrix.

The effective meson-baryon Lagrangian, Eqs.~(\ref{LagrphiB1}, \ref{LagrphiB2}), has been
used at different levels of sophistication in the literature. While only the Weinberg-Tomozawa term
from the covariant derivative in Eq.~(\ref{LagrphiB1}) is taken, e.g., in \cite{OR}, the direct
and crossed Born terms are included in \cite{OM}. In \cite{KSW2} the Lagrangian
of second chiral order is added which yields additional contact interactions, whereas in 
\cite{BMW} the contact interactions and the direct Born term have been taken into account,
but the crossed Born term has been excluded.
In order to provide an estimate of the model-dependence of such approaches,
we will discuss four different choices for the amplitude 
$V_{j b, i a}(s, \Omega; \sigma, \sigma')$.

\begin{figure}
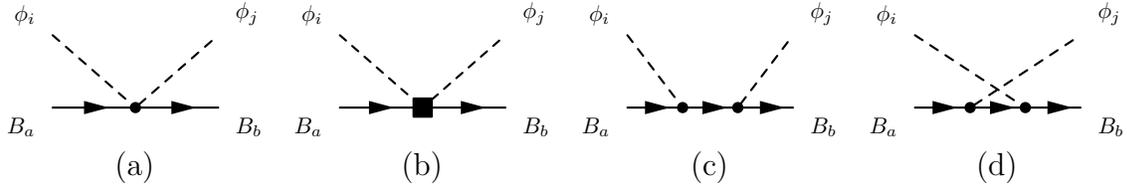

\centering
\begin{tabular}{cccc}
\includegraphics[width=0.2\textwidth]{feynps.1} & 
\includegraphics[width=0.2\textwidth]{feynps.2} &  
\includegraphics[width=0.2\textwidth]{feynps.3} &
\includegraphics[width=0.2\textwidth]{feynps.4} \\
(a) & (b) & (c) & (d)
\end{tabular}
\caption{Shown are the $\mathcal{O}(p^1)$ (a) and $\mathcal{O}(p^2)$ (b) contact 
         interactions as well as the direct (c) and crossed (d) Born terms. 
         Solid and dashed lines represent baryons and pseudoscalar mesons, respectively.}
\label{fig:feyns}
\end{figure}

First, only the leading order contact (Weinberg-Tomozawa) term is taken 
into account, see Figure~\ref{fig:feyns}a. 
Subsequently, the contact interactions from the Lagrangian of second chiral order,
$\Lagr_{\phi B}^{(2)}$, are included, see Fig.~\ref{fig:feyns}b. 
In the third and fourth approach we add successively the direct 
(Fig.~\ref{fig:feyns}c) and crossed (Fig.~\ref{fig:feyns}d) Born diagrams. 
For brevity, we will refer to these variants as ``WT'' (Weinberg-Tomozawa),
``c'' (additional contact terms), ``$s$'' (including $s$-channel Born
diagram) and ``$u$'' (including $u$-channel Born diagram), respectively.

It turns out that already the  inclusion of the 
next-to-leading order contact terms, which have been neglected in many previous coupled-channel analyses \cite{OR, OM, LK, JOORM}, improves the 
agreement of our results with the well-measured $K^- p$ threshold branching ratios
and the shape of the $\pi \Sigma$ mass spectrum, whereas the Born diagrams 
Fig.~\ref{fig:feyns}c,d yield only small numerical changes.
The explicit expressions for the diagrams, Figs.   
\ref{fig:feyns}a,c,d, can be found in \cite{OM}, but for completeness we display the formulae 
of all those contributions in the appendix.


Since we are primarily concerned with a narrow center-of-mass energy region around
the $\bar{K} N$ threshold, it is sufficient to restrict ourselves to the
$s$-wave (matrix) amplitude $V(s)$ given by
\beq \label{eq:prj_swave}
V(s) = \frac{1}{ 8 \pi} \sum_{\sigma} \int d \Omega \; V(s,\Omega;\sigma, \sigma) ,
\eeq
where we have averaged over the spin $\sigma = \pm 1/2$ of the baryons and $s$ is
the invariant energy squared.
We work in the physical basis assigning each particle its physical mass. This scheme 
produces the correct thresholds of the different channels, and it is 
consistent at the order at which the driving amplitudes $V$ are calculated.

For each partial wave $l$ unitarity imposes a restriction on
the (inverse) $T$-matrix above the pertinent thresholds
\beq \label{unit}
\mbox{Im} T^{-1}_l = - \frac{|\mathbf{q}_{cm}|}{8 \pi \sqrt{s}}
\eeq
with $\mathbf{q}_{cm}$ being the three-momentum in the
center-of-mass frame of the channel under consideration.
Hence
the imaginary part of $T_l^{-1}$ is identical with the imaginary part
of the basic scalar loop integral $\tilde{G}$ above threshold,
\beq
\tilde{G}(q^2) = \int \frac{d^d p}{(2 \pi)^d}
\frac{i}{[ (q-p)^2 - M_B^2 + i \epsilon]
   [ p^2 - m_\phi^2 + i \epsilon] } ~ ,
\eeq
where $M_B$ and $m_\phi$ are the physical masses of
the baryon and the meson, respectively \cite{OM, BMW, MO}.
In dimensional regularization we obtain for the finite part 
$G$ of $\tilde{G}$,
\beqa
G(q^2) & = & a({\mu}) + \frac{1}{32 \pi^2 q^2} \Bigg\{ q^2
             \left[ \ln\Big(\frac{m_\phi^2 }{\mu^2}\Big) +
             \ln\Big(\frac{M_B^2 }{\mu^2}\Big) -2 \right]\no \\
       &   & + (m_\phi^2 - M_B^2)  \ln\left(\frac{m_\phi^2 }{M_B^2}\right)
             - 8 \sqrt{q^2} \,|\mathbf{q}_{cm}| \ \mbox{artanh}
             \left(\frac{2 \sqrt{q^2} \ |\mathbf{q}_{\scriptstyle{cm}}|}{
             (m_\phi + M_B )^2 - q^2} \right) \Bigg\}\ ,
\eeqa
where $\mu$ is the regularization scale.
The subtraction constant $a({\mu})$ cancels the scale dependence of
the chiral logarithms and simulates higher order contributions with
the value of $a({\mu})$ depending on the respective channel.

To the order we are working the inverse of the $T$ matrix can be written as
(suppressing the subscript $l\ ( = 0)$ for brevity)
\beq \label{invers}
T^{-1} = V^{-1} + G \ ,
\eeq
which yields after inversion
\beq \label{V}
T = [1 + V \cdot G]^{-1} \; V .
\eeq
Eq. (\ref{V}) is a matrix equation with the diagonal matrix $G$
collecting the loop integrals in each channel.
This amounts to a summation of a bubble chain to all orders in the $s$-channel,
equivalent to solving a Bethe-Salpeter equation with $V$ as driving term,
where all momenta in $V$ are set to their on-shell values. This so-called 
on-shell scheme reduces the full Bethe-Salpeter equation to the simple matrix equation 
(\ref{V}). 

However, in the presence of the crossed Born term (Fig.~\ref{fig:feyns}d)  
this simplification must be treated with care due to the appearance of
unphysical subthreshold cuts. 
In the unphysical region below the threshold of a given channel 
the propagator of the intermediate baryon in the crossed Born term leads to divergences 
(in the $SU(2)$ case this fact is known as the 
nucleon cut \cite{Hoe}), which correspond to logarithmic singularities in the $s$-wave amplitude.
Within the coupled channel formalism the subthreshold cuts 
of heavier virtual meson-baryon pairs can contribute above
the thresholds of physical processes involving lighter meson-baryon systems.
Since this is an artifact of the 
on-shell formalism and would not be present in a full field theoretical 
calculation, we eliminate the unphysical subthreshold cuts by matching the contribution of the crossed 
Born diagram to a constant value below a certain invariant energy $\sqrt{s_0}$. 
We have convinced ourselves that our conclusions do not depend on 
the specific choice of $\sqrt{s_0}$ as long as it is not too close to the singularities.
As a matter of fact, the contribution of the crossed Born diagram to the full $s$-wave 
interaction kernel $V$ turns out to be numerically small.

\subsection{Coulomb interaction}

The Coulomb interaction has been shown to yield  significant
contributions to the elastic $K^- p$ scattering amplitude up to kaon laboratory momenta of
100-150 MeV/$c$ \cite{JackDal}. Close to $K^- p$ threshold the electromagnetic
interactions are thus important as well and should not be neglected
as in previous coupled channel calculations \cite{OR, KSW2, OM, LK, JOORM}.
The quantum-mechanical Coulomb scattering amplitude for point charges 
can be calculated exactly and reads \cite{LL}
\beq \label{eq:fCoul}
f_{K^- p \to K^- p}^{\textrm{coul}} = 
\frac{1}{2 \,\mathbf{q}_{cm}^2 \, a_B \sin^2{(\theta_{cm}/2)}} \ 
\frac{\Gamma(1 - i/(|\mathbf{q}_{cm}| \, a_B))}
     {\Gamma(1 + i/(|\mathbf{q}_{cm}| \, a_B))} \ 
\exp{\left[\frac{2 i}{|\mathbf{q}_{cm}| \, a_B} \ln{\sin{\frac{\theta_{cm}}{2}}} \right]} \ ,
\eeq
where $a_B = 84$~fm is the Bohr radius of the $K^- p$ system, while 
$\mathbf{q}_{cm}$ and $\theta_{cm}$ denote the center-of-mass three-momentum and 
scattering angle, respectively.
We account for the electromagnetic interaction by adding 
$f_{K^- p \to K^- p}^{\textrm{coul}}$ to the unitarized strong elastic $K^- p$ amplitude
\beq
f^{\textrm{str}}_{K^- p \to K^- p} = \frac{1}{8 \pi \sqrt{s}} 
                                     \ T_{K^- p \to K^- p}^{\textrm{str}}\ .
\eeq
The total elastic cross section is then obtained by performing the integration of $d\sigma/d\Omega = |f^{\textrm{coul}} + f^{\textrm{str}}|^2$ over the
center-of-mass scattering angle. Since this expression is divergent for forward scattering, a
cutoff for the scattering angle must be introduced. In the analysis of the scattering data \cite{Hum, Sak}, forward angles were suppressed by accepting only events with $\theta_{cm}$ larger than a minimum angle $\theta_{min}$. In practice the value employed in ref.~\cite{Hum, Sak} was $\cos{\theta_{min}} = 0.966$. We choose the same $\theta_{min}$ for a meaningful comparison with data. Some $K^- p$ angular distributions (though of very limited quality) were reported in ref.~\cite{Hum}. We have checked that our treatment of Coulomb effects reproduces the measured small-angle differential cross sections in the relevant momentum range. The dependence of our results on the infrared cutoff provided by $\theta_{min}$ will be discussed in the Section~\ref{sec:coul}.

The Coulomb potential vanishes at infinity as $1/r$ and leads to
an infrared divergent scattering amplitude 
for $\mathbf{q}_{cm} \to 0$.
In physical reality, however, the kaons are scattered 
off neutral hydrogen atoms rather than off protons and the range of the Coulomb 
interaction -- given by the Bohr radius of the hydrogen atom -- is therefore finite. Deviations from the 
pure Coulomb potential will be important, if the de Broglie wavelength of the kaons 
is of the order of the atomic radius, corresponding to kaon 
laboratory momenta of a few keV/$c$. The lowest experimentally accessible kaon momenta
are around 100~MeV/$c$, four orders of magnitude higher, so the electronic shielding of 
the Coulomb potential can be safely neglected.

Deviations from the point Coulomb scattering amplitude are expected when the wavelength of the 
incident kaon is comparable to the size of the proton. This translates into kaon momenta larger than
200 MeV/c. For such momenta $K^- p$ scattering is completely dominated by the strong interaction
since the Coulomb amplitude decreases as $1/\mathbf{q}_{cm}^2$. The corrections induced by finite size effects in the Coulomb amplitude are negligible in the relevant range of kaon energies. We will therefore work with the formula given in Eq.~(\ref{eq:fCoul}) combined with the small-angle cut as mentioned before.

\section{Results and Discussion}\label{sec:Res}

In this section we present and discuss the numerical results of our calculation. Low-energy
antikaon-nucleon scattering and reactions have been studied experimentally decades ago 
\cite{Hum, Sak, Kim, Kit, Eva, Cib}. The available data (admittedly with large errors) are mostly restricted to $K^-$ momenta above 100 MeV/c. On the other hand, there is 
the new and precise DEAR measurement of the strong interaction shift and width in kaonic 
hydrogen \cite{DEAR} as well as a similar recent analysis of the KEK collaboration
\cite{Ito}, which set constraints for the strong-interaction part of the elastic $K^- p$
amplitude at threshold. Further tight constraints are imposed by the accurately determined threshold branching ratios into the inelastic channels 
$\pi \Sigma$ and $\pi^0 \Lambda$ \cite{Now, Tov}:
\beqa \label{BRdef}
\gamma & = & \frac{\Gamma(K^- p \to \pi^+ \Sigma^-)}{\Gamma(K^- p \to \pi^- \Sigma^+)}
         = 2.36 \pm 0.04 , \no \\
R_c & = & \frac{\Gamma(K^- p \to \pi^+ \Sigma^- , \ \pi^- \Sigma^+)}
          {\Gamma(K^- p \to \textrm{\small all inelastic channels})}
      = 0.664 \pm 0.011 , \no \\
R_n & = & \frac{\Gamma(K^- p \to \pi^0 \Lambda)}{\Gamma(K^- p \to \textrm{\small neutral states})}
      = 0.189 \pm 0.015\, ,
\eeqa
and by the $\pi \Sigma$ invariant mass spectrum in the isospin $I = 0$ channel \cite{Hem}.

Our approach has six subtractions constants $a(\mu)$ in the different channels
and eight constants given within certain limited ranges: the decay constant $f$ and the 
higher order couplings $b_i$, $d_i$. As mentioned before, the $b_i$ and $d_i$
have been constrained by the analysis of \cite{BMW} which includes $\eta$ photoproduction as a
high quality data set. 
Since both pions and kaons are involved we choose to vary the decay constant
$f$ in the range given
by  its value in the chiral limit, $f = 88$~MeV \cite{GL2}, 
and the physical kaon decay constant $F_K = 112.7$~MeV.
Furthermore, only the ``$s$'' approach in our investigation, i.e. 
the one that involves the leading and next-to-leading order contact interactions as well as
the direct Born term, exactly coincides with the framework chosen in \cite{BMW}.
We can therefore expect moderate deviations in the numerical determination of the coupling 
constants from a fit to low-energy hadronic data.

In the first part of this section, we compare the four different approaches
described in the preceding section which follow from the successive inclusion of the 
diagrams in Fig.~\ref{fig:feyns} in the interaction kernel $V$.
It turns out that in all four cases the results cannot be brought into 
simultaneous satisfactory agreement with the elastic $K^- p$ elastic cross section 
and the kaonic hydrogen data from the DEAR experiment \cite{DEAR},
although the inclusion of the Coulomb interaction ameliorates the situation compared 
to previous coupled-channel calculations.
In order to examine how well the four approaches under consideration 
agree with the scattering data, we first 
exclude the DEAR data from the fit and ``predict''
the strong-interaction shift and width in kaonic hydrogen based on the rest 
of the low-energy scattering data. 
We note that the ``$u$'' approach yields the fit with the smallest overall $\chi^2$ value, but 
only slightly larger values are obtained in the ``c'' and ``$s$'' versions,
whereas the ``WT'' model, based only on the leading Weinberg-Tomozawa term, has a 
significantly larger $\chi^2$.
However, one should keep in mind that the ``WT'' approach has less parameters.

As a second step, we then investigate the changes of the results when the DEAR data are 
included in the fit. 
For this purpose it is sufficient to restrict ourselves to the ``$u$'' ansatz
involving all the diagrams in Fig.~\ref{fig:feyns}---i.e. the entire set of
next-to-leading order contributions to the $s$-wave amplitude---since 
qualitatively similar results are obtained in the ``c'' and ``$s$'' models.

The third part of this section is devoted to the detailed discussion of 
Coulomb corrections in the elastic $K^- p$ scattering cross section,
and we conclude with a study of the relevant resonance poles in the complex energy plane.

\subsection{Comparison of the different approaches}

We have first performed an overall $\chi^2$ fit to the available low energy $\bar{K} N$ data 
excluding the strong level shift and width of kaonic hydrogen. In order to emphasize 
the importance of the precisely measured threshold branching ratios, the
$\chi^2$ value of each observable has been divided by the number of pertinent
data points.
The resulting numerical 
values of the parameters are compiled in Table~\ref{Comppar}. We point out that the subtraction constants $a(\mu)$ translate into values close to $-2$ in the 
framework of \cite{OM} and are therefore---according to the authors---``of natural size''.
Our subtraction constants
in the important channels $\pi \Sigma$, $\bar{K} N$ 
are also roughly compatible with the numbers following from the matching condition to the 
crossed amplitude as advocated in \cite{LK}.
In the ``WT'' and ``c'' approaches, the value of the pseudoscalar decay constant $f$ tends
towards the physical kaon decay constant, whereas it is lowered by the inclusion of $s$- and 
$u$-channel Born terms. The low energy constants $b_i$, $d_i$ are roughly compatible with the 
numbers obtained from $\eta$ photoproduction by employing a closely related coupled
channel approach \cite{BMW}.
We note that the $b_i$ parameters in the ``u'' fit correspond at tree level to the $KN$
sigma terms $\sigma_{K N}^{(1)} (0) = 305$ MeV and
$\sigma_{K N}^{(2)} (0) = 181$ MeV. These numbers are in fair agreement with the values
for the tree level contributions presented in \cite{B} (including a $\pi N$ sigma term $\sigma_{\pi N} (0) \simeq 30$ MeV at tree level).

\begin{figure}
\centering
\begin{tabular}{p{0.5cm}rp{1.0cm}r}
 &
\begin{overpic}[height=0.20\textheight,clip]{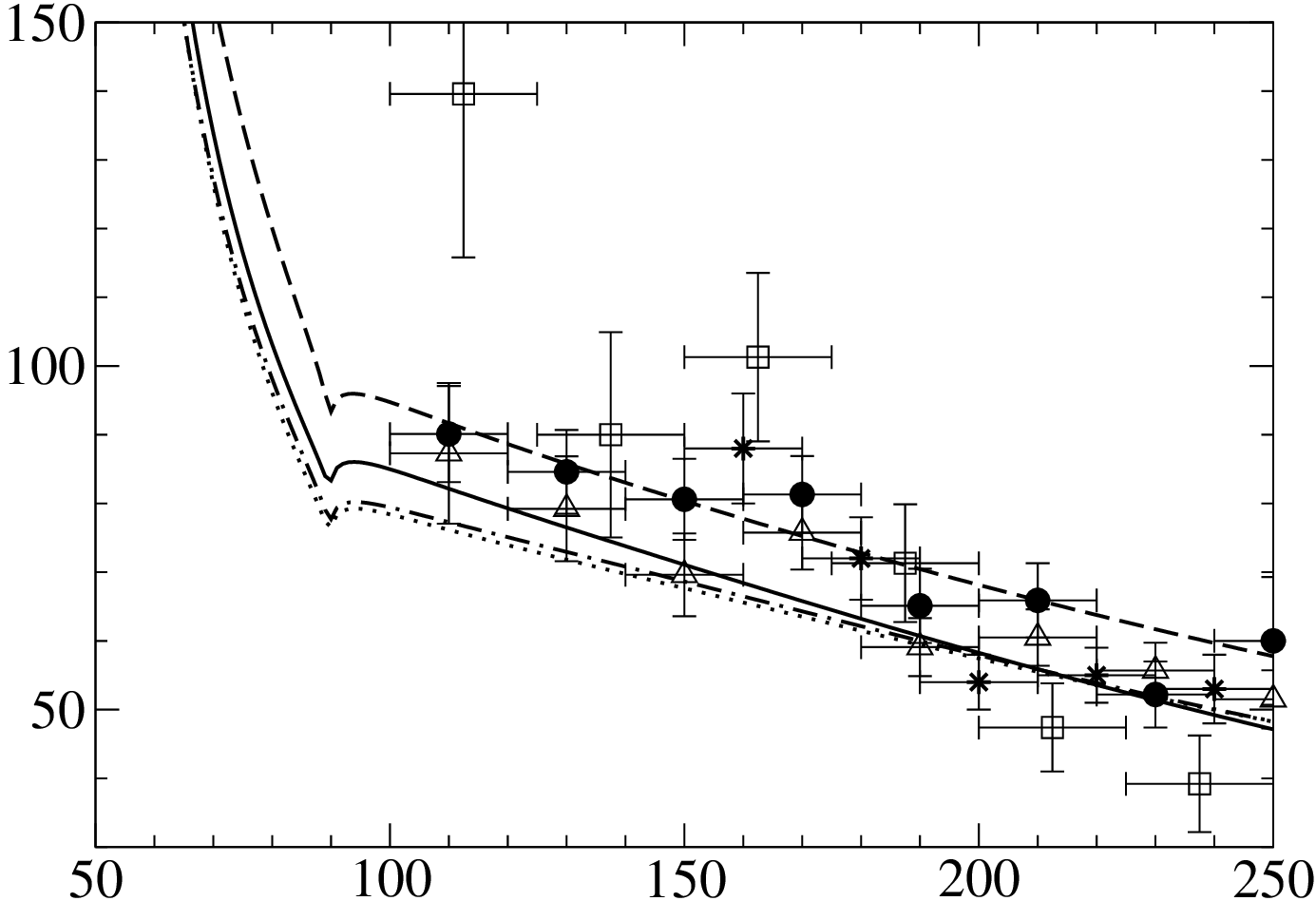}
  \put(-12,9){\rotatebox{90}{{\scalebox{0.9}{$\sigma [K^- p \to K^- p]$ (mb)}}}}
\end{overpic} & &
\begin{overpic}[height=0.20\textheight,clip]{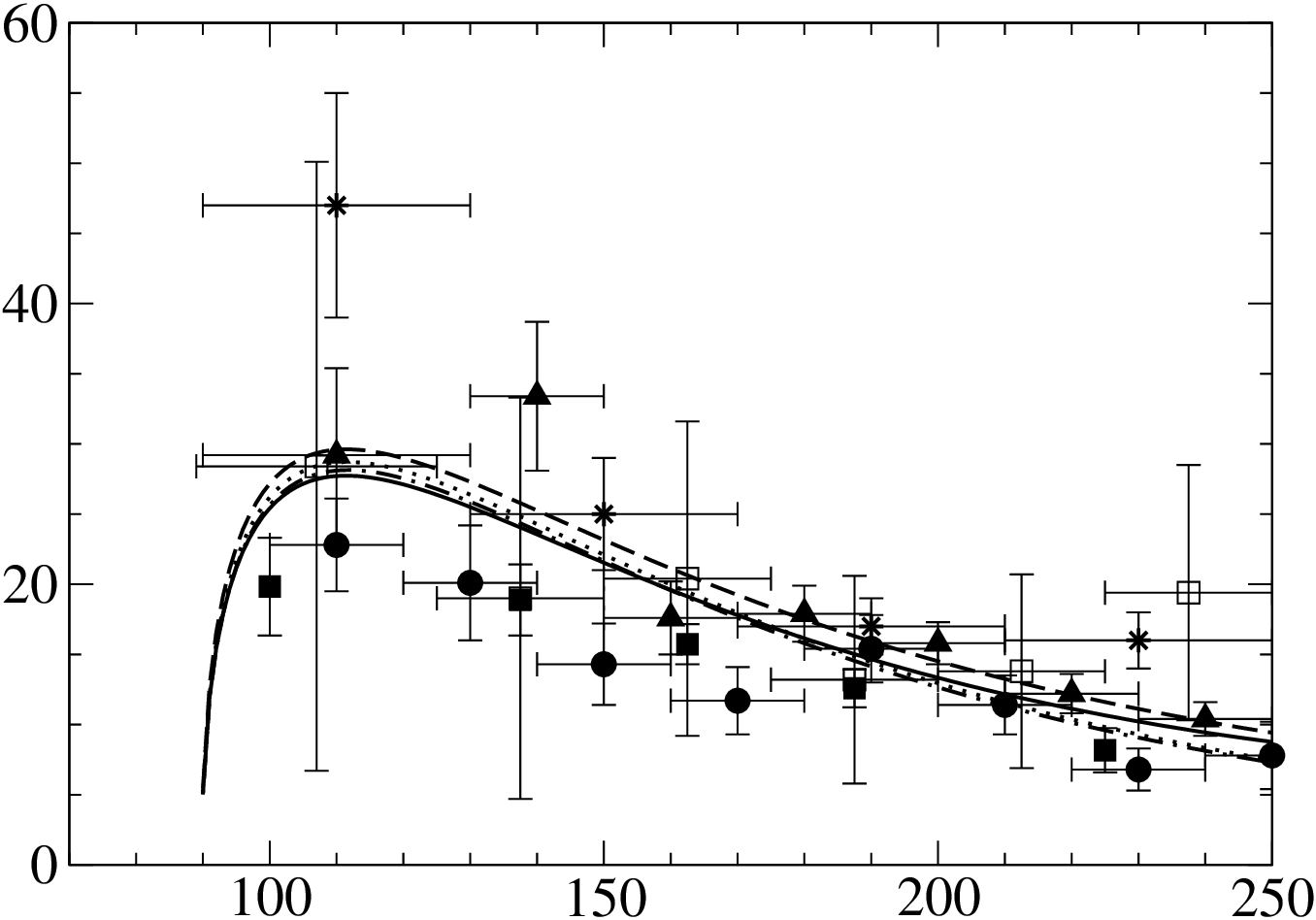}
  \put(-12,9){\rotatebox{90}{{\scalebox{0.9}{$\sigma [K^- p \to \bar{K}^0 n]$ (mb)}}}}
\end{overpic} \\[0.02\textheight]
 &
\begin{overpic}[height=0.20\textheight,clip]{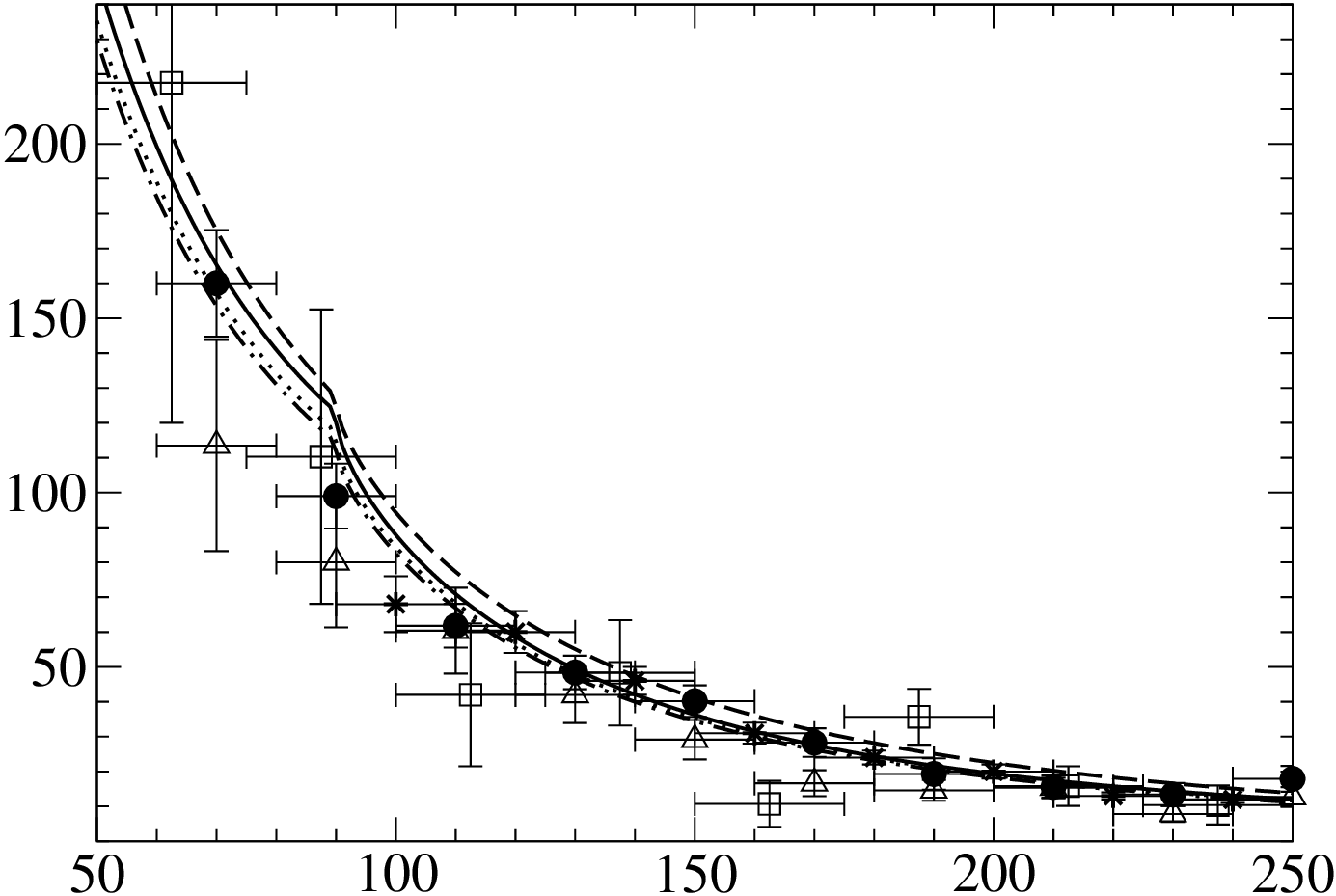}
  \put(-10,9){\rotatebox{90}{{\scalebox{0.9}{$\sigma [K^- p \to \pi^+ \Sigma^-]$ (mb)}}}}
\end{overpic} & &
\begin{overpic}[height=0.20\textheight,clip]{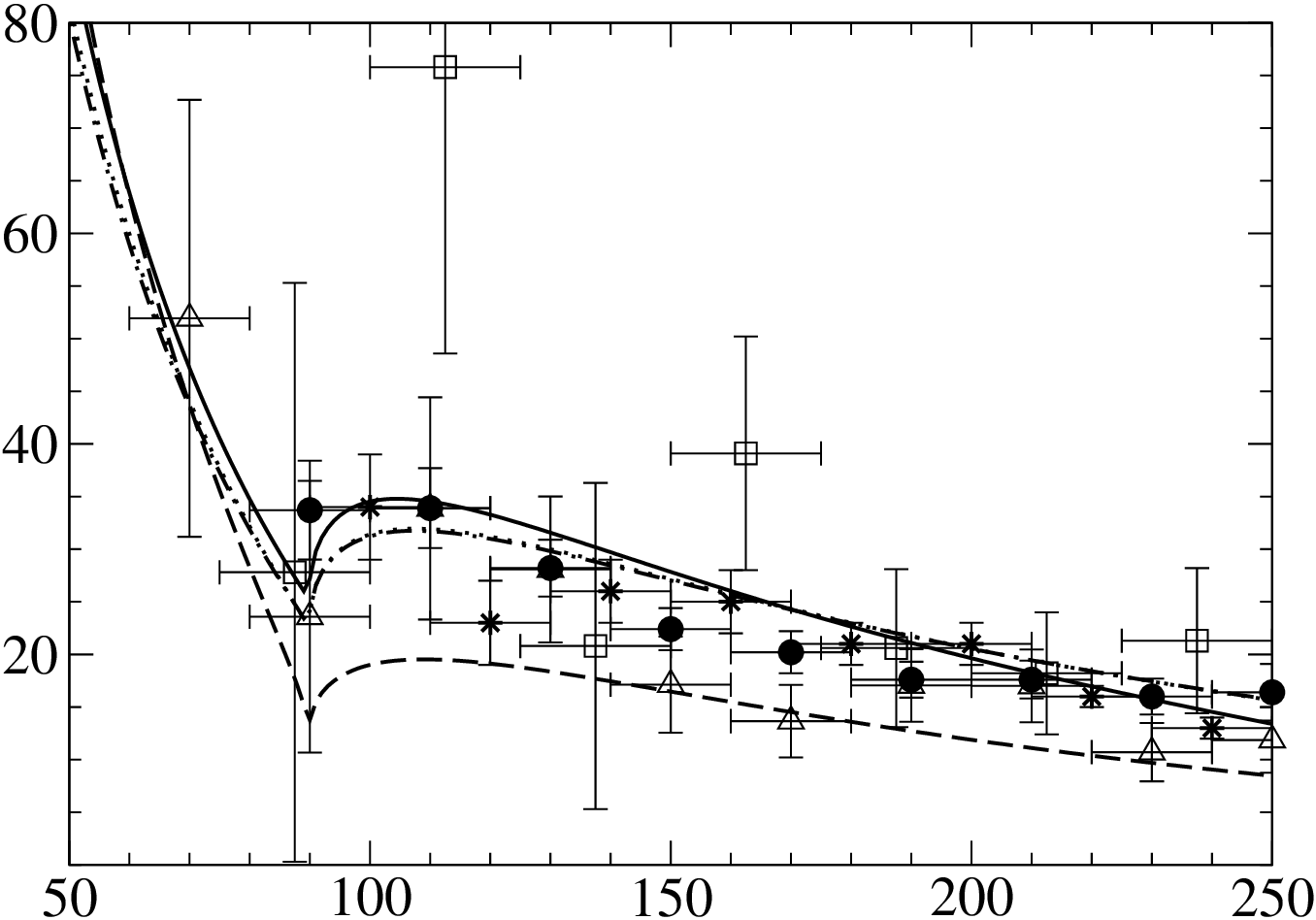}
  \put(-12,9){\rotatebox{90}{{\scalebox{0.9}{$\sigma [K^- p \to \pi^- \Sigma^+]$ (mb)}}}}
\end{overpic} \\[0.02\textheight]
 &
\begin{overpic}[height=0.20\textheight,clip]{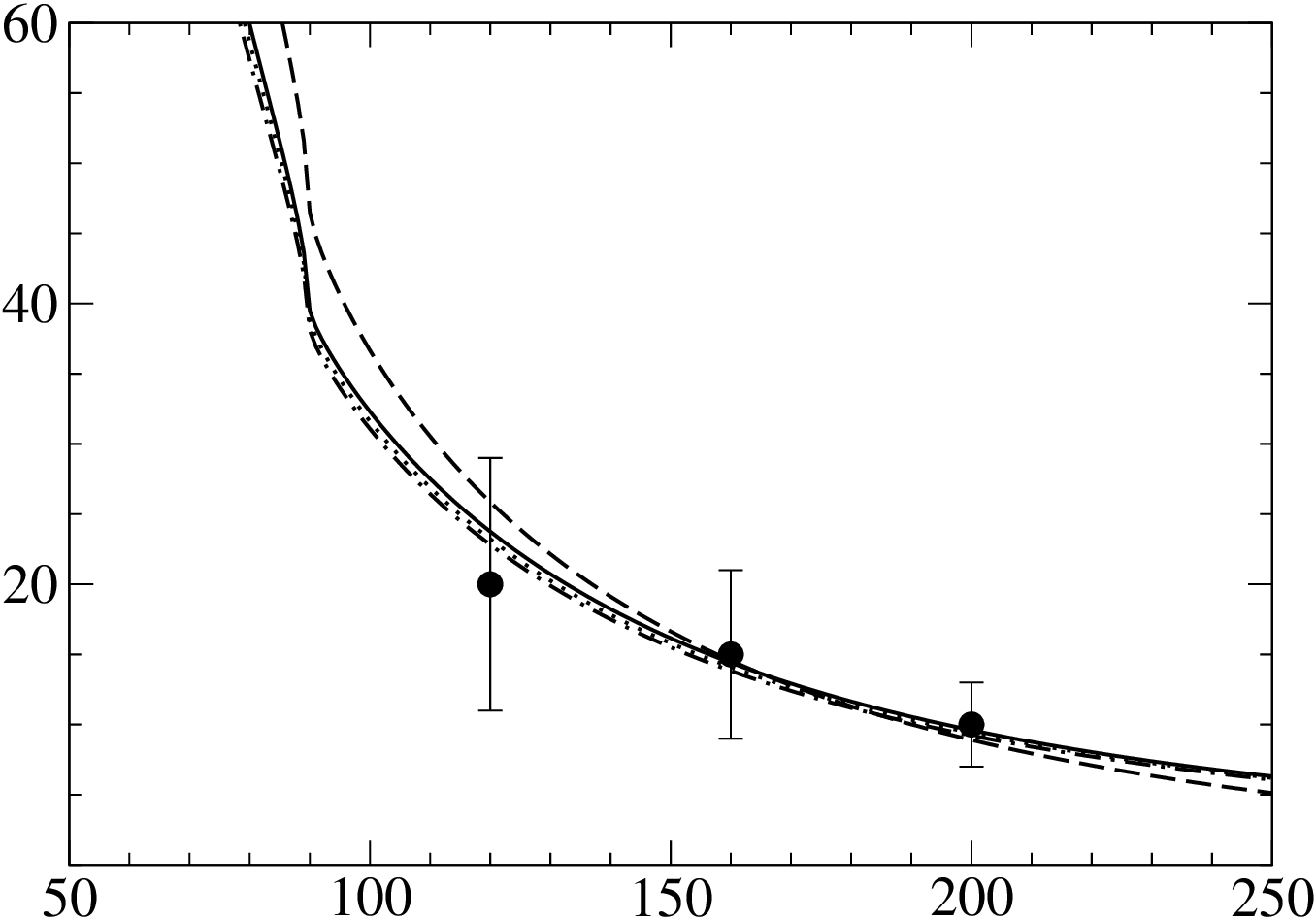}
  \put(-15,9){\rotatebox{90}{{\scalebox{0.9}{$\sigma [K^- p \to \pi^0 \Sigma^0]$ (mb)}}}}
  \put(5,-9){\scalebox{0.9}{incident $K^-$ lab.\  momentum (MeV/c)}}
\end{overpic} & &
\begin{overpic}[height=0.20\textheight,clip]{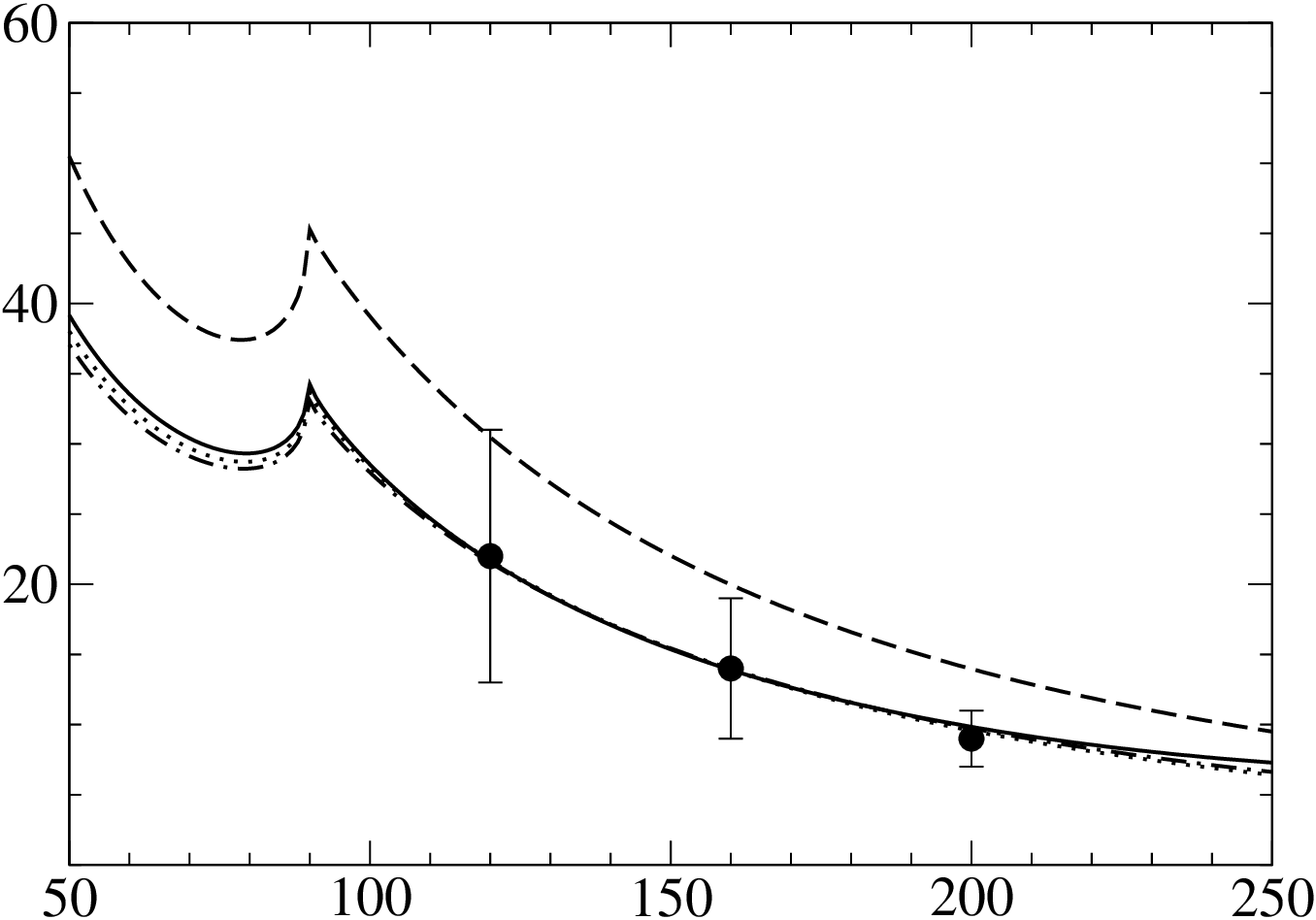}
  \put(-12,7){\rotatebox{90}{{\scalebox{0.9}{$\sigma [K^- p \to \pi^0 \Lambda]$ (mb)}}}}
  \put(5,-9){\scalebox{0.9}{incident $K^-$ lab.\  momentum (MeV/c)}}
\end{overpic}
\end{tabular}
\vspace{2.0ex}
\caption{Total cross sections for $K^- p$ scattering into various channels. The data
         are taken from  \cite{Hum} (empty squares), \cite{Sak} (empty triangles),
         \cite{Kim} (filled circles), \cite{Kit} (filled squares), 
         \cite{Eva} (filled triangles), \cite{Cib} (stars). The dashed, dotted, dot-dashed
         and solid lines correspond to the approaches ``WT'', ``c'', ``$s$'' and ``$u$'',
         respectively.}
\label{CompCS}
\end{figure}

\begin{figure}
\centering
\begin{tabular}{cp{1cm}c}
\begin{overpic}[width=0.40\textwidth,clip]{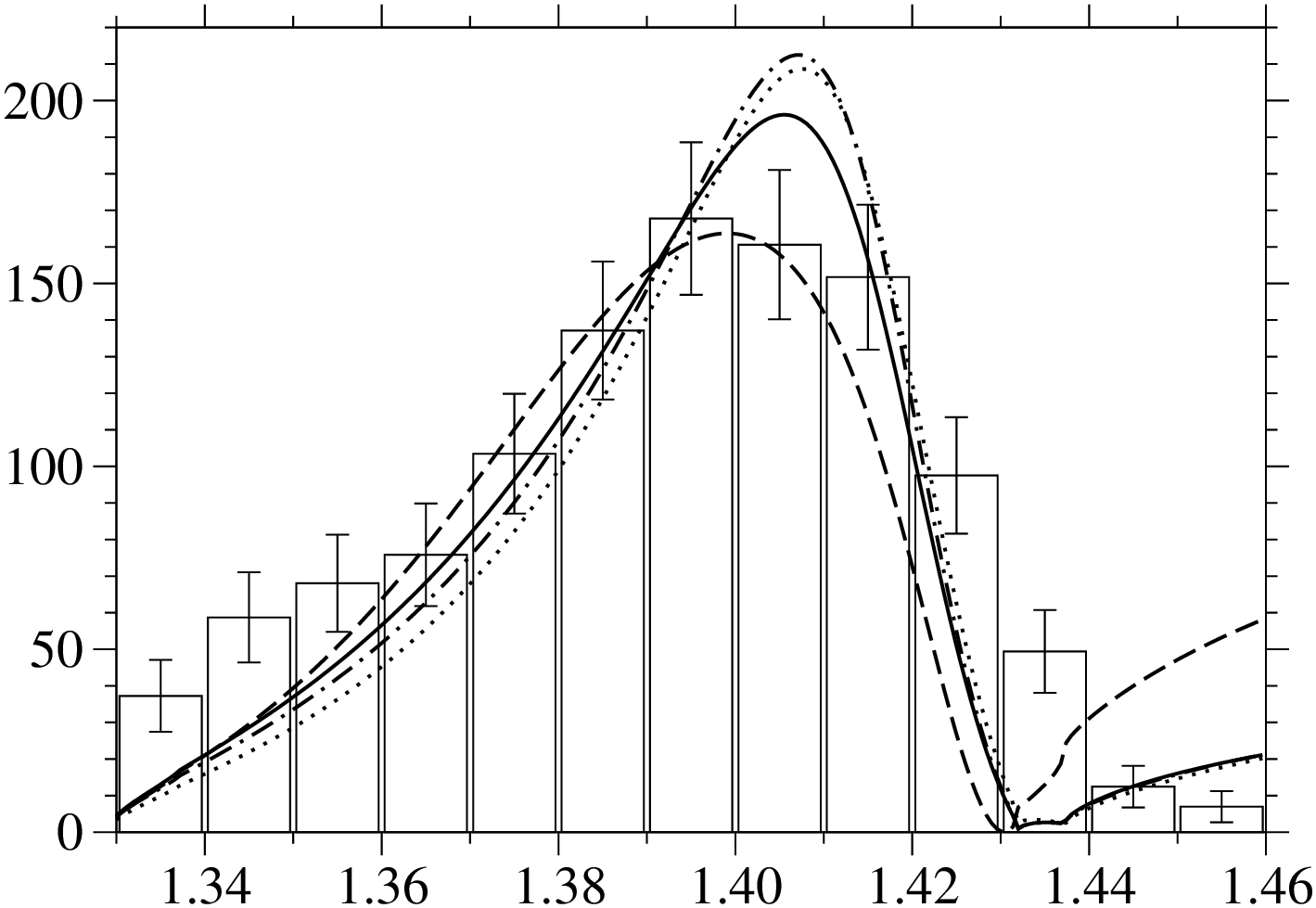}
  \put(-11,19){\rotatebox{90}{{\scalebox{0.9}{arbitrary units}}}}
  \put(40,-8){\scalebox{0.9}{$\sqrt{s}$ \quad (GeV)}}
\end{overpic}
& &
\begin{overpic}[width=0.40\textwidth,clip]{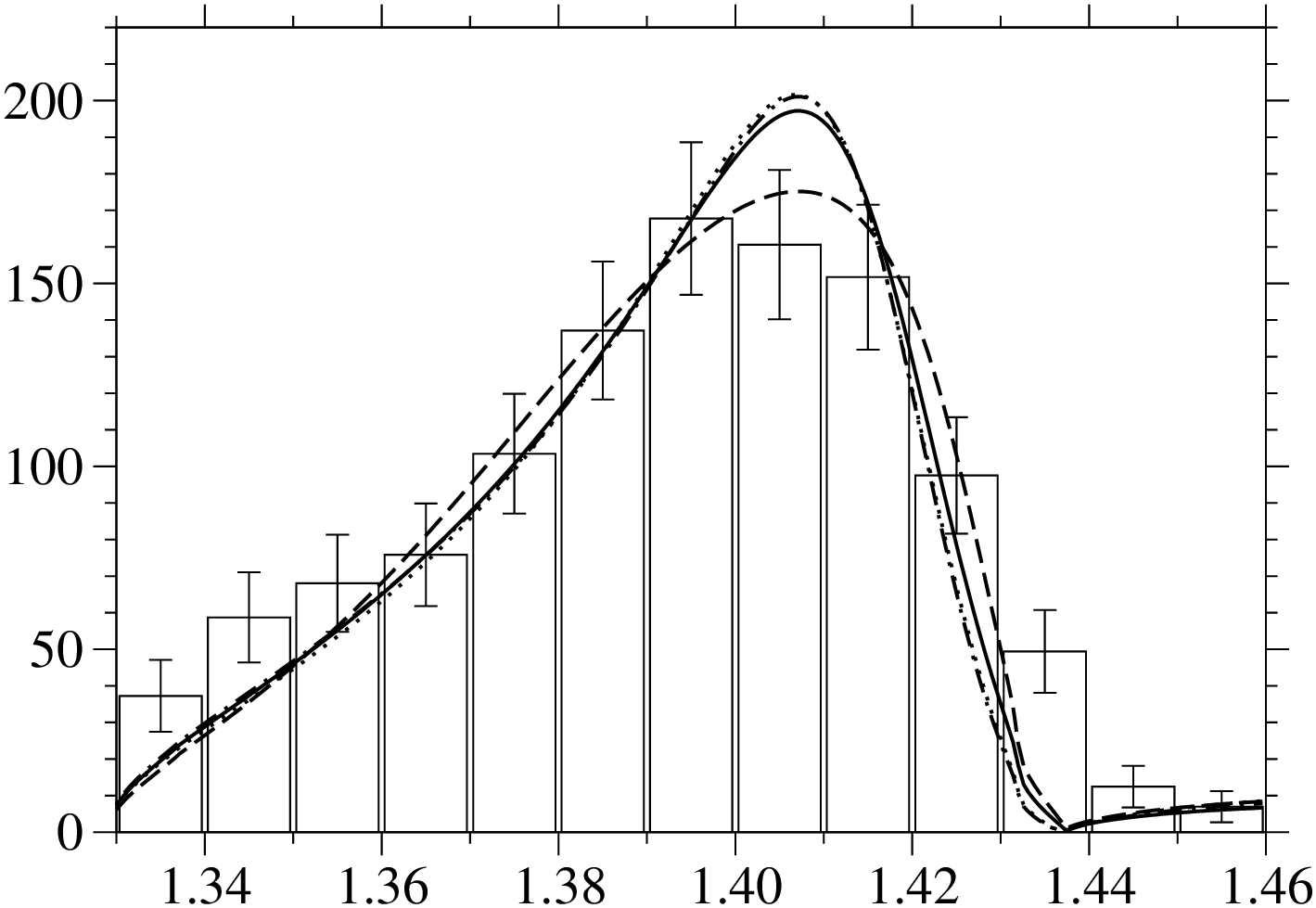}
  \put(-11,19){\rotatebox{90}{{\scalebox{0.9}{arbitrary units}}}}
  \put(40,-8){\scalebox{0.9}{$\sqrt{s}$ \quad (GeV)}}
\end{overpic}
\\[3.5ex]
(a) & & (b) \\
\end{tabular}
\caption{$\pi^- \Sigma^+$ event distribution from \cite{Hem},
         where statistical errors have been supplemented following \cite{DD}.
         The curves in diagram (a) where obtained by assuming a $\pi \Sigma$ invariant 
         mass spectrum with $I=0$; the curves in diagram (b) result from the ansatz
         advocated in \cite{OM}.
         The dashed, dotted, dot-dashed and solid lines correspond to the 
         ``WT'', ``c'', ``$s$'' and ``$u$'' approach, respectively.}
\label{CompPiSigma}
\end{figure}

In Fig.~\ref{CompCS} we show the results for the elastic and
inelastic cross sections of $K^- p$ scattering. The four lines correspond to 
the four different approaches under consideration, all of them in
good agreement with the experimental data.
The $\pi \Sigma$ mass spectrum in the isospin $I = 0$ channel is displayed in 
Fig.~\ref{CompPiSigma}a. It is well reproduced by the approaches  which include 
the additional $\mathcal{O}(p^2)$ contact terms (``c'', ``$s$'', ``$u$''), whereas the ``WT''
approach fails to explain the experimental data points for higher invariant energies.
Following ref.~\cite{OM} the experimental data displayed in 
Fig.~\ref{CompPiSigma} can be regarded as a $\pi^- \Sigma^+$ event distribution 
originating from a generic $s$-wave $I=0$ source. This source is assumed to be dominated 
by $\bar{K} N$ and $\pi \Sigma$ $I=0$ states which are multiplied by energy-independent 
coefficients $r_1$ and $r_2$, respectively. Since the experimental data 
are not normalized, only the ratio $r_1 / r_2$ is of significance. Utilizing this ansatz, 
we obtain the curves in Fig.~\ref{CompPiSigma}b and observe that now all four approaches reproduce the 
experimental spectrum almost equally well. But the number of free parameters has been
increased by one and---consequently---the quality 
of the fits has improved.
The ratios $r_1/r_2$ range between 1.40 for the ``WT'' and 0.96 for the ``c'' approach. 
Within this scenario 
an $I=0$ source with roughly equal portions of initial $\bar{K} N$ and $\pi \Sigma$ states thus 
seems to be favored by all schemes, with a tendency toward higher $\bar{K} N$ shares.

The results for the threshold branching ratios are compiled in Table~\ref{CompBR}. 
Independently of the chosen approach, the quantities $\gamma$ and $R_n$ agree well 
with the experimental numbers. For the branching ratio $R_c$ into charged final states the situation is different.
Whereas the ``c'', ``$s$'', and ``$u$'' fits are in perfect agreement with the
experimental error bars, the ``WT'' result happens to be too small in magnitude
emphasizing again the importance of $\mathcal{O}(p^2)$ contact terms.

\begin{figure}
\centering
\begin{overpic}[width=0.40\textwidth]{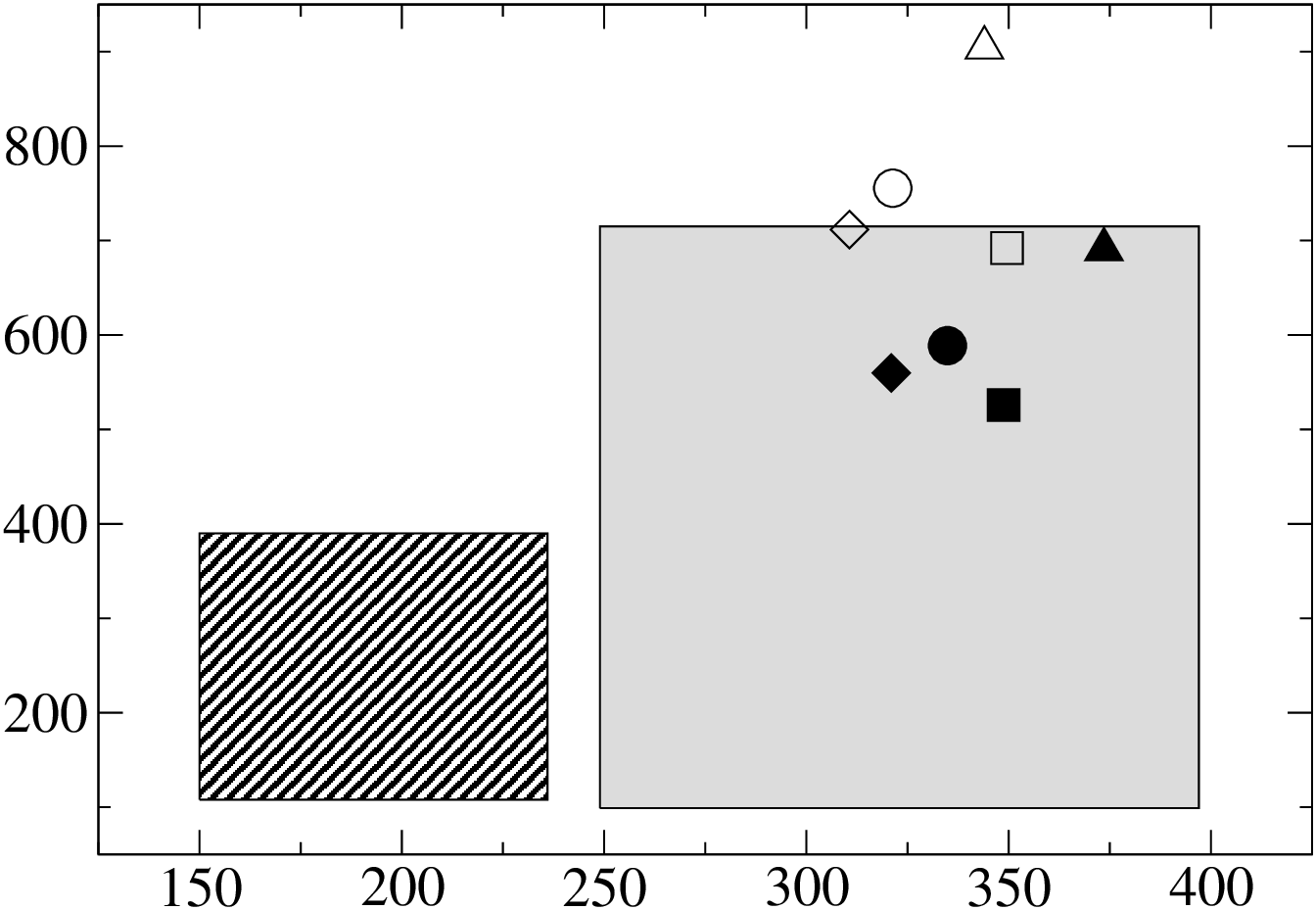}
  \put(-10,27){\rotatebox{90}{\scalebox{0.9}{$\Gamma$ (eV)}}}
  \put(45,-8){\scalebox{0.9}{$\Delta E^s$ (eV)}}
\end{overpic}
\vspace{1.5ex}
\caption{Predictions for the strong interaction shift and width of kaonic hydrogen
         from the different approaches both by using the Deser-Trueman formula, 
         Eq.~(\ref{eq:Des}), (empty symbols) and by including isospin breaking corrections, 
         Eq.~(\ref{eq:Rus}), (filled symbols).
         The ``WT'', ``c'', ``$s$'' and ``$u$'' approach is depicted by triangles, diamonds,
         squares and circles, respectively.
         The DEAR data are represented by the shaded
         box \cite{DEAR}, the KEK data by the light gray box \cite{Ito}.}
\label{CompDEAR}
\end{figure}

Having so far omitted the strong interaction shift and 
width in kaonic hydrogen from the fits, we can now predict these observables for the 
different approaches. To this end we employ the result of \cite{MRR} relating the ground state strong energy shift $\Delta E$ and width $\Gamma$
of kaonic hydrogen to the $K^- p$ scattering length $a_{K^- p}$ in the presence of
electromagnetic corrections:
\beq \label{eq:Rus}
\Delta E - \frac{i}{2} \Gamma = -2 \alpha^3 \mu_{c}^2 a_{K^- p} \ 
    [1 - 2 \alpha \mu_{c} (\ln{\alpha} -1) a_{K^- p}] \ .
\eeq
The reduced mass of the $K^- p$ system is denoted by $\mu_c$, $\alpha$ is the 
fine-structure constant, and the scattering length $a_{K^- p}$ is given by the strong 
interaction $T$ matrix at threshold
\beq
a_{K^- p} = \frac{1}{8 \pi \sqrt{s}} \ T_{K^- p \to K^- p}(s) |_{s = (m_{K^-} + M_p)^2} \ .
\eeq
In order to demonstrate the importance of the electromagnetic corrections calculated 
in \cite{MRR}, we compare Eq. (\ref{eq:Rus}) with the predictions derived from the well-known
Deser-Trueman formula \cite{DT}
\beq \label{eq:Des}
\Delta E - \frac{i}{2} \Gamma = -2 \alpha^3 \mu_{c}^2 a_{K^- p} 
\eeq
and the kaonic hydrogen data from the DEAR \cite{DEAR}
and KEK \cite{Ito} experiments in Fig.~\ref{CompDEAR}; the pertinent numerical values are displayed in
Table~\ref{CompShiWi}. The shifts and widths 
corresponding to the different approaches agree all with the error ranges given in 
\cite{Ito} if Eq.~(\ref{eq:Rus}) is utilized. In contrast, both the shift and the 
width of the new DEAR experiment cannot be accommodated by the coupled-channels approaches constrained by scattering and reaction cross sections,
although the electromagnetic corrections given in \cite{MRR} reduce the 
width $\Gamma$ by a significant amount.
As can be seen in Fig.~\ref{CompDEAR} the disagreement is reduced by the inclusion of 
higher order contact terms (approaches ``c'', ``$s$'', ``$u$'').

In summary we note that the approaches which include the $\mathcal{O}(p^2)$ contact terms
(``c'', ``$s$'', ``$u$'') describe all available low energy hadronic scattering
data excluding kaonic hydrogen experiments at DEAR. The fits to the $K^- p \to \pi^- \Sigma^+$ cross 
section and the (related) branching ratio $R_c$ which we obtain within the ``WT'' approach
are not of the same high quality. In this case a decent description of the $\pi \Sigma$
mass spectrum in Fig.~\ref{CompPiSigma} can only be achieved by utilizing the ad-hoc parametrization
suggested in \cite{OM}, but not by a simple isospin zero $\pi \Sigma$ invariant mass
distribution. One should keep in mind however that the leading order ``WT'' framework is oversimplified. It does not involve the coupling constants $b_i$, $d_i$ which turn out to be important in the more complete approaches.


\begin{table}
\centering
\begin{tabular}{|ll|c|c|c|c|}
\hline
 & & ``WT'' & ``c'' & ``$s$'' & ``$u$'' \\
\hline
$a_{\bar{K} N}$ & $(10^{-3})$ & -0.38 & -1.64 & -2.13 & -2.16 \\
\hline
$a_{\pi \Lambda}$& $(10^{-3})$& 0.21 & 5.43 & -2.32 & -6.34 \\
\hline
$a_{\pi \Sigma}$& $(10^{-3})$& 2.69 & -1.01 & -2.16 & -1.24 \\
\hline
$a_{\eta \Lambda}$& $(10^{-3})$& 4.73 & 2.36 & -0.53 & -1.99 \\
\hline
$a_{\eta \Sigma}$& $(10^{-3})$& 5.56 & 1.72 & 3.55 & -2.75 \\
\hline
$a_{K \Xi}$& $(10^{-3})$& -4.38 & 2.91 & 0.32 & -4.37 \\
\hline
$f$& (MeV) & 111.2 & 111.6 & 103.6 & 103.3 \\
\hline
$b_0$& (GeV$^{-1}$) & --- & -0.24 & -0.27 & -0.31 \\
\hline
$b_D$& (GeV$^{-1}$) & --- & 0.03 & 0.00 & 0.00 \\
\hline
$b_F$& (GeV$^{-1}$) & --- & -0.02 & -0.12 & -0.13 \\
\hline
$d_1$& (GeV$^{-1}$) & --- & -0.15 & -0.15 & -0.16 \\
\hline
$d_2$& (GeV$^{-1}$) & --- & 0.11 & 0.11 & 0.12 \\
\hline
$d_3$& (GeV$^{-1}$) & --- & 0.28 & 0.31 & 0.25 \\
\hline
$d_4$& (GeV$^{-1}$) & --- & -0.32 & -0.31 & -0.23 \\
\hline
\end{tabular}
\caption{Numerical values of the parameters for the different approaches.
         The subtraction constants are taken at $\mu = 1$~GeV.}
\label{Comppar}
\end{table}

\begin{table}
\centering
\begin{tabular}{|c|c|c|c|c||c|}
\hline
 & ``WT'' & ``c'' & ``$s$'' & ``$u$'' & Exp. \cite{Now,Tov} \\
\hline
$\gamma$ & 2.35  & 2.36  & 2.36  & 2.36  & $2.36 \pm 0.04$ \\
\hline
$R_c$    & 0.635 & 0.655 & 0.655 & 0.661 & $0.664 \pm 0.011$ \\
\hline
$R_n$    & 0.203 & 0.189 & 0.187 & 0.189 & $0.189 \pm 0.015$ \\
\hline
\end{tabular}
\caption{Threshold branching ratios as defined in the text,
         resulting from the different approaches.}
\label{CompBR}
\end{table}

\begin{table}
\centering
\begin{tabular}{|ll|c|c|c|c|}
\hline
& & ``WT'' & ``c'' & ``$s$'' & ``$u$'' \\
\hline
$a_{K^{-} p}$ & (fm) & $-0.83+1.10i $ & $-0.75+0.86i $ & $-0.85+0.84i $ & $-0.78+0.92i $ \\
\hline
$\Delta E_{D}$ & (eV) & 344 & 311 & 350 & 321 \\
\hline
$\Gamma_{D}$ & (eV) & 904 & 712 & 692 & 755 \\
\hline
$\Delta E_{c}$ & (eV) & 374 & 321 & 349 & 335 \\
\hline
$\Gamma_{c}$ & (eV) & 691 & 560 & 526 & 589 \\
\hline
\end{tabular}
\caption{Shown are the $K^- p$ scattering lengths $a_{K^- p}$ as well as the
         strong interaction shift $\Delta E$ and width $\Gamma$ in kaonic hydrogen
         resulting from
         Eq.~(\ref{eq:Des}) (subscript $D$) and Eq.~(\ref{eq:Rus}) (subscript $c$).}
\label{CompShiWi}
\end{table}

\subsection{Inclusion of the DEAR data}

\begin{figure}
\centering
\begin{overpic}[width=0.40\textwidth]{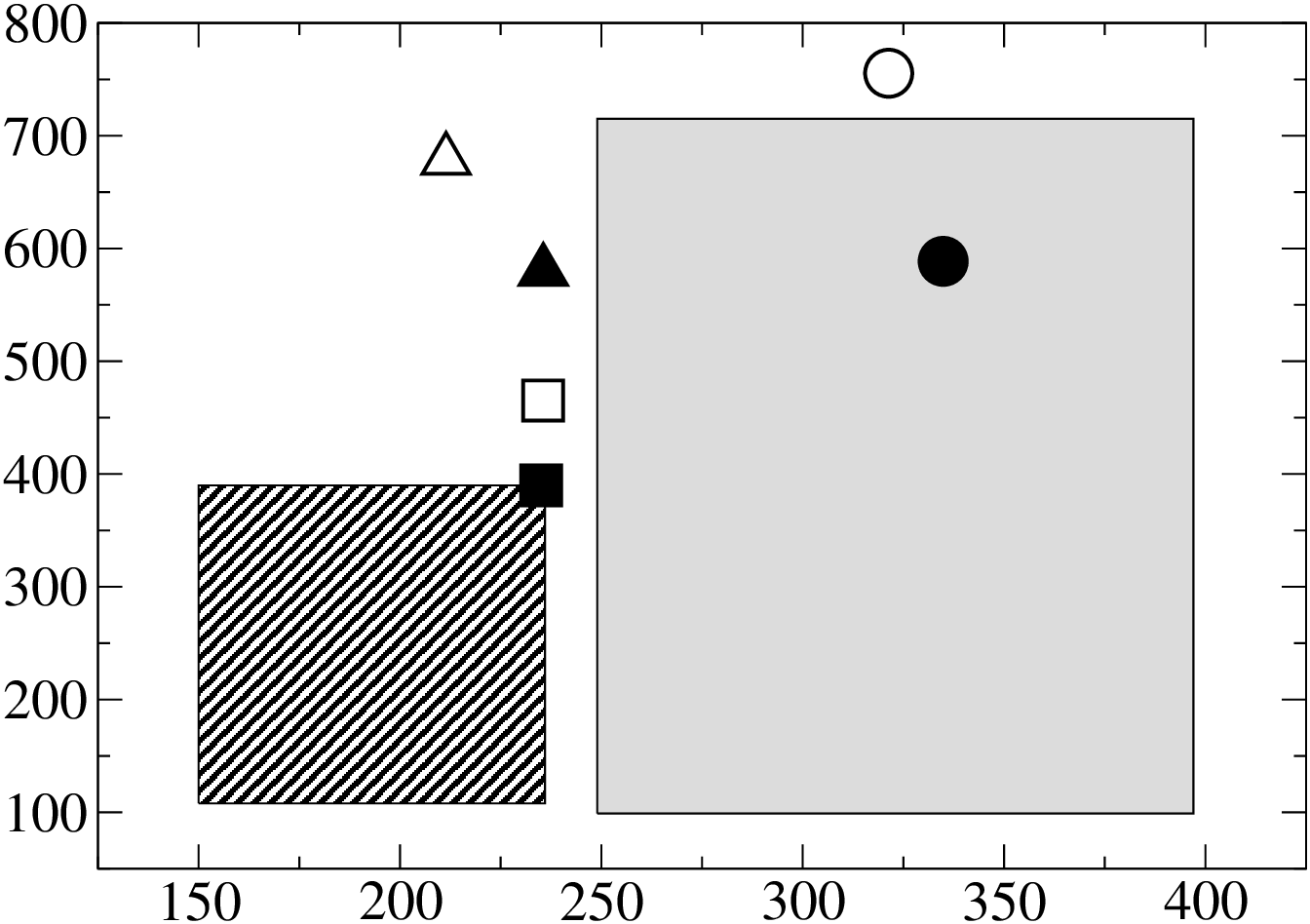}
  \put(-10,27){\rotatebox{90}{\scalebox{0.9}{$\Gamma$ (eV)}}}
  \put(45,-8){\scalebox{0.9}{$\Delta E^s$ (eV)}}
\end{overpic}
\vspace{1.5ex}
\caption{Results for the strong interaction shift and width of kaonic hydrogen
         from the fits ``1'', ``2'' and ``3'' depicted by circles, triangles and squares. 
         Empty symbols correspond to the Deser-Trueman formula, Eq.~(\ref{eq:Des}),
         full symbols to Eq.~(\ref{eq:Rus}), where isospin breaking corrections are 
         included. 
         The DEAR data are represented by the shaded
         box \cite{DEAR}, the KEK data by the light gray box \cite{Ito}.}
\label{UDEARDEAR}
\end{figure}

As already mentioned, the new high-precision DEAR data \cite{DEAR} set additional tight 
constraints on 
$\bar{K} N$ interactions. In this section we explore changes of our results when
the DEAR data are included in the fit. For brevity we restrict ourselves to the 
discussion of the ``$u$'' scheme, the one that has turned out most successful in the previous steps. Apart from yielding the least overall $\chi^2$ fit in the 
previous section, it includes the full set of next-to-leading order contributions to the
interaction kernel. We utilize this ``$u$'' fit and rename it ``1''. Forcing the fit to strictly remain 
within the error range given by the DEAR experiment we obtain result ``3'', see 
Fig.~\ref{UDEARDEAR}. 
The detailed numbers can be found in Table~\ref{UDEARShiWi}.
Fit ``2'' represents a compromise 
between fits ``1'' and ``3'' in the presence of the DEAR data.
(Note, however, that fit ``2" is not unique; and is presented here  only to illustrate
the changes seen in $K^- p$ scattering processes when the DEAR data are approached from the initial
fit ``1''.)
The numerical values of the parameters for the different fits are collected in Table~\ref{UDEARpar}.

\begin{figure}
\centering
\begin{tabular}{p{0.5cm}rp{1.0cm}r}
 &
\begin{overpic}[height=0.20\textheight,clip]{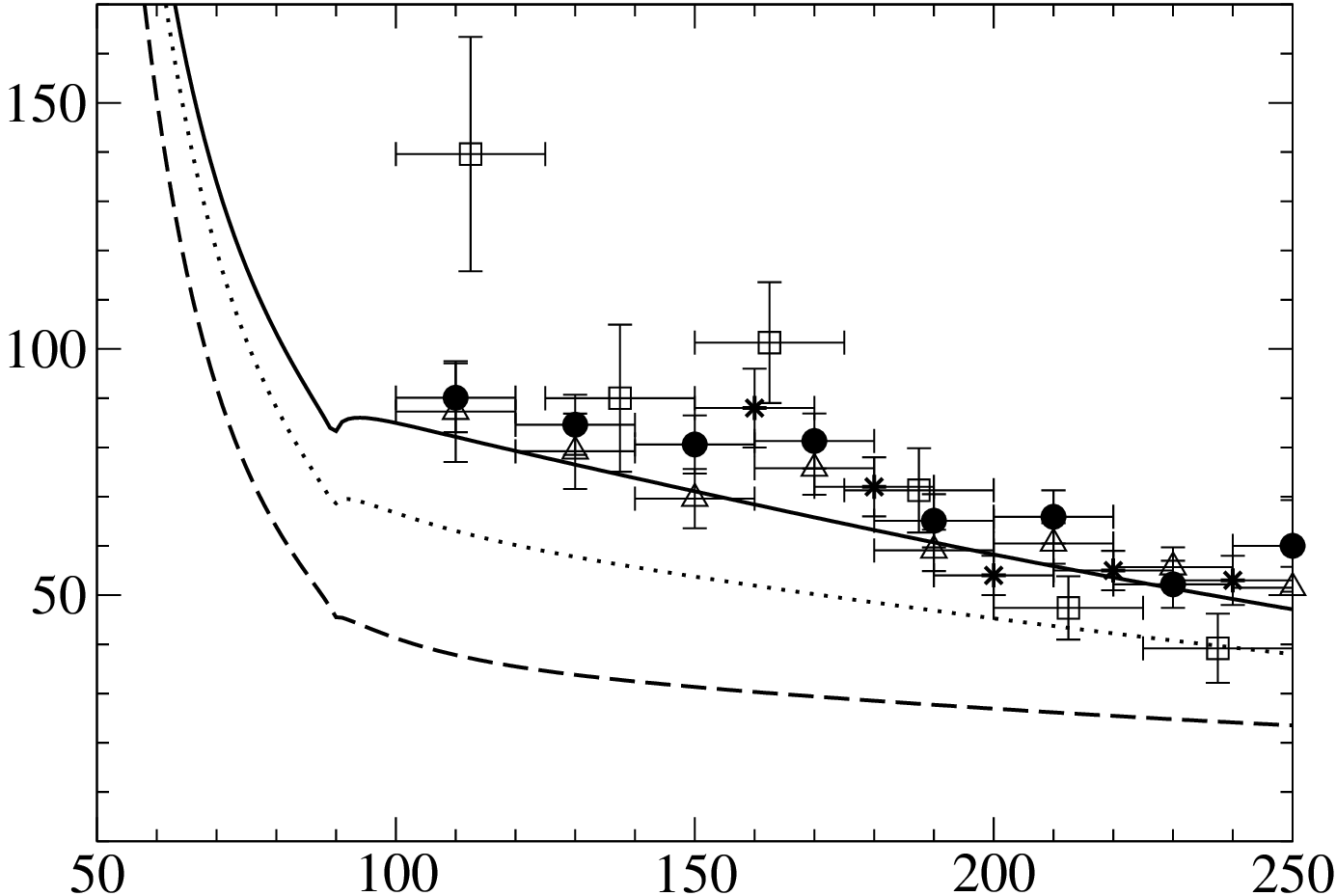}
  \put(-12,9){\rotatebox{90}{{\scalebox{0.9}{$\sigma [K^- p \to K^- p]$ (mb)}}}}
\end{overpic} & &
\begin{overpic}[height=0.20\textheight,clip]{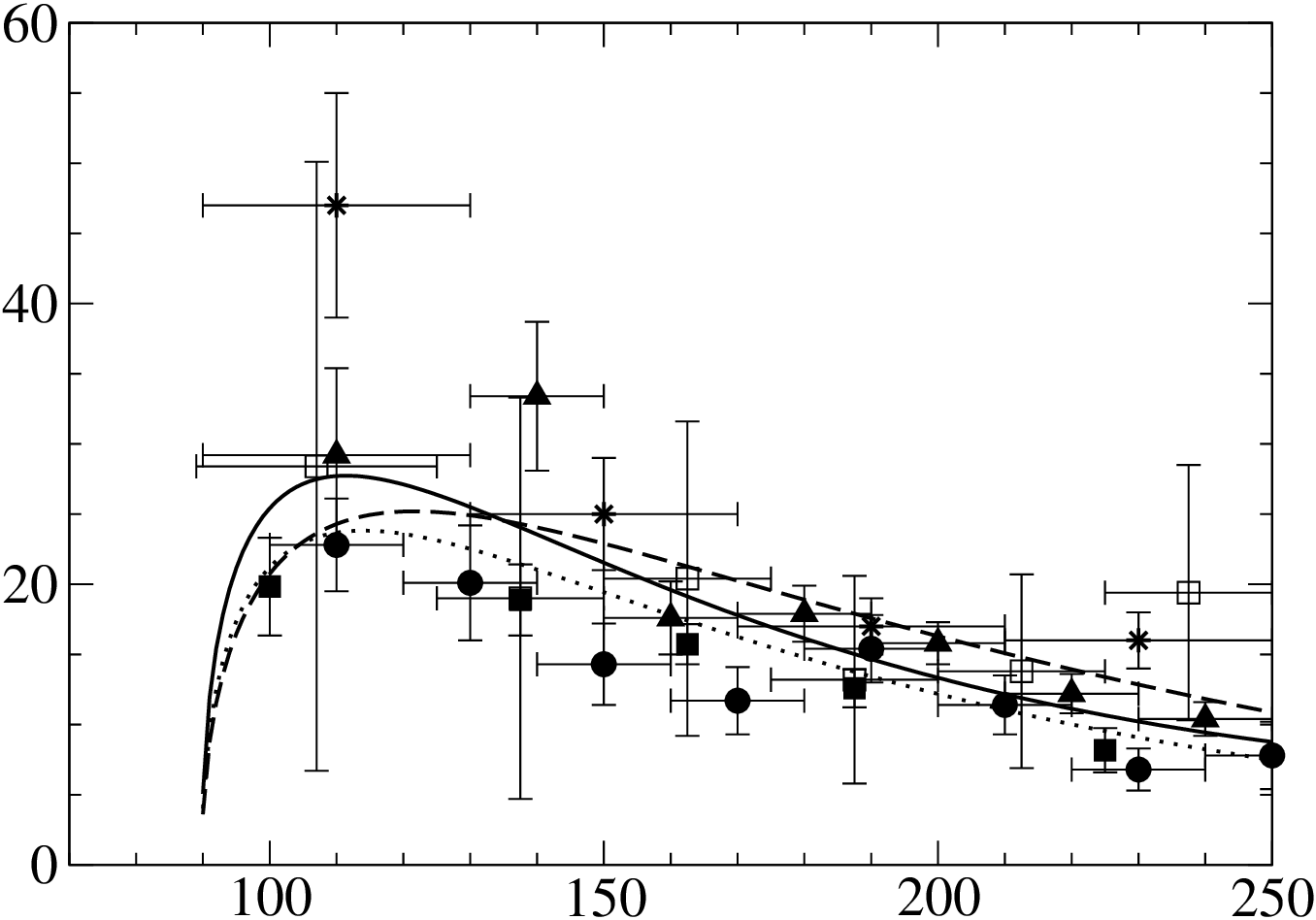}
  \put(-12,9){\rotatebox{90}{{\scalebox{0.9}{$\sigma [K^- p \to \bar{K}^0 n]$ (mb)}}}}
\end{overpic} \\[0.02\textheight]
 &
\begin{overpic}[height=0.20\textheight,clip]{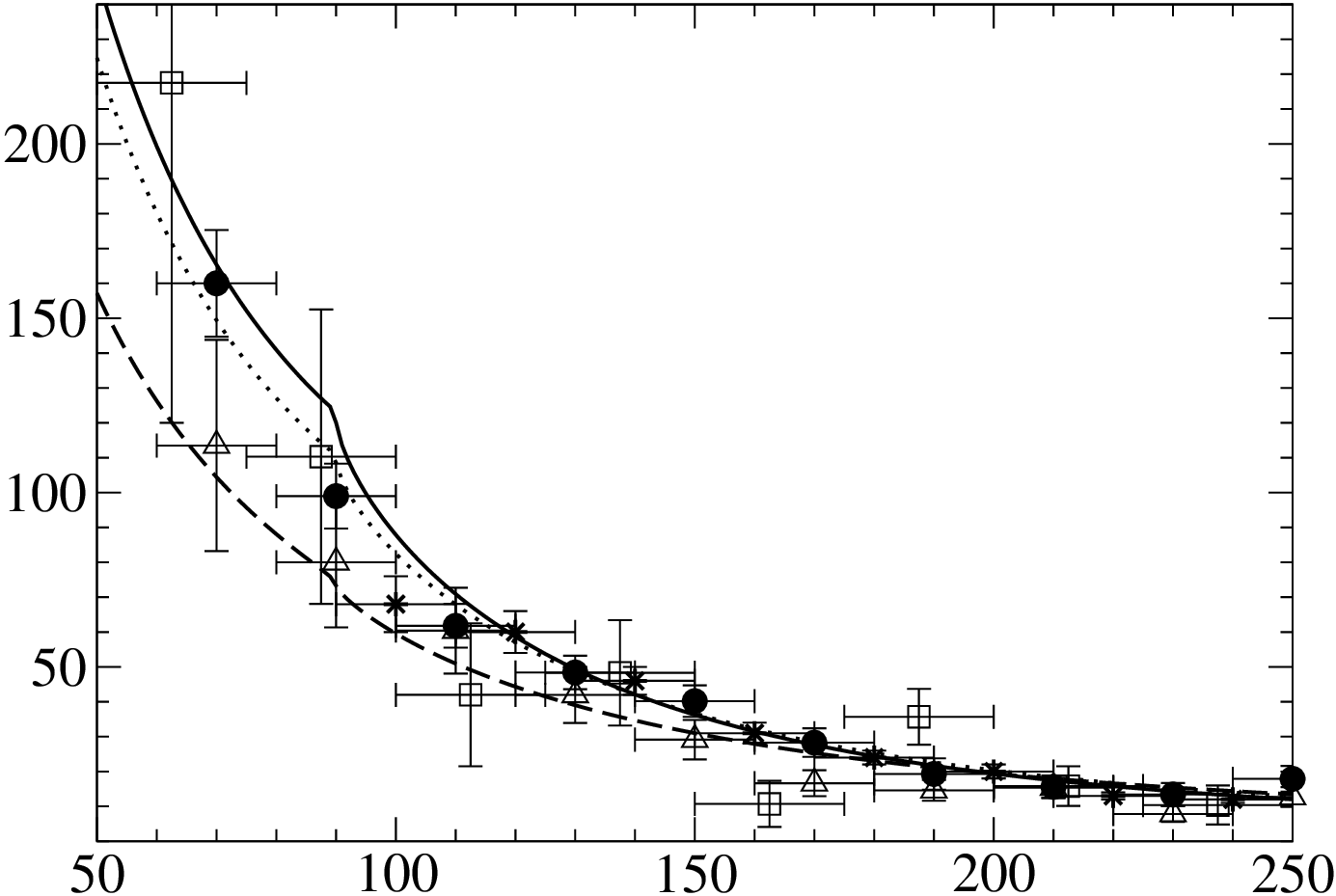}
  \put(-10,9){\rotatebox{90}{{\scalebox{0.9}{$\sigma [K^- p \to \pi^+ \Sigma^-]$ (mb)}}}}
\end{overpic} & &
\begin{overpic}[height=0.20\textheight,clip]{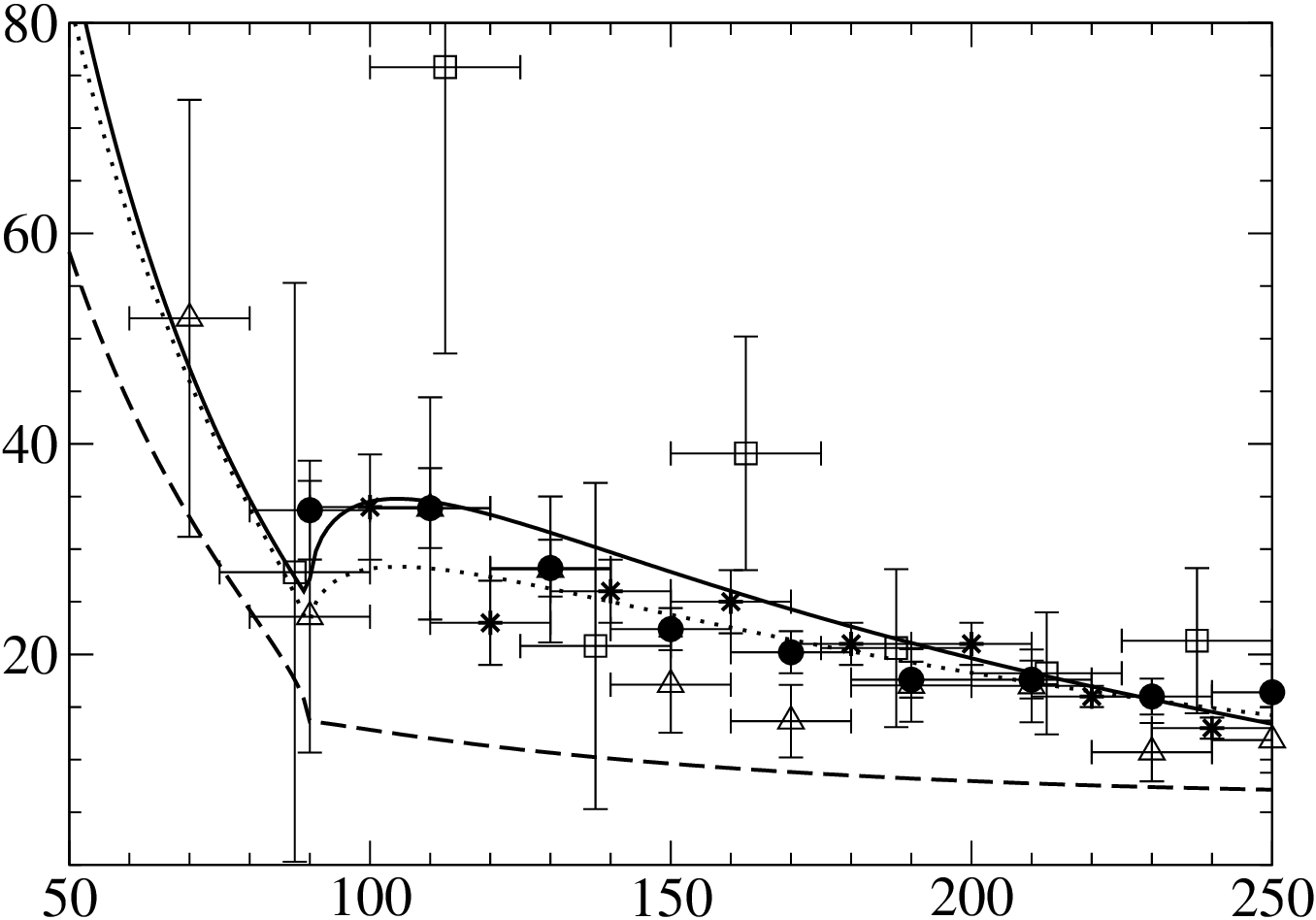}
  \put(-12,9){\rotatebox{90}{{\scalebox{0.9}{$\sigma [K^- p \to \pi^- \Sigma^+]$ (mb)}}}}
\end{overpic} \\[0.02\textheight]
 &
\begin{overpic}[height=0.20\textheight,clip]{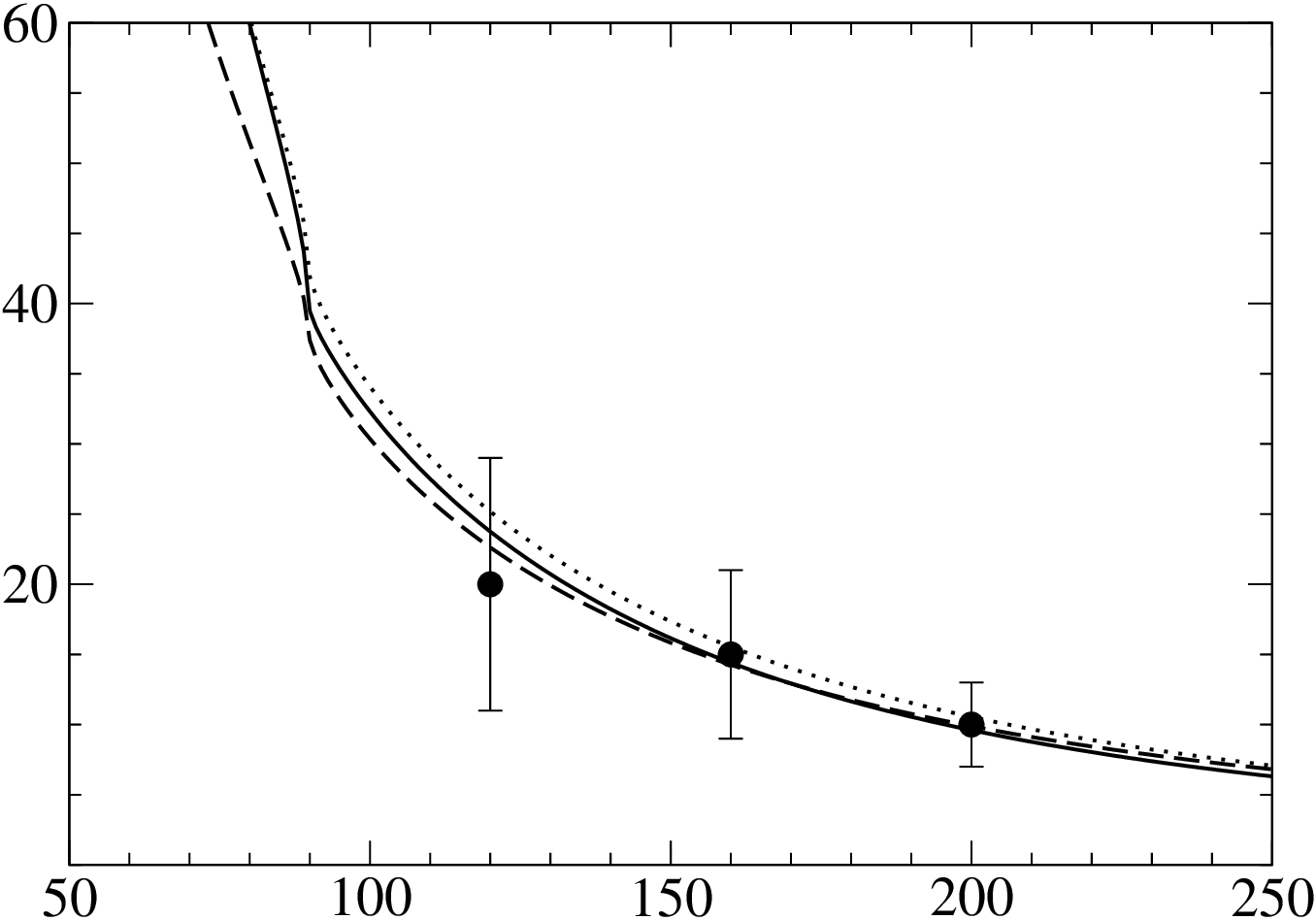}
  \put(-15,9){\rotatebox{90}{{\scalebox{0.9}{$\sigma [K^- p \to \pi^0 \Sigma^0]$ (mb)}}}}
  \put(5,-9){\scalebox{0.9}{incident $K^-$ lab.\  momentum (MeV/c)}}
\end{overpic} & &
\begin{overpic}[height=0.20\textheight,clip]{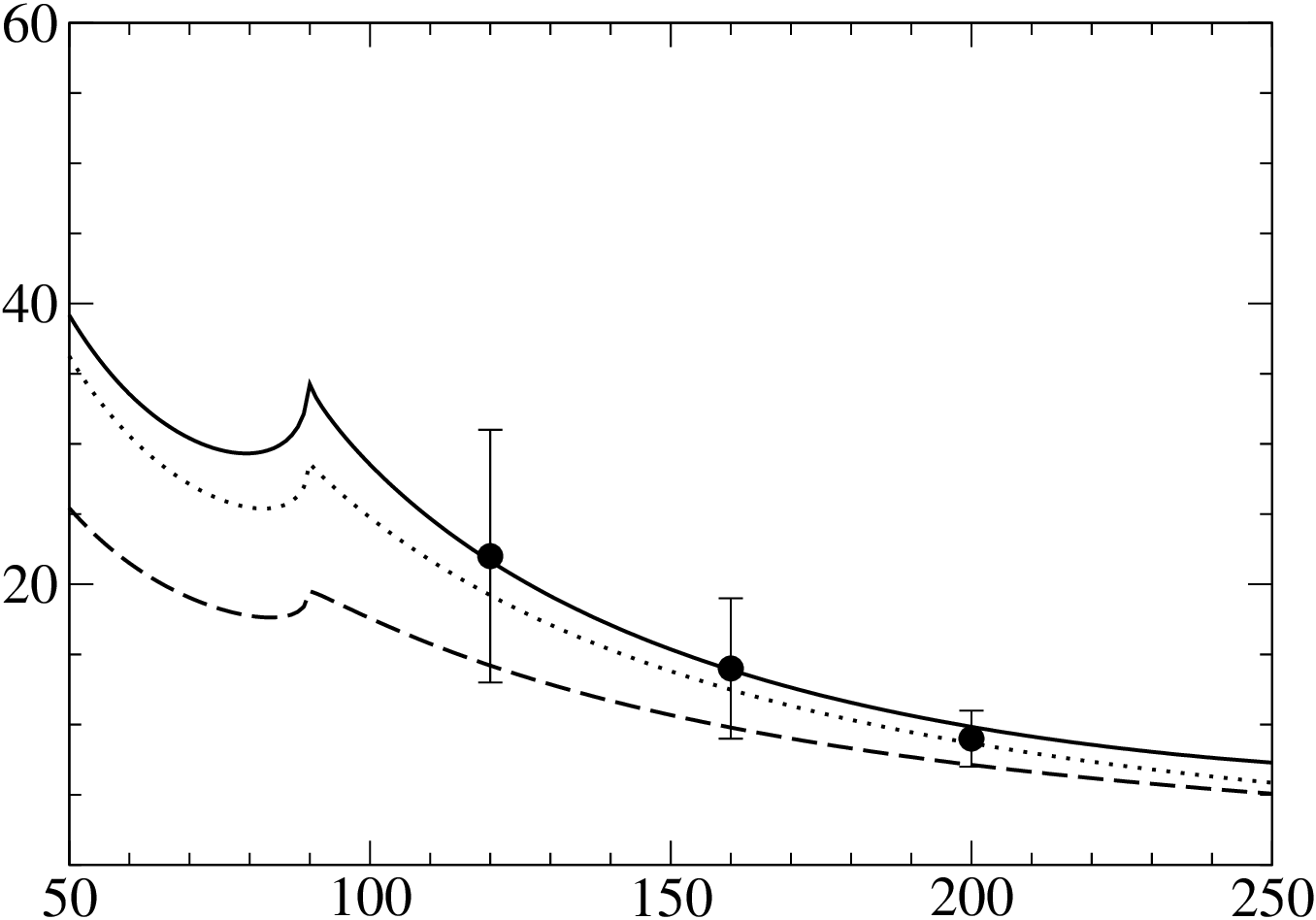}
  \put(-12,7){\rotatebox{90}{{\scalebox{0.9}{$\sigma [K^- p \to \pi^0 \Lambda]$ (mb)}}}}
  \put(5,-9){\scalebox{0.9}{incident $K^-$ lab.\  momentum (MeV/c)}}
\end{overpic}
\end{tabular}
\vspace{2.0ex}
\caption{Total cross sections of $K^- p$ scattering into various channels. The data
         are taken from  \cite{Hum} (empty squares), \cite{Sak} (empty triangles),
         \cite{Kim} (filled circles), \cite{Kit} (filled squares), 
         \cite{Eva} (filled triangles), \cite{Cib} (stars). The solid, dotted and dashed
         lines represent the fits ``1'', ``2'' and ``3'', respectively.}
\label{UDEARCS}
\end{figure}

Total cross sections of $K^- p$ scattering into various channels are shown in 
Fig.~\ref{UDEARCS}. Deviations between fit ``3'' (which satisfies 
the DEAR constraints) and fit ``1'' (where these constraints have been omitted) are most pronounced in the elastic channel $K^- p \to K^- p$. Approaching the DEAR data by going from fit ``1'' to 
fit ``2'' and eventually to fit ``3'' subsequently lowers the total elastic cross section in the 
whole energy range under consideration and produces results which lie below
the experimental data points (not without mentioning again that these data sets themselves scatter over a wide range). While there may be questions about the detailed treatment of Coulomb corrections, these
effects can safely be neglected for kaon momenta above 200~MeV/$c$. Our findings suggest that within coupled-channels schemes constrained by large amounts of data, 
the new accurate DEAR results and the old elastic $K^- p$ scattering cross sections at low energy
cannot be simultaneously accommodated. 

The results in the inelastic $K^- p$ scattering channels are not altered significantly 
by the inclusion of 
the DEAR data, with the exception of the reaction $K^- p \to \pi^{\pm} \Sigma^{\mp}$ where the 
cross sections resulting from fit ``3'' are slightly reduced. However, the threshold
values of these curves also enter the branching ratios $\gamma$ and $R_c$. While $\gamma$, 
i.e. the branching ratio of the two charged $\pi \Sigma$ channels, remains within experimental
errors, cf.\ Table~\ref{UDEARBR}, the value of $R_c$ drops substantially below the 
experimental boundary when moving from fit ``1'' to fit ``3''. This fact raises another
consistency issue. While the elastic
$K^- p$ scattering data close to threshold include the Coulomb interaction,
both the branching ratio $R_c$ and the observables measured at DEAR represent exclusively
effects of the strong interaction. These observables therefore provide a cleaner consistency check 
than elastic $K^- p$ scattering. The branching ratio $R_n$, on the other hand, involves neutral 
channels. It turns out to be uncritical and it is well reproduced by any of the fits.

\begin{figure}
\centering
\begin{tabular}{cp{1cm}c}
\begin{overpic}[width=0.40\textwidth,clip]{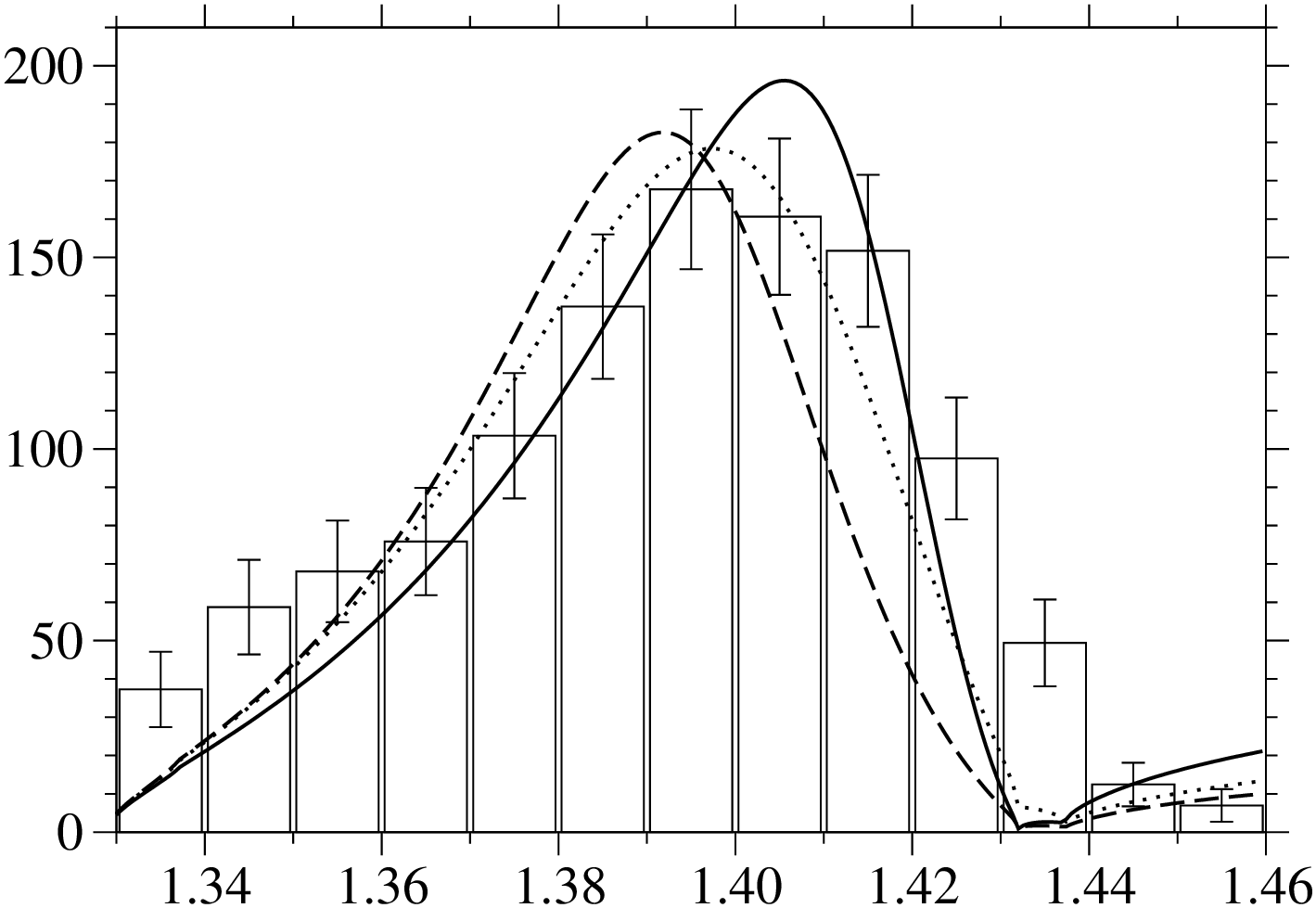}
  \put(-11,17){\rotatebox{90}{{\scalebox{0.9}{arbitrary units}}}}
  \put(45,-8){\scalebox{0.9}{$\sqrt{s}$ \quad (GeV)}}
\end{overpic}
& &
\begin{overpic}[width=0.40\textwidth,clip]{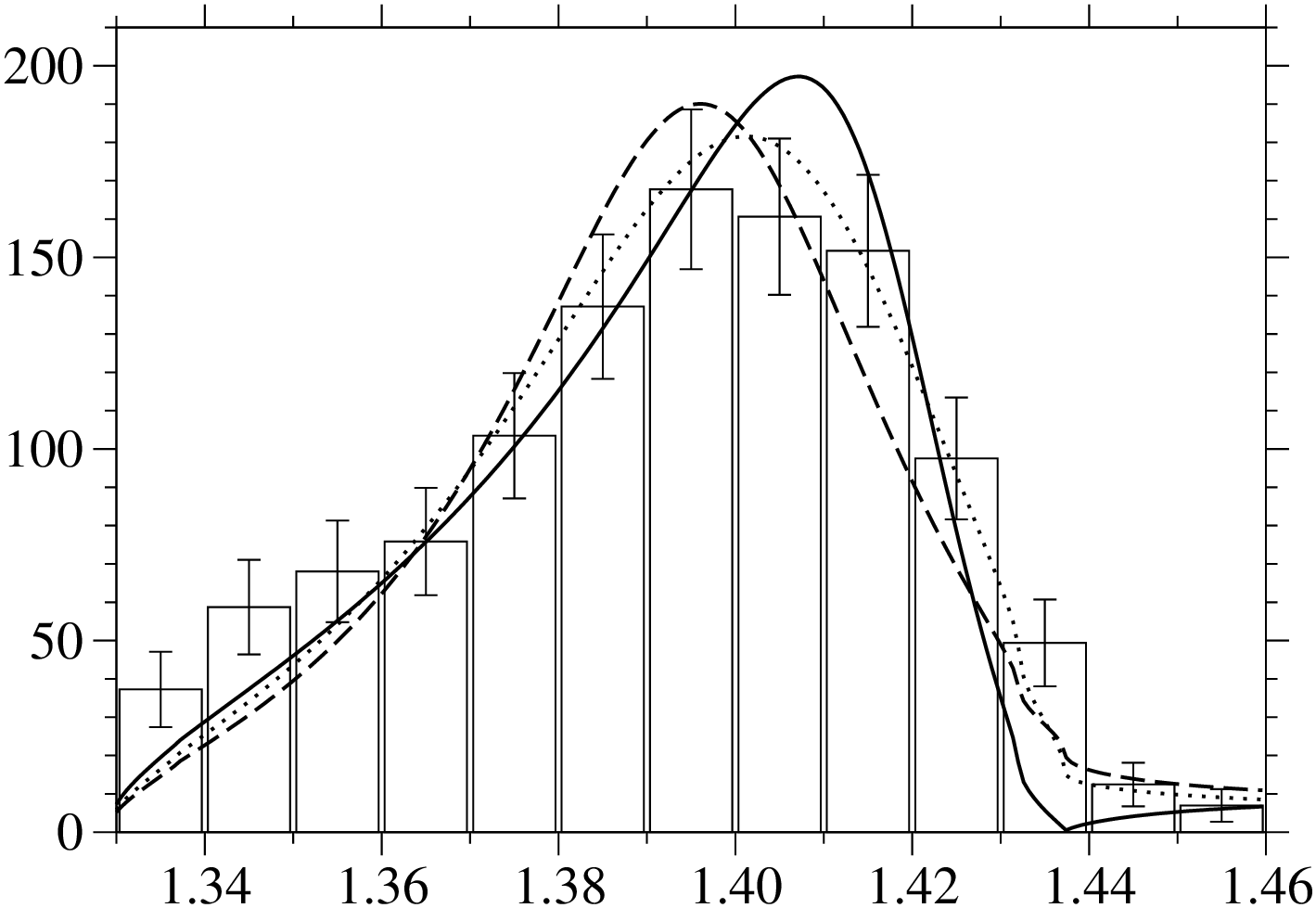}
  \put(-11,17){\rotatebox{90}{{\scalebox{0.9}{arbitrary units}}}}
  \put(45,-8){\scalebox{0.9}{$\sqrt{s}$ \quad (GeV)}}
\end{overpic}
\\[3.5ex]
(a) & & (b)
\end{tabular}
\caption{$\pi^- \Sigma^+$ event distribution from \cite{Hem},
         where statistical errors have been supplemented following \cite{DD}.
         The curves in diagram (a) where obtained by assuming a $\pi \Sigma$ invariant 
         mass spectrum with $I=0$; the curves in diagram (b) result from the ansatz
         advocated in \cite{OM}.
         The solid, dotted and dashed lines correspond to the fits ``1'', ``2'' and 
         ``3'', respectively.}
\label{UDEARPiSigma}
\end{figure}

If the $\pi^- \Sigma^+$ event distribution from \cite{Hem} is interpreted as a pure $I=0$
$\pi \Sigma$ invariant mass spectrum, approaching the DEAR data results in shifting the peak 
of the curve to lower energies and therefore worsening the fit to the data, cf.\ 
Fig.~\ref{UDEARPiSigma}a. The assumption of a generic $I=0$ source made up of an admixture 
of $\bar{K} N$ and $\pi \Sigma$ states \cite{OM} improves the fit 
by introducing one additional parameter, see Fig.~\ref{UDEARPiSigma}b. One striking feature is 
that the resulting ratio $r_1/r_2 = 2.23$ for fit ``3'' deviates substantially from those 
fits for which the DEAR data were not taken into account, and corresponds to a source that 
is dominated by $\bar{K} N$ states also below threshold.

\begin{figure}
\centering
\begin{tabular}{p{0.4cm}cp{1.0cm}c}
 &
\begin{overpic}[width=0.40\textwidth,clip]{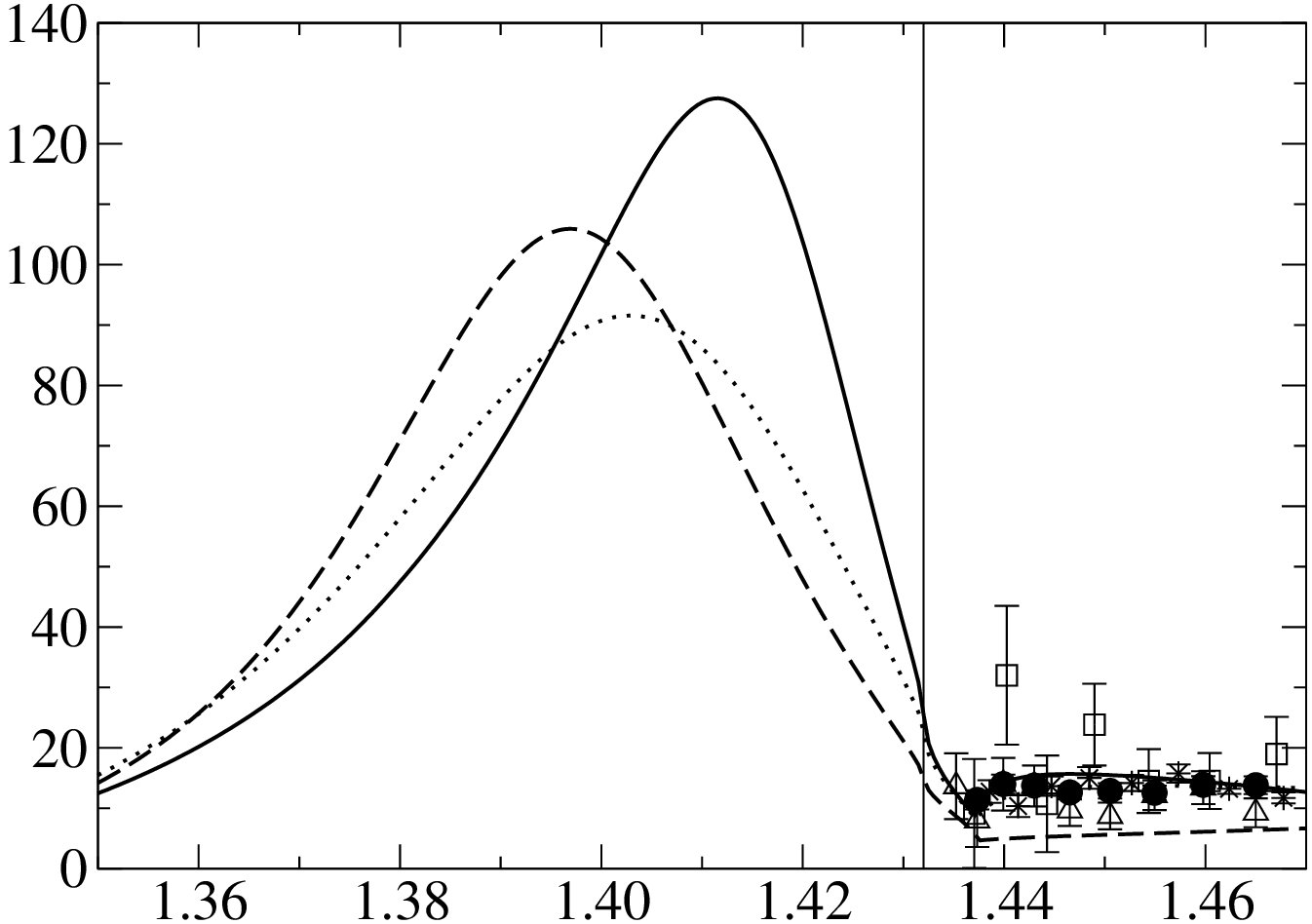}
  \put(-11,-4){\rotatebox{90}{{\scalebox{0.8}{
    $4\,|\mathbf{q}_{cm}^{K^- p}| \sqrt{s} \ \sigma_{K^- p \to \pi^- \Sigma^+}$
(mb GeV$^2$)}}}}
  \put(45,-8){\scalebox{0.9}{$\sqrt{s}$ \quad (GeV)}}
\end{overpic}
& &
\begin{overpic}[width=0.40\textwidth,clip]{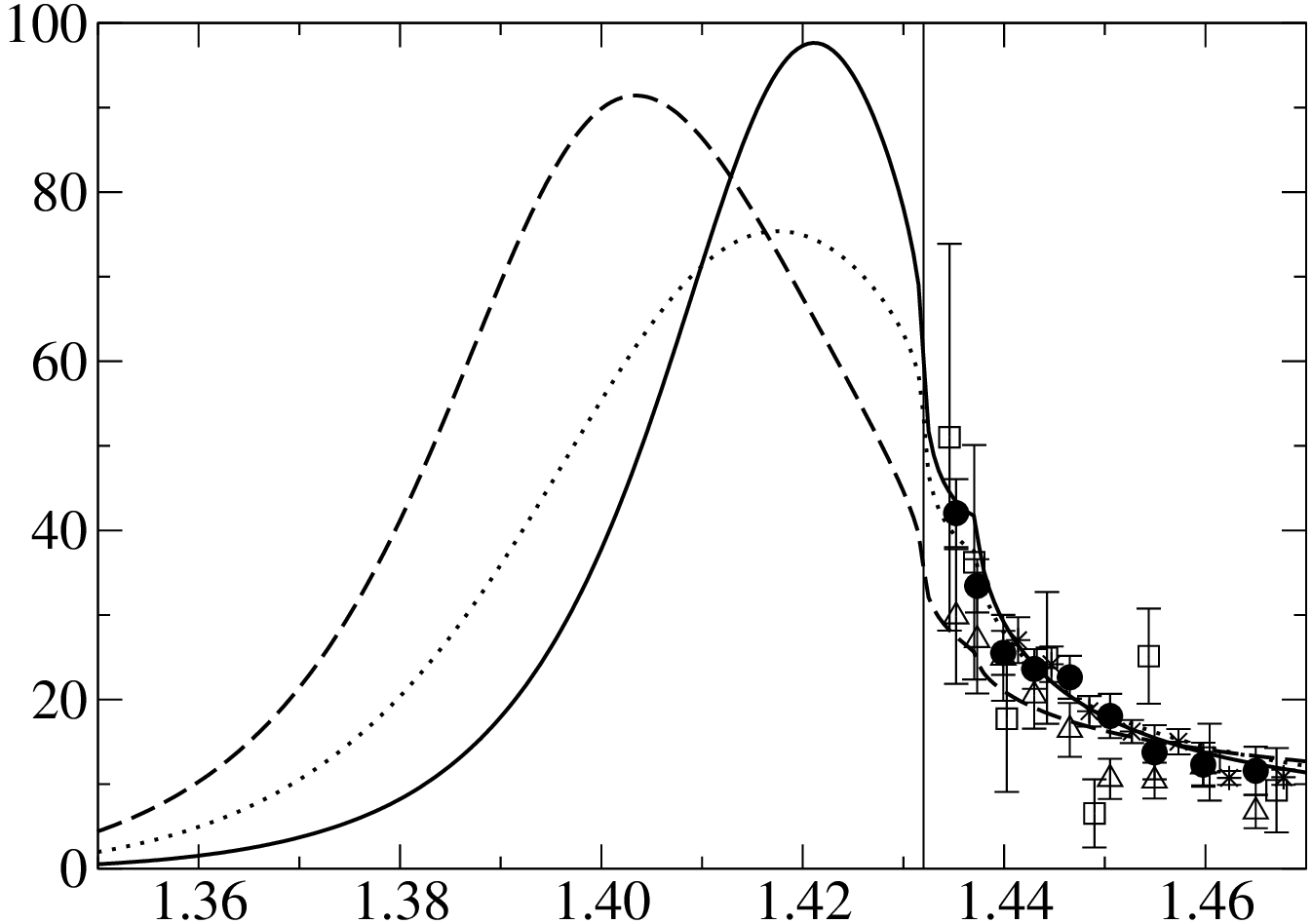}
  \put(-11,-4){\rotatebox{90}{{\scalebox{0.8}{
    $4\,|\mathbf{q}_{cm}^{K^- p}| \sqrt{s} \ \sigma_{K^- p \to \pi^+ \Sigma^-}$
(mb GeV$^2$)}}}}
  \put(45,-8){\scalebox{0.9}{$\sqrt{s}$ \quad (GeV)}}
\end{overpic}
\\[3.5ex]
& (a) & & (b)
\end{tabular}
\caption{Shown are the cross sections for $K^- p \to \pi^- \Sigma^+$ (a) and 
         $K^- p \to \pi^+ \Sigma^-$ (b) multiplied by $4 |\mathbf{q}_{cm}^{K^- p}| \sqrt{s}$
         and continued below $K^- p$ threshold (vertical line). The experimental data 
         points are the same as in Fig.~\ref{UDEARCS}, but have been modified accordingly.
         The solid, dotted and dashed lines correspond to the fits ``1'', ``2'' and 
         ``3'', respectively.}
\label{UDEARsubthr}
\end{figure}

As pointed out in \cite{LS} $\Lambda(1405)$ photoproduction, which has been
investigated experimentally at SPring-8 \cite{Ahn} and at ELSA,
could serve as a tool to constrain $\bar{K} N$ dynamics below threshold. 
If $t$-channel exchange of $K^-$ mesons can be isolated, 
it should be possible to extract from the process $\gamma p \to K^+ \pi \Sigma$
the $K^- p \to \pi \Sigma$ amplitudes below the $K^- p$ threshold. This statement is also 
of interest for the present investigation, since the fits which either in- or exclude the DEAR data 
yield different predictions for these amplitudes. In Fig.~\ref{UDEARsubthr} we plot the 
quantity $4\,|\mathbf{q}_{cm}^{K^- p}| \sqrt{s} \ \sigma_{K^- p \to \pi^\mp \Sigma^\pm}(s)$
continued below threshold, and the experimental cross section data above the $K^- p$ threshold
have been normalized
accordingly. 
In the case of fit ``3'', the one consistent with the DEAR data,
the shape of the curve is altered
for both final states $\pi^- \Sigma^+$ and $\pi^+ \Sigma^-$. 
Compared to fit ``1'' the peak position is shifted to lower energies, while the width 
is considerably increased.
This difference can be examined experimentally once the necessary $t$-channel analysis and normalization of the SPring-8 results 
\cite{Ahn} has been performed. These data cover an energy range from the $\pi \Sigma$ 
threshold up to energies above the $K^- p$ threshold, where consistency with 
existing cross section data can be tested.
In conclusion, the SPring-8/ELSA experiments may provide a
further important consistency check of scattering data and the DEAR results
within our framework.

\begin{figure}
\centering
\begin{tabular}{cp{1cm}c}
\begin{overpic}[width=0.40\textwidth,clip]{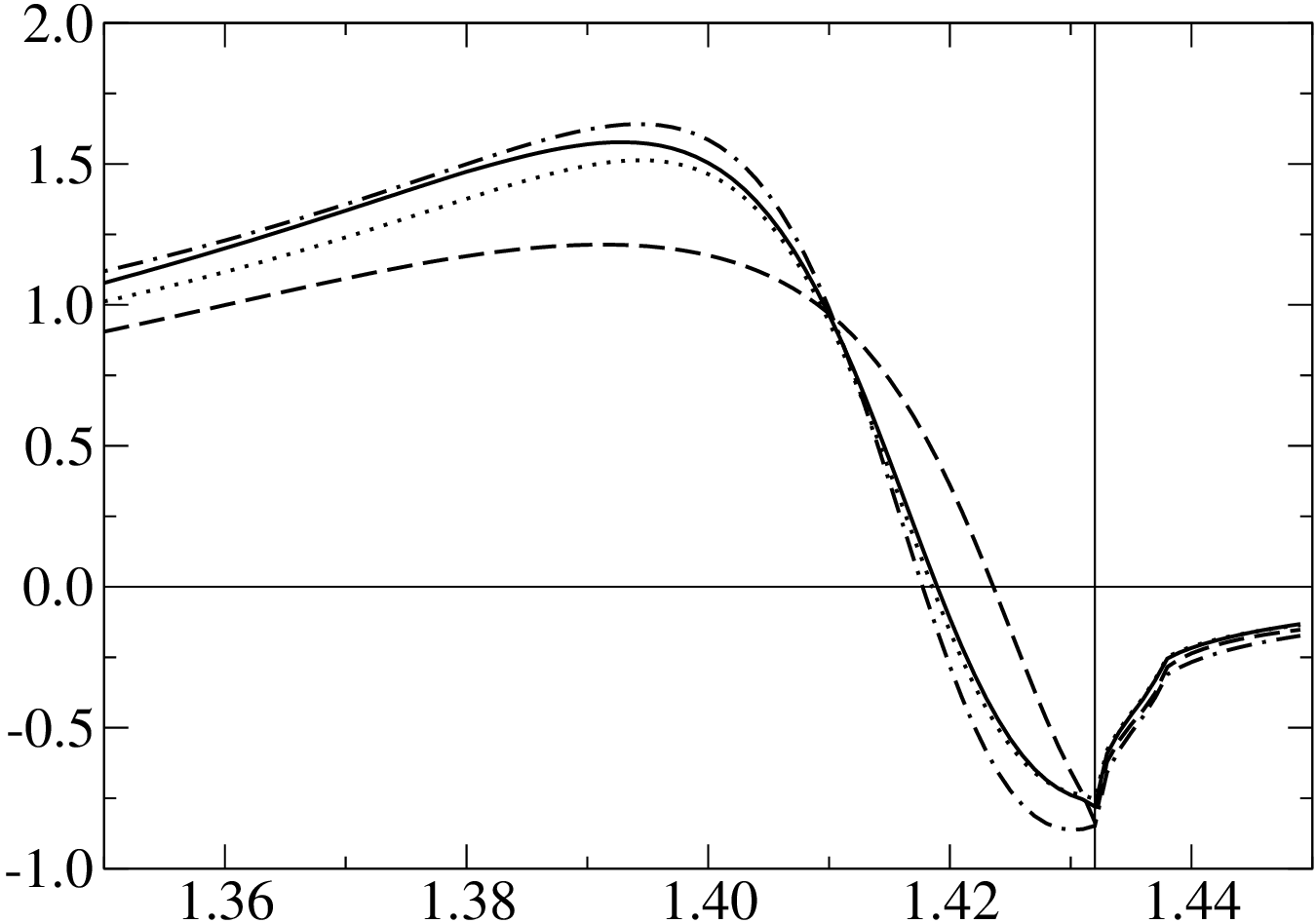}
  \put(-11,14){\rotatebox{90}{{\scalebox{0.9}{{$\Re{f^{\textrm{str}}_{K^- p \to K^- p}}$ (fm)}}}}}
  \put(40,-8){\scalebox{0.9}{$\sqrt{s}$ \quad (GeV)}}
\end{overpic}
& &
\begin{overpic}[width=0.40\textwidth,clip]{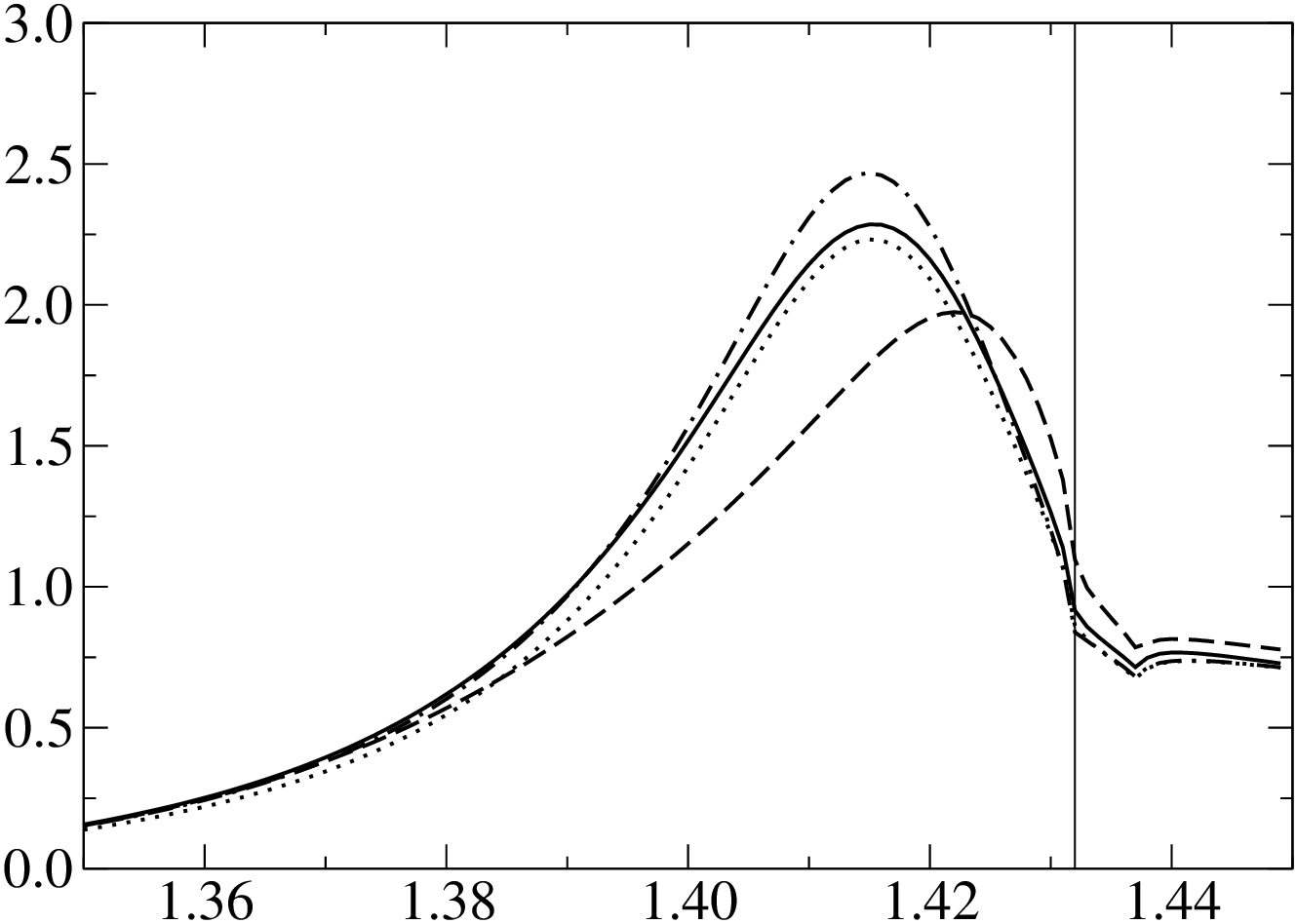}
  \put(-11,14){\rotatebox{90}{{\scalebox{0.9}{{$\Im{f^{\textrm{str}}_{K^- p \to K^- p}}$ (fm)}}}}}
  \put(40,-8){\scalebox{0.9}{$\sqrt{s}$ \quad (GeV)}}
\end{overpic}
\\[3.5ex]
(a) & & (b)
\end{tabular}
\caption{Real (left panel) and imaginary part (right panel) of the strong interaction
         elastic $K^- p$ amplitude. The dashed, dotted, dot-dashed
         and solid lines correspond to the approaches ``WT'', ``c'', ``$s$'' and ``$u$'',
         respectively. The $K^- p$ threshold is indicated by the vertical line.}
\label{Compfstr}
\end{figure}

\begin{figure}
\centering
\begin{tabular}{cp{1cm}c}
\begin{overpic}[width=0.40\textwidth,clip]{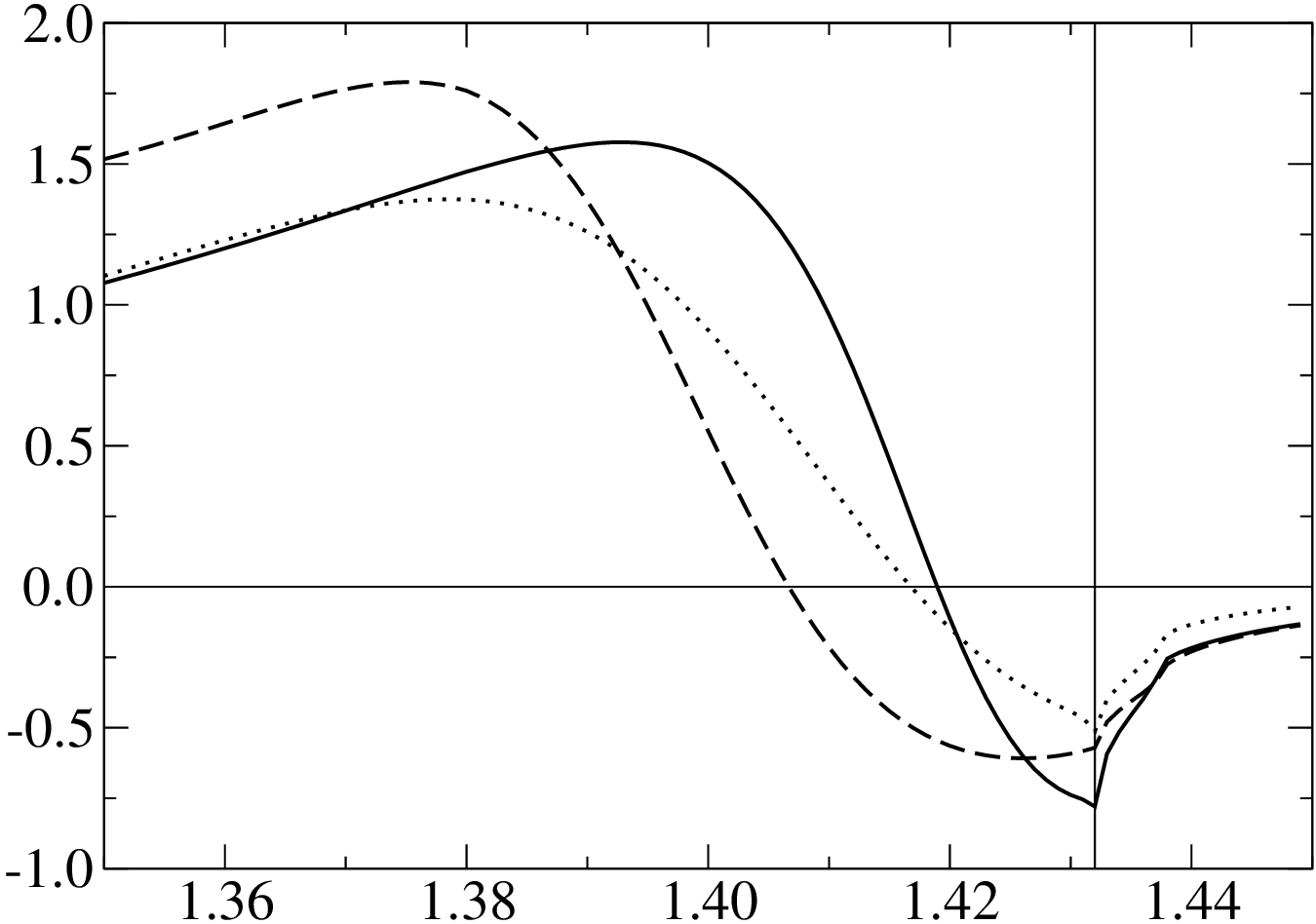}
  \put(-11,14){\rotatebox{90}{{\scalebox{0.9}{$\Re{f^{\textrm{str}}_{K^- p \to K^- p}}$ (fm)}}}}
  \put(40,-8){\scalebox{0.9}{$\sqrt{s}$ \quad (GeV)}}
\end{overpic}
& &
\begin{overpic}[width=0.40\textwidth,clip]{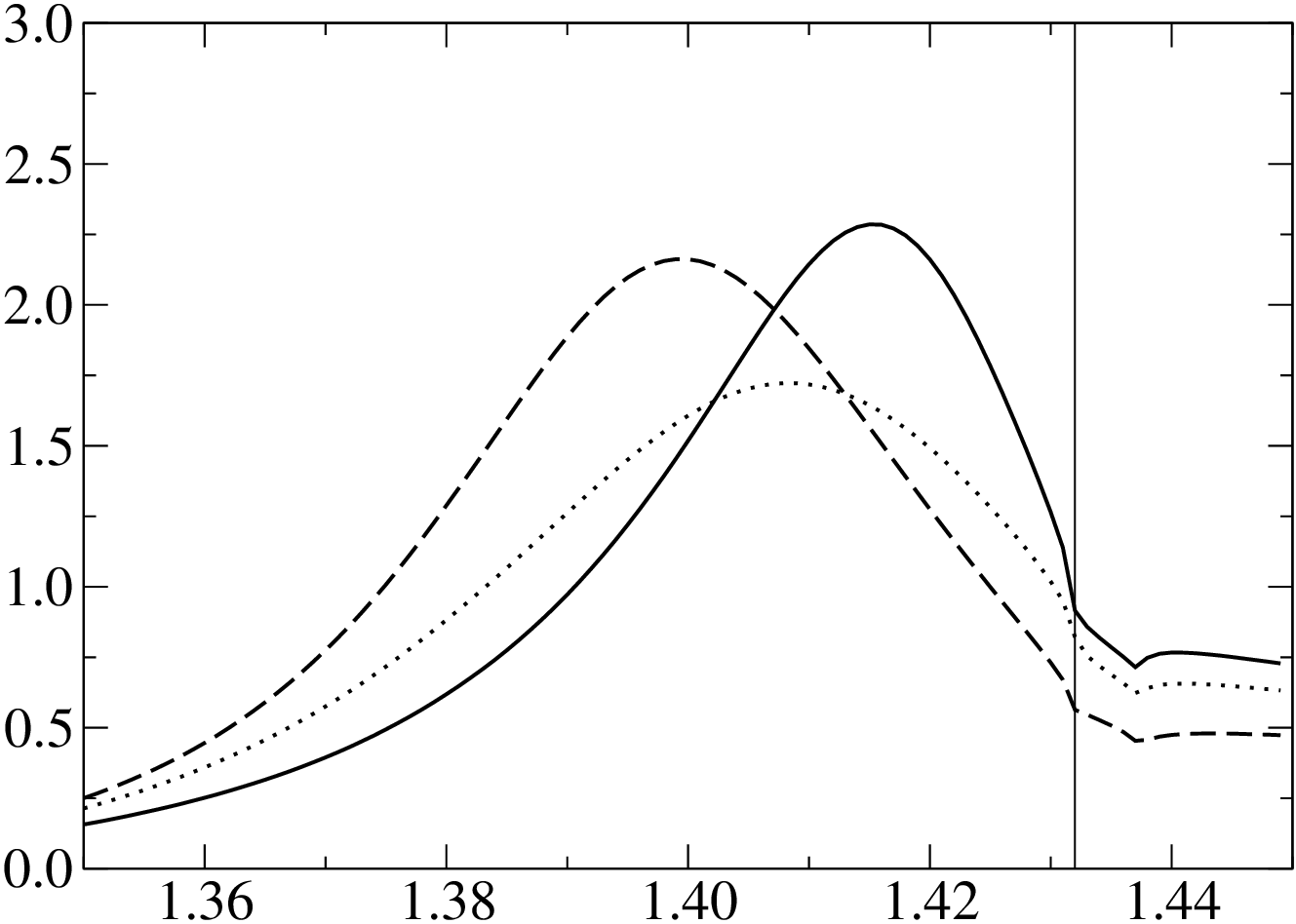}
  \put(-11,14){\rotatebox{90}{{\scalebox{0.9}{$\Im{f^{\textrm{str}}_{K^- p \to K^- p}}$ (fm)}}}}
  \put(40,-8){\scalebox{0.9}{$\sqrt{s}$ \quad (GeV)}}
\end{overpic}
\\[3.5ex]
(a) & & (b)
\end{tabular}
\caption{Real (left panel) and imaginary part (right panel) of the strong interaction
         elastic $K^- p$ amplitude. The solid, dotted and dashed
         lines represent the fits ``1'', ``2'' and ``3'', respectively.
         The $K^- p$ threshold is indicated by the vertical line.}
\label{wDEARfstr}
\end{figure}

It is instructive to investigate the real and imaginary parts of the elastic 
$K^- p \to K^- p$ scattering
amplitude below threshold (see Figs.~\ref{Compfstr}, \ref{wDEARfstr}). The important role of 
next-to-leading order dynamics (the ``c'', ``$s$'' and ``$u$'' versions) as compared to the 
leading order driven only by the Weinberg-Tomozawa term (the ``WT'' version) becomes visible 
in Fig.~\ref{Compfstr}. The influence of the additional constraint imposed by the DEAR threshold 
data is seen in Fig.~\ref{wDEARfstr}. It has a pronounced effect in shifting the $\Lambda$(1405) 
resonance spectrum further down in $\sqrt{s}$, primarily by enforcing a smaller imaginary part 
of $f_{K^- p \to K^- p}$ at threshold.


\begin{table}
\centering
\begin{tabular}{|ll|c|c|c|}
\hline
& & ``1'' & ``2'' & ``3'' \\
\hline
$a_{K^{-} p}$ & (fm) & $-0.78+0.92i $ & $-0.51+0.82i $ & $-0.57+0.56i $ \\
\hline
$\Delta E_{D}$ & (eV) & 321 & 211 & 236 \\
\hline
$\Gamma_{D}$ & (eV) & 755 & 678 & 465 \\
\hline
$\Delta E_{c}$ & (eV) & 335 & 236 & 235 \\
\hline
$\Gamma_{c}$ & (eV) & 589 & 580 & 390 \\
\hline
\end{tabular}
\caption{Shown are the $K^- p$ scattering lengths $a_{K^- p}$ as well as the
         strong interaction shift $\Delta E$ and width $\Gamma$ in kaonic hydrogen
         resulting from
         Eq.~(\ref{eq:Des}) (subscript $D$) and Eq.~(\ref{eq:Rus}) (subscript $c$).}
\label{UDEARShiWi}
\end{table}

\begin{table}
\centering
\begin{tabular}{|ll|c|c|c|c|}
\hline
 & & ``1'' & ``2'' & ``3'' \\
\hline
$a_{\bar{K} N}$ & $(10^{-3})$ & -2.16 & -0.95 & -2.62 \\
\hline
$a_{\pi \Lambda}$& $(10^{-3})$& -6.34 & 0.59 & 11.46 \\
\hline
$a_{\pi \Sigma}$& $(10^{-3})$& -1.24 & -1.80 & -3.06 \\
\hline
$a_{\eta \Lambda}$& $(10^{-3})$& -1.99 & -2.92 & 5.10 \\
\hline
$a_{\eta \Sigma}$& $(10^{-3})$& -2.75 & -0.98 & -4.26 \\
\hline
$a_{K \Xi}$& $(10^{-3})$& -4.37 & -2.90 & 3.69 \\
\hline
$f$& (MeV) & 103.3 & 103.1 & 94.4 \\
\hline
$b_0$& (GeV$^{-1}$) & -0.31 & -0.36 & -0.20 \\
\hline
$b_D$& (GeV$^{-1}$) & 0.00 & 0.00 & 0.14 \\
\hline
$b_F$& (GeV$^{-1}$) & -0.13 & -0.13 & -0.11 \\
\hline
$d_1$& (GeV$^{-1}$) & -0.16 & -0.11 & -0.30 \\
\hline
$d_2$& (GeV$^{-1}$) & 0.12 & 0.05 & 0.02 \\
\hline
$d_3$& (GeV$^{-1}$) & 0.25 & 0.31 & 0.39 \\
\hline
$d_4$& (GeV$^{-1}$) & -0.23 & -0.32 & -0.35 \\
\hline
\end{tabular}
\caption{Numerical values of the parameters for the different fits described in the text.
         The subtraction constants are taken at $\mu = 1$~GeV.}
\label{UDEARpar}
\end{table}

\begin{table}
\centering
\begin{tabular}{|c|c|c|c||c|}
\hline
 & ``1'' & ``2'' & ``3'' & Exp. \cite{Now,Tov} \\
\hline
$\gamma$ & 2.36  & 2.35  & 2.38  & $2.36 \pm 0.04$ \\
\hline
$R_c$    & 0.661 & 0.653 & 0.631 & $0.664 \pm 0.011$ \\
\hline
$R_n$    & 0.189 & 0.194 & 0.176 & $0.189 \pm 0.015$ \\
\hline
\end{tabular}
\caption{Threshold branching ratios as defined in the text,
         resulting from the different fits.}
\label{UDEARBR}
\end{table}

\subsection{$\mbox{\boldmath$K^- n$ and $K^+ p$}$ scattering}

\begin{figure}
\centering
\begin{tabular}{cp{1cm}c}
\begin{overpic}[width=0.40\textwidth,clip]{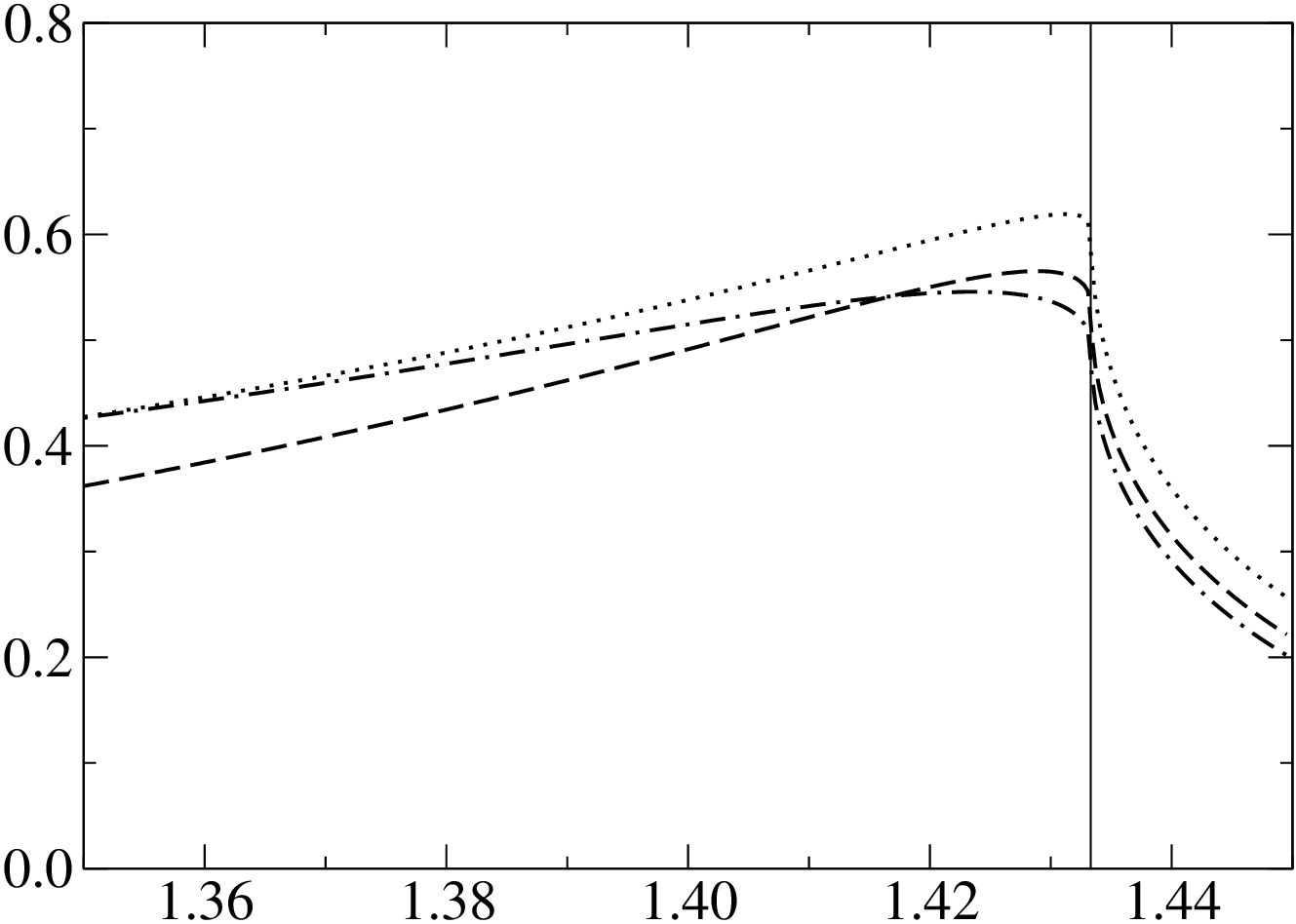}
  \put(-11,14){\rotatebox{90}{{\scalebox{0.9}{{$\Re{f^{\textrm{str}}_{K^- n \to K^- n}}$ (fm)}}}}}
  \put(40,-8){\scalebox{0.9}{$\sqrt{s}$ \quad (GeV)}}
\end{overpic}
& &
\begin{overpic}[width=0.40\textwidth,clip]{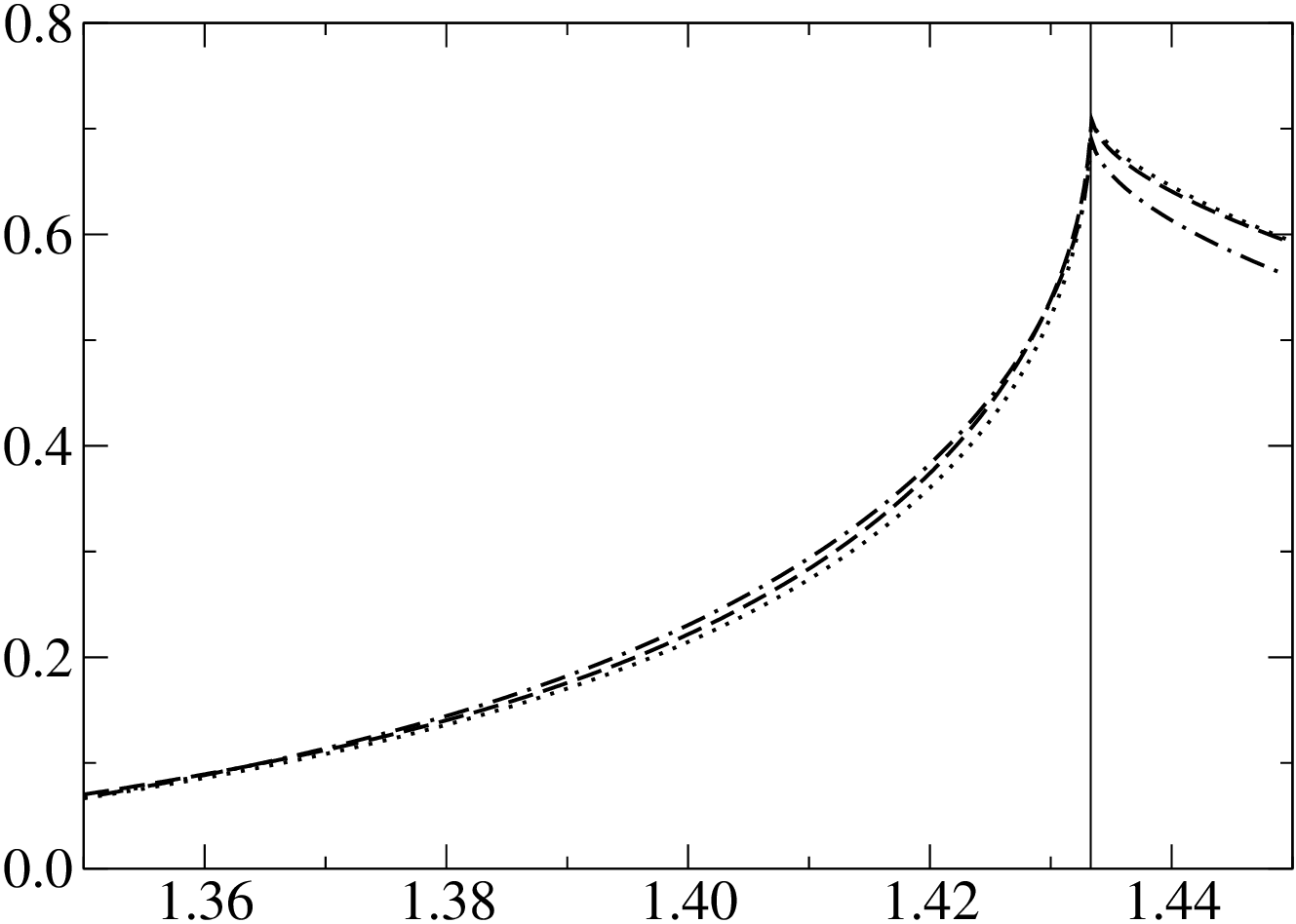}
  \put(-11,14){\rotatebox{90}{{\scalebox{0.9}{{$\Im{f^{\textrm{str}}_{K^- n \to K^- n}}$ (fm)}}}}}
  \put(40,-8){\scalebox{0.9}{$\sqrt{s}$ \quad (GeV)}}
\end{overpic}
\\[3.5ex]
(a) & & (b)
\end{tabular}
\caption{Real (left panel) and imaginary part (right panel) of the strong interaction ($s$-wave)
         elastic $K^- n$ amplitude. The dashed, dotted and dot-dashed lines
         correspond to the approaches ``WT'', ``c'' and ``$s$'', respectively.
         The $K^- n$ threshold is indicated by the vertical line.}
\label{CompKmnfstr}
\end{figure}

Once the parameters have been fixed from $K^- p$ data, the same approach 
provides predictions for $K^- n$ scattering, since no new unknown constants
appear. The real and imaginary parts of the elastic $s$-wave $K^- n$ scattering amplitude
are presented in Fig.~\ref{CompKmnfstr} for the ``WT'', ``c'' and ``s'' frameworks.
All three versions yield similar results and the predicted scattering
lengths given in Table~\ref{CompKmnSL} are consistent with the empirical value
$a_{K^- n} \sim 0.4 + i\,0.6$~fm \cite{ADM} within errors. 

We have refrained 
from presenting results for the ``$u$''-approach for $K^- n \to K^- n$.
The reason is the appearance of an unphysical subthreshold cut in the $\eta \Sigma^-$
channel at 1.426~GeV just below the $K^- n$ threshold. 
As already mentioned in Sec.~\ref{sec:cc}, this is an artifact of the 
on-shell formalism which would not be present in a full field theoretical calculation.
The previously applied procedure of eliminating the unphysical subthreshold 
cut by matching the contribution of the crossed 
Born diagram to a constant value below a certain invariant energy $\sqrt{s_0}$ 
does not work here since
the singularity at 1.426~GeV is just 7~MeV away from $K^- n$ threshold.

Finally, we note that in 
the $K^+$-proton channel, the different approaches (``WT'', ``c'' and ``3'') 
yield scattering lengths in the range 
$a_{K^+ p} \simeq -(0.26 \dots 0.36)$~fm, consistent with the empirical 
$a_{K^+ p} \simeq -0.33$~fm \cite{ADM}.


\begin{table}
\centering
\begin{tabular}{|ll|c|c|c|}
\hline
& & ``WT'' & ``c'' & ``$s$'' \\
\hline
$a_{K^{-} n}$ & (fm) & $0.53+0.72i $ & $0.61+0.71i $ & $0.49+0.70i $ \\
\hline
\end{tabular}
\caption{$K^- n$ scattering lengths for the approaches ``WT'', ``c'' and ``$s$''.}
\label{CompKmnSL}
\end{table}

\subsection{Coulomb effects}  \label{sec:coul}

For small incident kaon momenta close to $\bar{K} N$ threshold, 
the elastic $K^- p$ scattering cross section
receives sizable contributions from both the strong and the electromagnetic interaction. 
Coulomb interactions are taken into account by utilizing the quantum
mechanical Coulomb scattering amplitude Eq.~(\ref{eq:fCoul}). Due to the infinite-range
nature of the Coulomb potential, the scattering amplitude is infrared divergent in the 
limit of small incident momenta as well as small scattering angles. 

\begin{figure}
\centering
\begin{tabular}{p{0.5cm}cp{1.0cm}c}
 &
\begin{overpic}[height=0.20\textheight,clip]{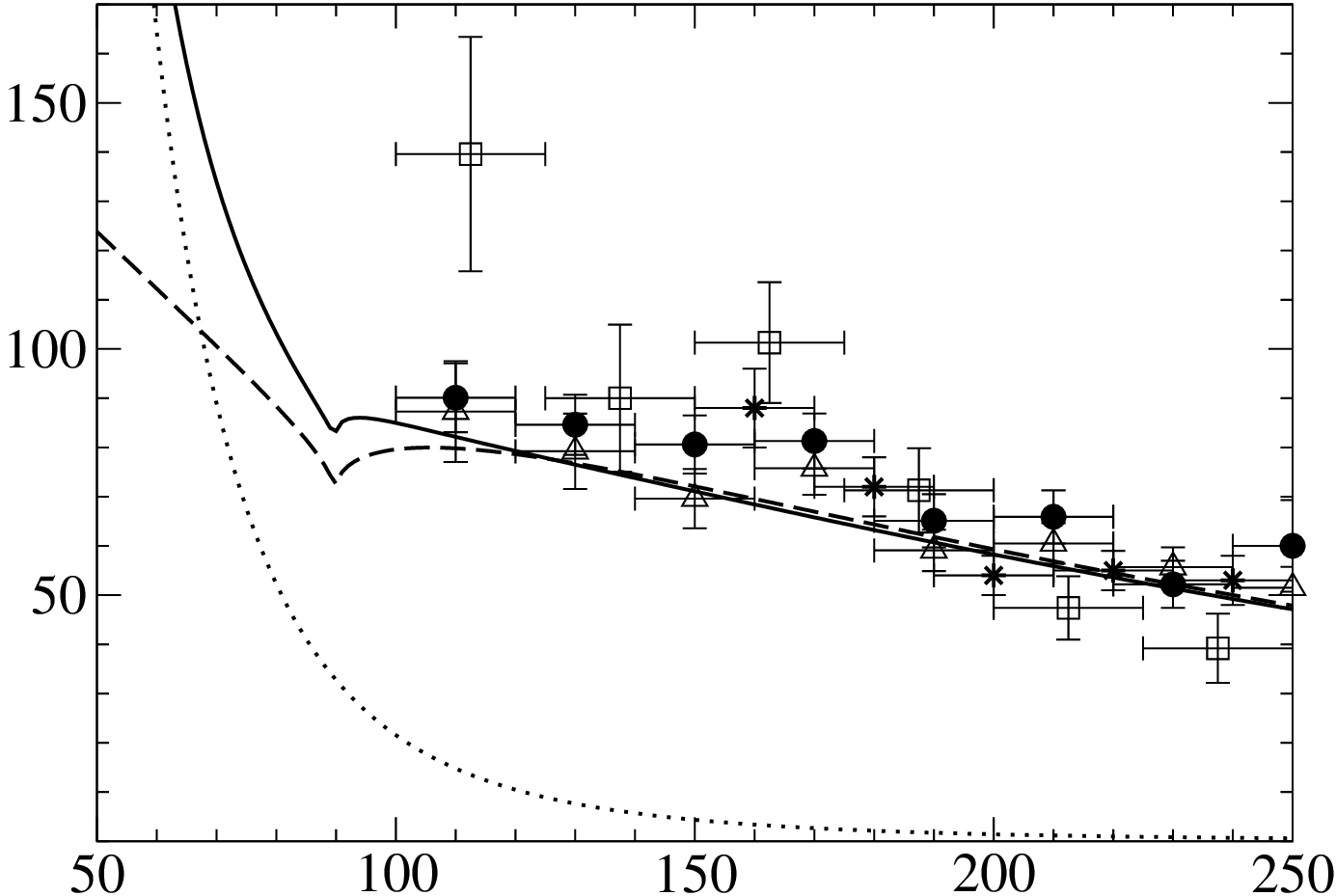}
  \put(-12,7){\rotatebox{90}{{\scalebox{0.9}{$\sigma [K^- p \to K^- p]$ (mb)}}}}
  \put(8,-9){\scalebox{0.9}{incident $K^-$ lab.\ momentum (MeV/c)}}
\end{overpic} & &
\begin{overpic}[height=0.20\textheight,clip]{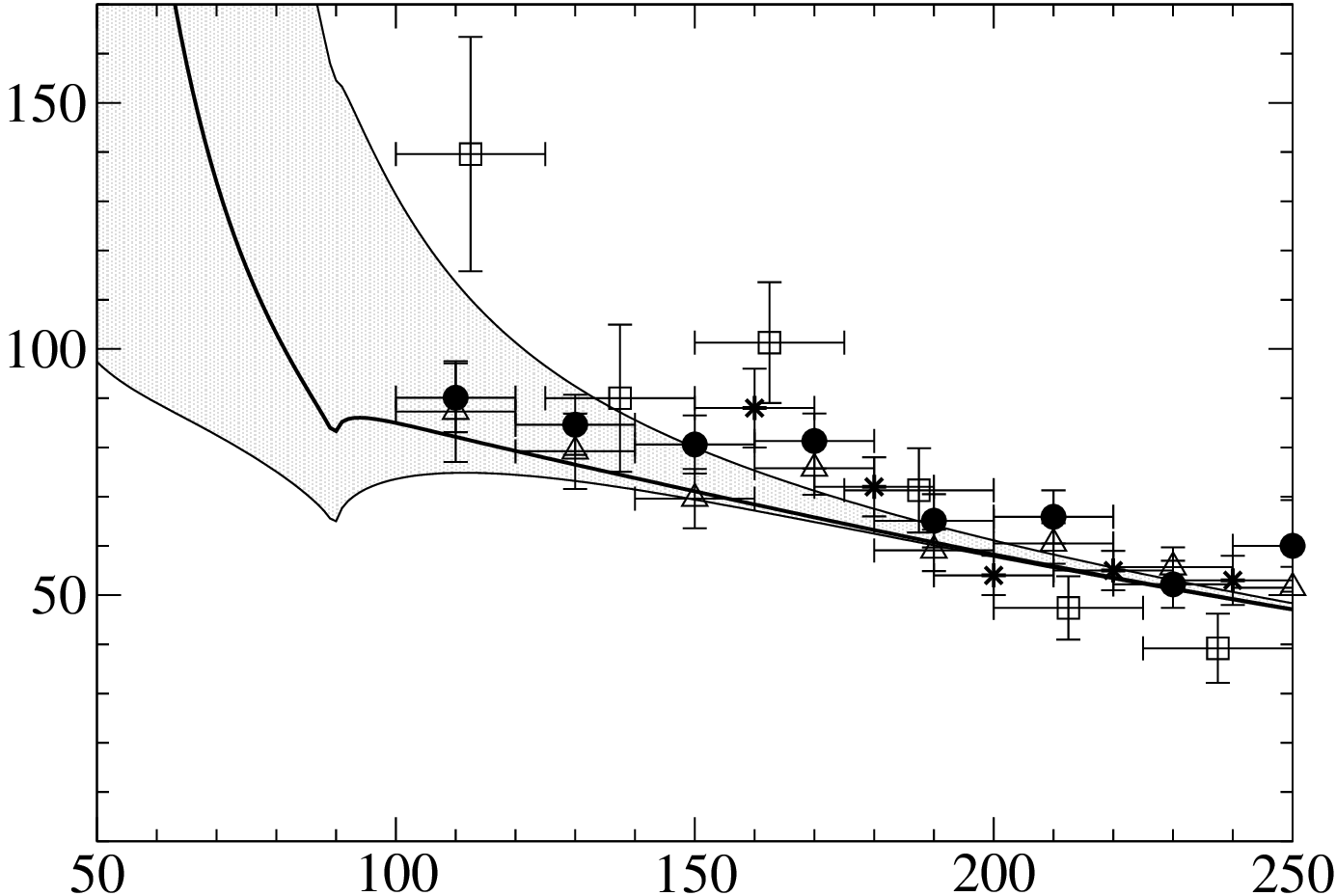}
  \put(-12,7){\rotatebox{90}{{\scalebox{0.9}{$\sigma [K^- p \to K^- p]$ (mb)}}}}
  \put(8,-9){\scalebox{0.9}{incident $K^-$ lab.\ momentum (MeV/c)}}
\end{overpic} \\[4.0ex]
 & (a) & & (b)
\end{tabular}
\caption{Left: Contributions to the total elastic cross sections of $K^- p$ scattering from 
         Coulomb interaction (dotted), strong interaction (dashed) and their coherent sum (solid).
         Right: Dependence on the small-angle cutoff excluding small center-of-mass angles. The lower boundary of the band
         corresponds to $\cos{\theta_{\textrm{min}}} = 0.7$, the upper one to
         $\cos{\theta_{\textrm{min}}} = 0.99$; the solid line represents the value
         established by the experiments \cite{Hum, Sak}, which we also use in our calculations:
         $\cos{\theta_{\textrm{min}}} = 0.966$.}
\label{Coulomb}
\end{figure}

As explained in section 2.3, the divergence at $\mathbf{q}_{cm} = 0$ can be ignored in the energy 
regime accessible by the scattering experiments. However, when performing the
integration over the center-of-mass scattering angle in order to calculate the total elastic cross section 
a cutoff in the angle must be introduced. Two of the experiments that have produced data 
at the lowest kaon momenta, exclude forward scattering angles and consider
only the range $-1 \leq \cos{\theta_{cm}} \leq 0.966$
\cite{Hum, Sak}. We choose to work with the same angle cutoff in order to perform consistent comparisons. The contributions of 
the Coulomb and the strong interaction as well as their coherent sum are displayed in an exemplary case for fit~``1'' in Fig.~\ref{Coulomb}a. While the corrections due to the Coulomb interaction 
are completely negligible for kaon laboratory momenta greater than 150~MeV/$c$, they start 
becoming important below 100~MeV/$c$.

In Fig.~\ref{Coulomb}b we show the dependence of our results on the small-angle cut.
The gray band indicates the variation between 
$\cos{\theta_{\textrm{min}}} = 0.7$, so that the Coulomb amplitude is highly suppressed,
and $\cos{\theta_{\textrm{min}}} = 0.99$ where it is sizable. The curves have been 
normalized to the solid angle covered by the experiments \cite{Hum, Sak}. 
For large 
incident kaon momenta (above 150 MeV/c) where the strong $s$-wave amplitude dominates, the 
omission of forward scattering angles makes the elastic cross section decrease by only a few
percent when compared with the integration over the full solid angle. It is therefore
justified to compare our results directly with all experimental data, given their large error spread.


\subsection{Resonance poles}

Finally, we turn our attention to the poles of the strong interaction $T$ matrix in the complex
$W \equiv \sqrt{s}$ plane. These poles are usually classified according to their isospin
and we keep the notation of $I=0$ and $I=1$ poles even though we work in the 
physical basis where isospin is broken by the physical masses of the particles. 
Although we observe two poles in the unphysical sheet which is directly connected 
to the physical region between the $\pi \Sigma$ and $\bar{K} N$ thresholds,
their positions depend strongly on the chosen approach. As a matter of fact,
the formation of a pronounced double pole structure close to the real axis as
reported in \cite{JOORM} occurs only in the ``WT'' model. When
next-to-leading order corrections are taken into account the second pole is shifted further away 
from the real axis and its contribution to the physical region tends to dissolve in the background.

\begin{figure}
\centering
\begin{tabular}{cp{1.0cm}c}
\begin{overpic}[height=0.20\textheight,clip]{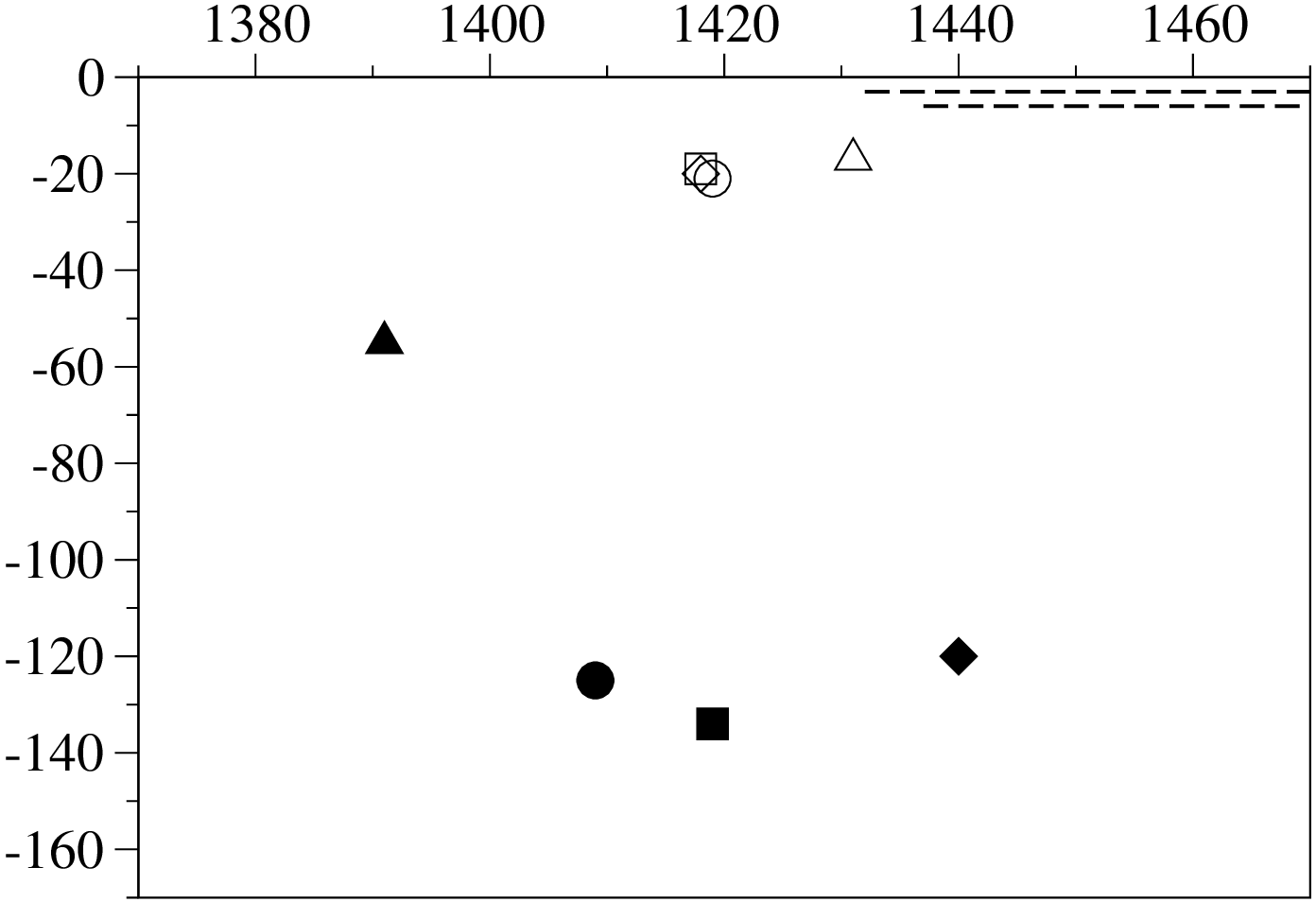}
  \put(-10,13){\rotatebox{90}{{\scalebox{0.9}{Im$(W)$ \quad (MeV)}}}}
  \put(34,74){\scalebox{0.9}{Re$(W)$ \quad (MeV)}}
\end{overpic} & &
\begin{overpic}[height=0.20\textheight,clip]{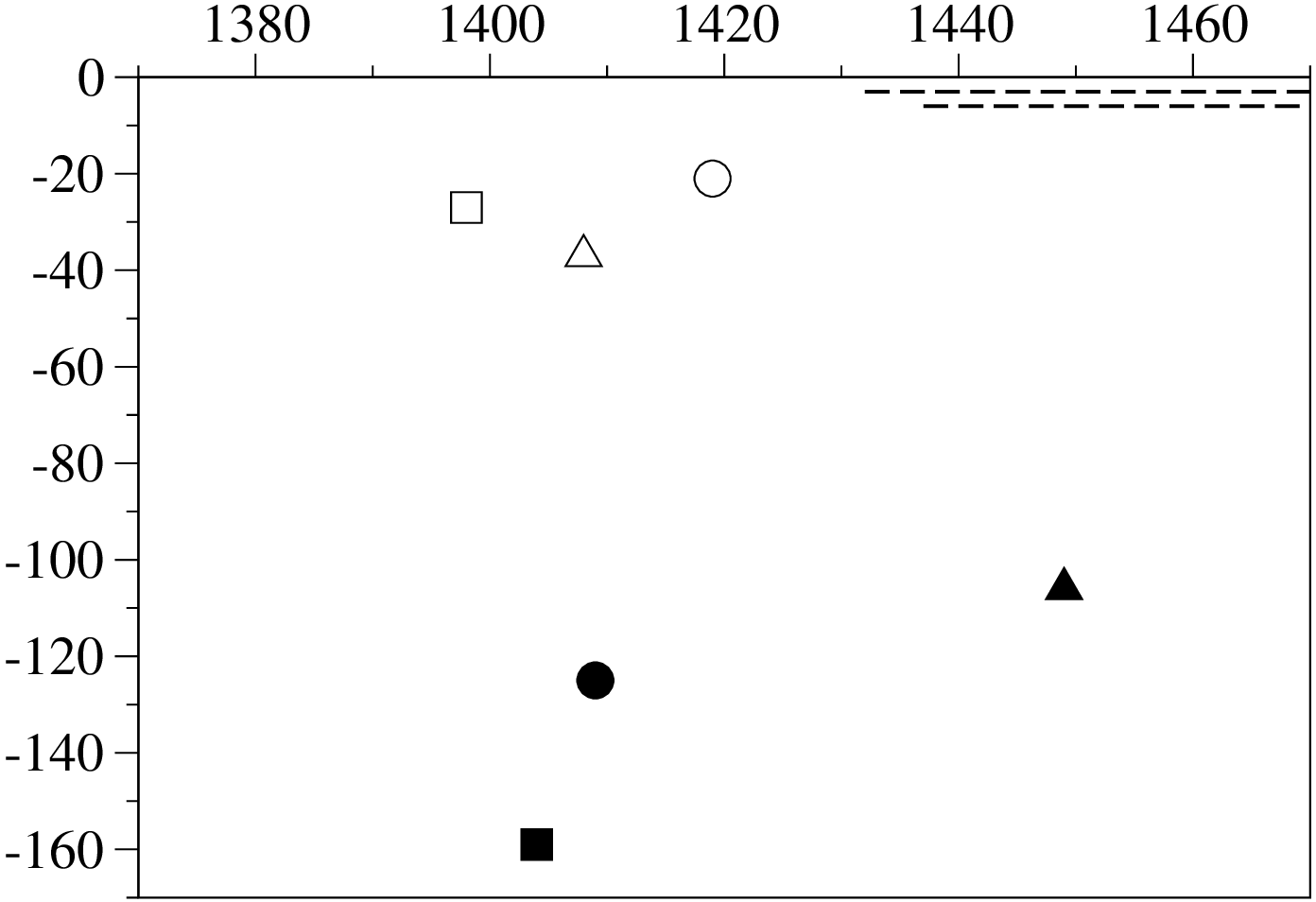}
  \put(-10,13){\rotatebox{90}{{\scalebox{0.9}{Im$(W)$ \quad (MeV)}}}}
  \put(34,74){\scalebox{0.9}{Re$(W)$ \quad (MeV)}}
\end{overpic} \\[1.0ex]
(a) & & (b)
\end{tabular}
\caption{Left: Pole positions of the $T$ matrix in the complex $W$ plane. The triangles,
         diamonds, squares and circles correspond to the ``WT'', ``c'', ``$s$'' and ``$u$'' 
         approach, respectively. The dashed lines represent the $K^- p$ and $\bar{K}^0 n$
         cuts, respectively.
         Right:  Pole positions of the $T$ matrix in the complex $W$ plane. The circles, 
         triangles and squares correspond to the fits ``1'', ``2'' and ``3'', respectively.}
\label{fig:1405poles}
\end{figure}

Following \cite{JOORM} we define 
complex parameters $g_i$ representing the contribution to the coupling strength at the pole from the 
channel with index $i$.\footnote{Note that our definition of the $T$ matrix differs from that 
in \cite{JOORM} by a factor of $-2 \sqrt{M_{a} M_{b}}$.}
They can be extracted by the residue of the $T$ matrix at the position $W_0$ of the pole
\beq
g_i g_j = \mathrm{Res}_{W_0} T_{ij} \ .
\eeq
Since the $T$ matrix is merely defined up to an arbitrary complex phase, only the modulus
of $g_i$ is meaningful. In Table~\ref{tab:1405poles} we show the positions and coupling
strengths of the $\Lambda(1405)$ poles classified according to the different approaches
and fits.
For clarity, the pole positions are also depicted in Fig.~\ref{fig:1405poles}.
A result common to all approaches is the fact that the pole which couples strongly 
to the $I=0$ $\bar{K} N$ state (see open symbols in Fig.~\ref{fig:1405poles}) is located closer to the real axis, in agreement with \cite{JOORM}. 
The inclusion of higher order
contact terms slightly lowers its real part, whereas its position is practically not affected
when the direct and crossed Born terms are taken into account (approaches ``$s$'' and ``$u$''),
cf.\ Fig.~\ref{fig:1405poles}a.
The second pole which couples strongly to the $\pi \Sigma$ channels is shifted 
drastically by going from the ``WT'' to the ``c'' approach. It moves up in energy (even 
above the $\bar{K} N$ thresholds) and away from the real axis. 
The inclusion of the $s$- and $u$-channel Born terms successively brings the pole to lower 
energies again. In the ``$u$'' approach it is almost in line with the first pole,
but located at quite some distance from the real axis, not supporting a 
pronounced double pole structure close to the real axis.

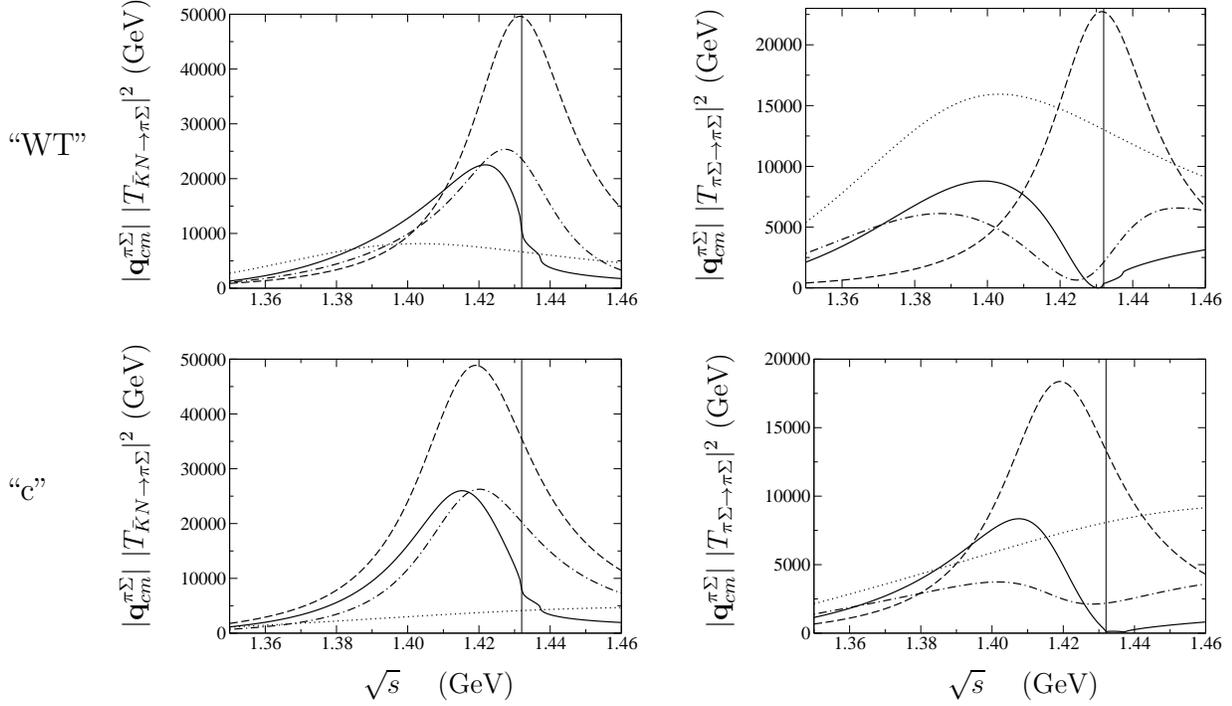
\begin{figure}
\centering
\begin{tabular}{p{1.8cm}rp{0.7cm}r}
 &
\begin{overpic}[height=0.17\textheight,clip]{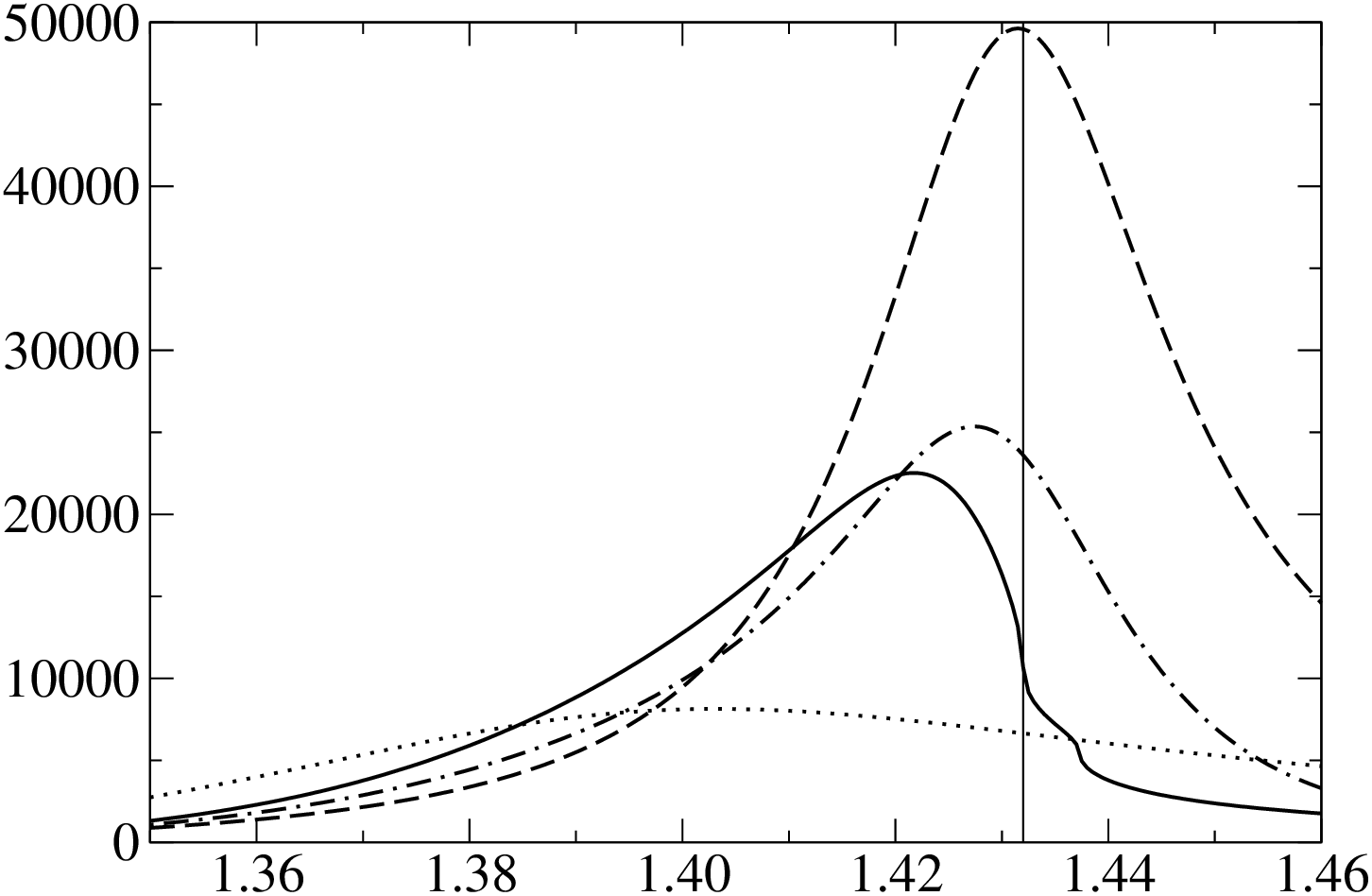}
  \put(-13,3){\rotatebox{90}{{\scalebox{0.9}{
    $|\mathbf{q}_{cm}^{\pi \Sigma}| \ |T_{\bar{K} N \to \pi \Sigma}|^2$ (GeV)}}}}
  \put(-38,33){``WT''}
\end{overpic} & &
\begin{overpic}[height=0.17\textheight,clip]{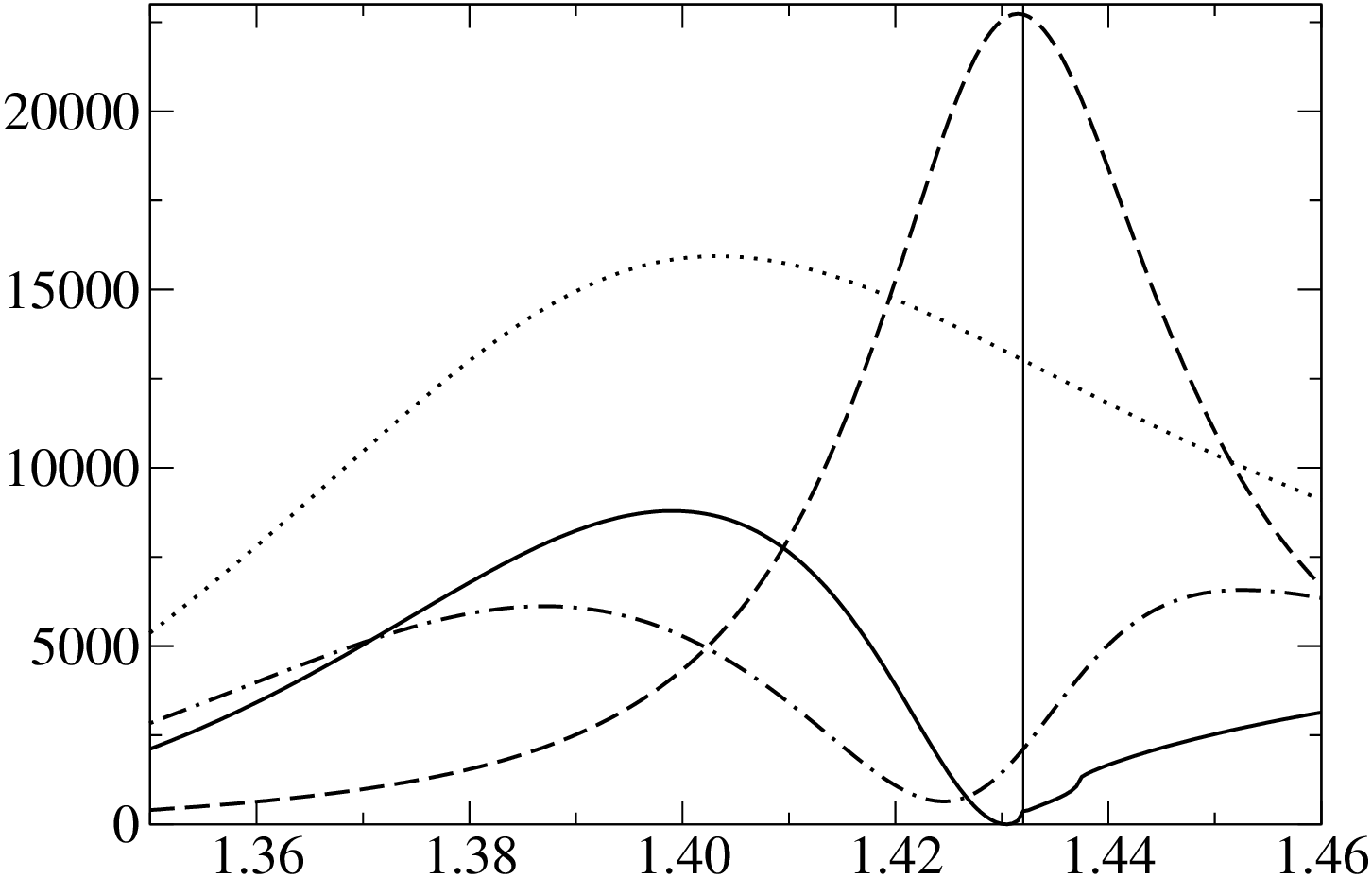}
  \put(-13,3){\rotatebox{90}{{\scalebox{0.9}{
    $|\mathbf{q}_{cm}^{\pi \Sigma}| \ |T_{\pi \Sigma \to \pi \Sigma}|^2$ (GeV)}}}}
\end{overpic} \\[0.02\textheight]
 &
\begin{overpic}[height=0.17\textheight,clip]{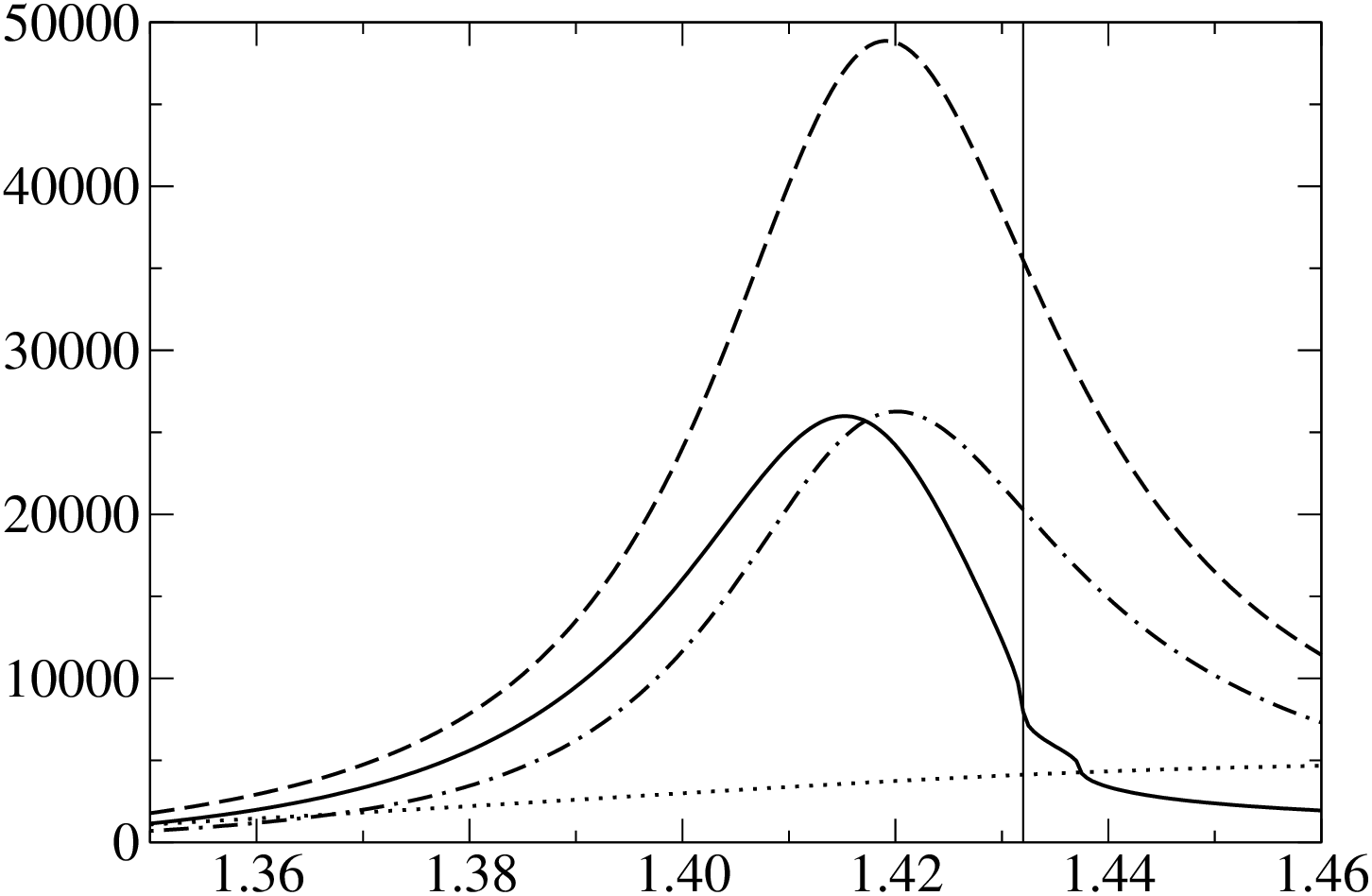}
  \put(-13,3){\rotatebox{90}{{\scalebox{0.9}{
    $|\mathbf{q}_{cm}^{\pi \Sigma}| \ |T_{\bar{K} N \to \pi \Sigma}|^2$ (GeV)}}}}
  \put(40,-9){\scalebox{0.9}{$\sqrt{s}$ \quad (GeV)}}
  \put(-38,33){``c''}
\end{overpic} & &
\begin{overpic}[height=0.17\textheight,clip]{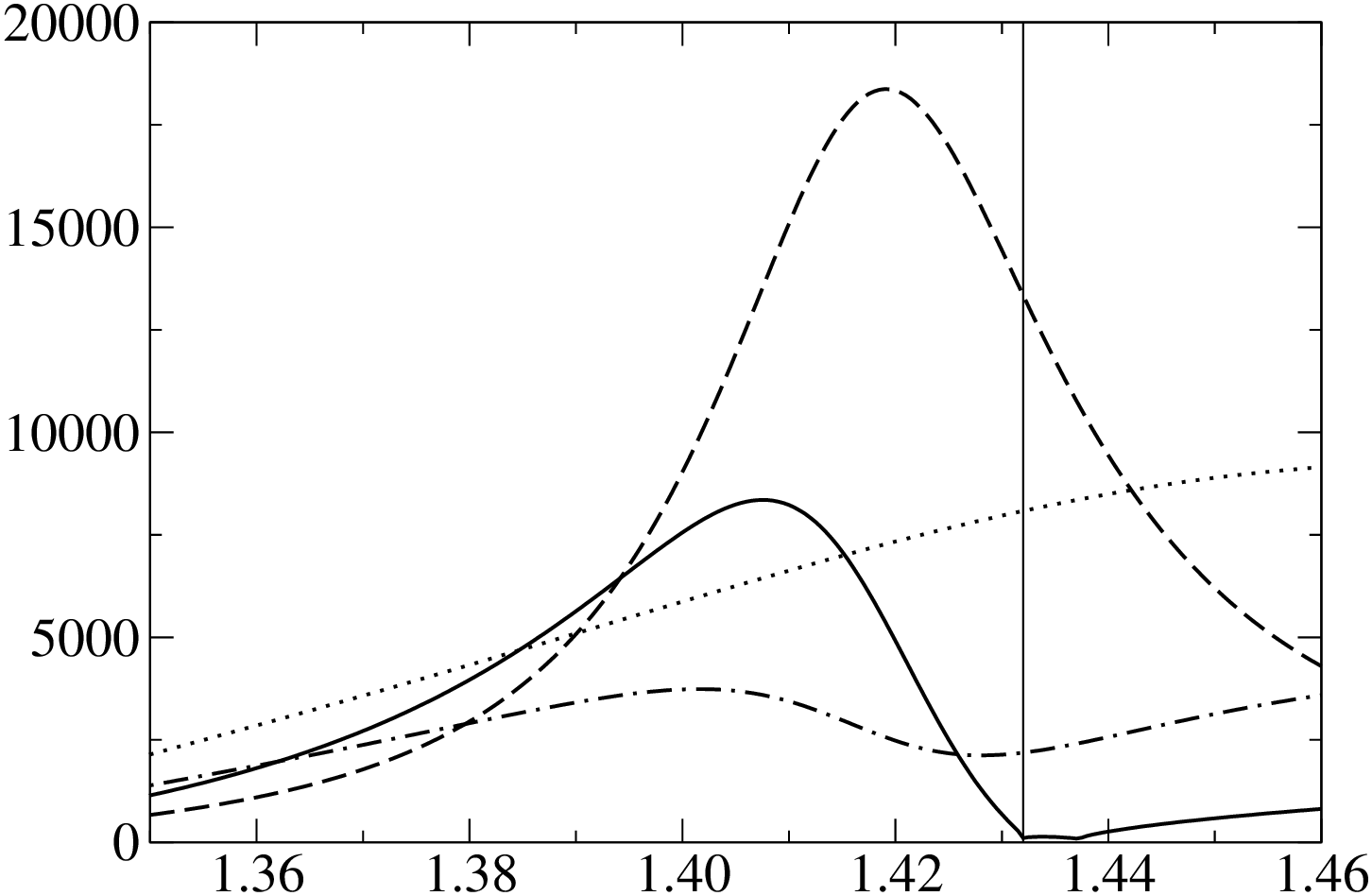}
  \put(-13,3){\rotatebox{90}{{\scalebox{0.9}{
    $|\mathbf{q}_{cm}^{\pi \Sigma}| \ |T_{\pi \Sigma \to \pi \Sigma}|^2$ (GeV)}}}}
  \put(40,-9){\scalebox{0.9}{$\sqrt{s}$ \quad (GeV)}}
\end{overpic}
\end{tabular} 
\vspace{2.0ex}
\caption{Comparison of the pole model described in the text and the full coupled channel
         $T$ matrix for the approaches ``WT'' (upper figures) and ``c'' (lower figures). 
         We plot the quantities
         $|\mathbf{q}_{cm}^{\pi \Sigma}| \ |T_{\bar{K} N \to \pi \Sigma}|^2$ (left column) and 
         $|\mathbf{q}_{cm}^{\pi \Sigma}| \ |T_{\pi \Sigma \to \pi \Sigma}|^2$ (right column)
         for $I=0$ meson-baryon states. The lines represent the contribution of the first pole
         (dashed), the second pole (dotted), both (dot-dashed), and the full coupled-channels
         result (solid).}
\label{CompTpT}
\end{figure}

In order to study the interplay of the two poles and their effect on
the real axis and thus on physical observables, we construct a simple model as done in 
\cite{JOORM}. If the $T$ matrix were solely furnished by two $\Lambda(1405)$ poles,
it would be given by
\beq
T_{ij}^{\textrm{(poles)}} = \frac{g_{i}^{(1)} g_{j}^{(1)}}{W - W_{0}^{(1)}}
                          + \frac{g_{i}^{(2)} g_{j}^{(2)}}{W - W_{0}^{(2)}} \ ,
\eeq
where the superscripts (1), (2) refer to the pertinent poles.
In Fig.~\ref{CompTpT} we compare this simple pole amplitude to the full coupled channel
$T$ matrix in the relevant $I=0$ $\pi \Sigma \to \pi \Sigma$ and 
$\bar{K} N \to \pi \Sigma$ channels by plotting the quantity 
$|\mathbf{q}_{cm}^{\pi \Sigma}| \ |T_{\pi \Sigma, \bar{K} N \to \pi \Sigma}|^2$, where $T$
is given by either just one pole, both poles, or by the full coupled channel result.
For illustrative purposes we restrict ourselves to the ``WT'' and ``c'' models.
Both for the ``WT'' and ``c'' approach the $\bar{K} N \to \pi \Sigma$ amplitude is 
dominated by the pole close to the real axis, which couples most strongly to the 
$\bar{K} N$ states, and the full $T$ matrix element is well 
described by the pole model. For the process $\pi \Sigma \to \pi \Sigma$ the biggest portion of 
the ``WT'' result stems again from the pole contributions, reflecting the double pole 
structure close to the real axis. In contrast, the inclusion of $\mathcal{O}(p^2)$ 
contact terms in the ``c'' approach significantly reduces the influence of the second pole
which couples mainly to the $I=0$ combination of $\pi \Sigma$ channels, giving rise to a 
large background contribution to the amplitude. 
This is also observed for the ``$s$'' and ``$u$'' approaches.
Note also that due to interference effects the subthreshold peak of the 
$\pi \Sigma \to \pi \Sigma$ pole model amplitude appears at a similar
position in both the ``WT'' and the ``c'' approach, although the second pole happens to be 
located at quite different positions in the complex plane.

For completeness, we also show in Table~\ref{tab:1405poles} and Fig.~\ref{fig:1405poles} the pole positions and couplings extracted from the fits ``2'' 
and ``3'', obtained by approaching the constraints set by the DEAR experiment.
Again the pole with strong coupling to $\bar{K} N$ is located close to the real axis,
and its real part is decreased by going from fit ``1'' to fit ``3''. The position of the  
second pole varies a lot with a strong tendency to further depart from the real axis. Furthermore,
we note that the characteristic strong coupling of the second pole
to $\pi \Sigma$ states is equaled in magnitude by the 
coupling to the $\bar{K} N$ channel when the DEAR data are taken into account. 
The key feature, independent of the additional constraint imposed by the DEAR data, 
is that the contribution from the second pole on the real axis
dissolves in the background once next-to-leading order $({\cal O}(p^2))$ 
dynamics are turned on in addition to the leading Weinberg-Tomozawa term.

We conclude that although the different approaches yield
similar fits to all available experimental data, the pertinent pole structures are 
quite diverse. For the ``WT'' version we observe a pronounced double pole structure as 
described in \cite{JOORM}. For the schemes which include higher order contact 
terms, only the pole which couples most strongly to the $\bar{K} N$ state is located close
to the real axis. The influence of the second pole is substantially reduced and the 
$T$ matrix cannot be well approximated by the outlined pole model. Instead, background
contributions are important.
Our findings emphasize that the analytic continuation of partial waves in the 
complex energy plane depends sensitively on the basic dynamical input of the underlying 
chiral SU(3) Lagrangian.


\begin{table}
\centering
\begin{tabular}{|c|r@{$\,-\,$}l|c|c|c|c|}
\hline
 & \multicolumn{2}{c|}{}            & \multicolumn{4}{c|}{$|g_i|$} \\
 & \multicolumn{2}{c|}{$W_0$ (MeV)} & $\pi \Sigma$ & $\bar{K} N$ & $\eta \Lambda$ & $K \Xi$ \\ 
\hline
``WT''           & 1431 & $ 17 i$ & 2.46 & 3.64 & 1.82 & 0.49 \\
                 & 1391 & $ 55 i$ & 4.29 & 3.06 & 0.84 & 0.81 \\
\hline
``c''            & 1418 & $ 20 i$ & 2.59 & 4.22 & 1.98 & 0.72 \\
                 & 1440 & $120 i$ & 5.05 & 3.60 & 1.69 & 1.63 \\
\hline
``$s$''          & 1418 & $ 19 i$ & 2.46 & 4.29 & 2.13 & 0.60 \\
                 & 1419 & $134 i$ & 5.02 & 3.72 & 1.49 & 1.73 \\
\hline
``$u$'' = ``1''  & 1419 & $ 21 i$ & 2.68 & 4.42 & 2.27 & 0.36 \\
                 & 1409 & $125 i$ & 5.13 & 3.99 & 1.90 & 1.10 \\
\hline
``2''            & 1408 & $ 37 i$ & 4.19 & 5.55 & 3.28 & 0.49 \\
                 & 1449 & $106 i$ & 6.16 & 6.12 & 4.25 & 1.39 \\
\hline
``3''            & 1398 & $ 27 i$ & 3.08 & 4.86 & 2.58 & 1.55 \\
                 & 1404 & $159 i$ & 4.58 & 4.57 & 2.97 & 1.17 \\
\hline
\end{tabular}
\caption{Positions of the poles which are
         relevant for $\Lambda(1405)$ and their coupling strengths to isospin $I=0$ states.}
\label{tab:1405poles}
\end{table}

\section{Conclusions}\label{sec:Concl}

In the present work, we have critically examined and updated the analysis of the $\bar{K} N$ system
within the framework of coupled-channels approaches combined with chiral SU(3) dynamics. 
There is renewed interest in the investigation
of the $\bar{K} N$ channel in the light of the new accurate measurement
of the strong interaction shift and width of kaonic hydrogen at DEAR which
sets tight constraints. It is therefore worth investigating whether both the DEAR data and the $K^- p$ scattering
data can be accommodated by coupled-channels analyses, while at the same time trying to reduce
the inherent model dependence of these approaches wherein chiral effective field theory
is combined with a non-perturbative Bethe-Salpeter equation.
The driving terms for the Bethe-Salpeter equation are derived from the effective Lagrangian
and constitute a major source of model dependence.
Several variants of such approaches are commonly used in the literature,
e.g., only the Weinberg-Tomozawa term originating from the leading order Lagrangian
is taken into account, while in other works direct and crossed Born terms are added
or contact interactions of the next-to-leading chiral order are included.

In the present investigation, we have worked out the driving terms in four consecutive steps.
Starting from the Weinberg-Tomozawa term, we 
successively added contact interactions of second chiral order, the direct Born term
and the crossed Born term.

All four versions have in common that the agreement with $\bar{K} N$ scattering data
is partly spoilt once the new tight constraints imposed by the DEAR experiment are taken into account.
(We mention though that the results of these models fall within the larger
error ranges of the KEK experiment.)
The largest discrepancies are observed in the elastic $K^- p$ channel where
the calculated cross section is substantially lowered by inclusion of the DEAR data.
Coulomb effects ameliorate the situation at low kaon laboratory momenta below
100 MeV$/c$, but an offset to elastic $K^- p$ scattering data
remains. Moreover, electromagnetic corrections to the strong interaction shift
and width in kaonic hydrogen as given in \cite{MRR} reduce the discrepancy further, but
are not able to compensate the difference between the coupled-channels approaches
and the experimental data.
Further tight phenomenological constraints in the $\bar{K} N$ system are provided by the threshold
branching ratios which have been measured very precisely. Inclusion of the DEAR data
produces results for the branching ratio $R_c$ which are not in agreement with the
quoted experimental error ranges.

Another consistency check could be provided by studying the $K^- p \to \pi \Sigma$ amplitude
below the $K^- p$ threshold, since inclusion of the DEAR data amounts to a
substantial change in this amplitude. Experiments towards this direction
are currently analyzed at SPring-8/LEPS and at ELSA, where photoproduction of 
$\Lambda(1405)$ has been measured. If $K^-$ exchange in the $t$-channel can be isolated
from these data, the information gained would be very useful in order to set constraints for the $\bar{K} N$ scattering amplitude below threshold \cite{LS}.

The comparison between the different variants of the coupled-channels approaches
can be summarized as follows.
The quality of the fits to data is improved substantially by including
next-to-leading order contact interactions with
new parameters of the effective Lagrangian which we vary within reasonable ranges as explained in the
text. The inclusion of the Born terms, on the other hand, leads only to minor changes.
The treatment of the interaction kernel at subleading order also destroys the
pronounced double pole structure of the $\Lambda(1405)$ close to the real axis
as observed in \cite{JOORM}. Although we still see two poles in the relevant
unphysical sheet, the pole with a stronger coupling to the
$\pi \Sigma$ channel now moves far away from the real axis, losing its
importance for any physical observables.
As a consequence, the full partial wave amplitude for $\pi \Sigma \to \pi \Sigma $
is not approximated well by just these two poles
and the background contribution becomes important.

The updated, constrained analysis presented here is also of considerable interest in the discussion of possible deeply bound $K^-$- nuclear states \cite{DAY}. The amplitudes shown in Figs.~\ref{Compfstr}, \ref{wDEARfstr} suggest a complex, energy dependent subthreshold $\bar{K}$-nucleus potential which is attractive in the $\bar{K}N$ energy range below the $\Lambda(1405)$. Its imaginary part decreases as the energy is lowered towards the $\pi\Sigma$ threshold, an effect that has been pointed out previously in Refs.~\cite{WW}. This is a potential mechanism for supporting narrow bound $\bar{K}$ 
states at sufficiently large nuclear densities, but details concerning the strong energy dependence of the driving potentials require additional constraints and further investigation.

\section*{Acknowledgments}\label{sec:Ackno}

We thank M.~Lutz and U.-G.~Mei{\ss}ner for useful discussions.
Partial financial support by DFG and BMBF is gratefully acknowledged.
This research is part of the EU Integrated Infrastructure Initiative Hadronphysics under contract
 number RII3-CT-2004-506078.

\appendix

\section{Tree level amplitudes}\label{app:amp}

The amplitudes for the meson-baryon scattering processes 
$\phi_i B_{a}^{\sigma} \to \phi_j B_{b}^{\sigma'}$ (with spin indices $\sigma$, $\sigma'$)
corresponding to the tree level diagrams in Fig.~\ref{fig:feyns}a,\,c,\,d 
have already been given in \cite{OM}, but for completeness we present them here
in our notation along with the next-to-leading order contact term depicted in 
Fig.~\ref{fig:feyns}b. One obtains
\begin{multline}
V_{jb,ia}^{(a)} = \frac{1}{8 f^2} \ C_{jb,ia}^{(a)} \ N_a \, N_b \\
  \times ({\chi_{b}^{\sigma'}})^T 
  \biggl[ 2 \sqrt{s} - M_a - M_b + (2 \sqrt{s} + M_a + M_b) 
  \frac{\mathbf{q'} \cdot \mathbf{q} + i (\mathbf{q'} \times \mathbf{q}) \cdot 
        \boldsymbol{\sigma}}{N_{a}^2 N_{b}^2} \biggr] \chi_{a}^{\sigma} \ ,
\end{multline}
\begin{multline}
V_{jb,ia}^{(b)} = \frac{-1}{f^2} \bigl( C_{jb,ia}^{(b_1)} 
  - 2 (E_i E_j - \mathbf{q'} \cdot \mathbf{q}) \, C_{jb,ia}^{(b_2)} \bigr) 
  \, N_a \, N_b \, ({\chi_{b}^{\sigma'}})^T  \biggl[ 1 - 
  \frac{\mathbf{q'} \cdot \mathbf{q} + i (\mathbf{q'} \times \mathbf{q}) \cdot
        \boldsymbol{\sigma}}{N_{a}^2 N_{b}^2} \biggr] \chi_{a}^{\sigma} \ , 
\end{multline}
\begin{multline}
V_{jb,ia}^{(c)} = \frac{-1}{12 f^2} \sum_{c=1}^{8} C_{jb,c}^{(c)} \, C_{ia,c}^{(c)}
  \ N_a \, N_b \ \frac{1}{s - M_{c}^2} \\
  \times ({\chi_{b}^{\sigma'}})^T
  \bigg[ (\sqrt{s} - M_a) (s - (M_b + M_c) \sqrt{s} + M_b M_c) 
    \qquad \qquad \qquad \qquad \\
  + (\sqrt{s} + M_a) (s + (M_b + M_c) \sqrt{s} + M_b M_c)
  \frac{\mathbf{q'} \cdot \mathbf{q} + i (\mathbf{q'} \times \mathbf{q}) \cdot
        \boldsymbol{\sigma}}{N_{a}^2 N_{b}^2} \bigg] \chi_{a}^{\sigma} \ ,
\end{multline}
\begin{multline}
V_{jb,ia}^{(d)} = \frac{1}{12 f^2} \sum_{c=1}^{8} C_{ic,b}^{(c)} \, C_{jc,a}^{(c)}
  \ N_a \, N_b \ \frac{1}{u - M_{c}^2} \ ({\chi_{b}^{\sigma'}})^T \\
  \times \bigg[ u(\sqrt{s} + M_{c}^2) + \sqrt{s}(M_b (M_a + M_c) + M_a M_c)
    - M_b (M_a + M_c)(M_a + M_b) - M_{a}^2 M_c \qquad \quad \\
    + \Big(u(\sqrt{s} - M_{c}^2) + \sqrt{s}(M_b (M_a + M_c) + M_a M_c) 
         + M_b (M_a + M_c)(M_a + M_b) + M_{a}^2 M_c \Big) \\ 
  \times \frac{\mathbf{q'} \cdot \mathbf{q} + i (\mathbf{q'} \times \mathbf{q}) \cdot
               \boldsymbol{\sigma}}{N_{a}^2 N_{b}^2} \bigg] \chi_{a}^{\sigma} \ .
\end{multline}
The two-component Pauli-spinor of a baryon $B$ with spin $\sigma$ is symbolized by 
$\chi_{B}^{\sigma}$ while the pertinent normalization factor is given by
$N_B = \sqrt{E_B + M_B}$ and $E_x$ is the center-of-mass energy of particle $x$.
The center-of-mass three-momenta of the initial and final particles are denoted by 
$\mathbf{q}$ and $\mathbf{q'}$, respectively. The Mandelstam 
variable $u$ is given by $u = (p - k')^2$, where $p$ is the four-momentum of the initial baryon and 
$k'$ that of the final meson.
The coefficients $C_{jb,ia}^{(a)}$, $C_{jb,ia}^{(b_1)}$ and $C_{jb,ia}^{(b_2)}$, which 
are symmetric under the interchange of initial and final meson-baryon pairs, are 
compiled in Tables~\ref{clebschWT}, \ref{clebschC1} and \ref{clebschC2}, respectively,
whereas the non-zero axial vector couplings $C_{\phi B_1 , B_2}^{(c)}$ (which are
symmetric under the combined transformation $B_1 \leftrightarrow B_2$ and 
$\phi \leftrightarrow \bar{\phi}$) are given by 
%
\beq
{\renewcommand{\arraystretch}{2.0}
\begin{tabular}{l}
$C_{K^- p, \Lambda}^{(c)} = C_{\bar{K}^0 n, \Lambda}^{(c)} = C_{\eta \Xi^- , \Xi^-}^{(c)}
  = C_{\eta \Xi^0 , \Xi^0}^{(c)} = - D - 3 F \, ,$ \\
$\sqrt{2}\,  C_{K^- p, \Sigma^0}^{(c)} = - \sqrt{2}\,  C_{\bar{K}^0 n, \Sigma^0}^{(c)} 
  = C_{\bar{K}^0 p, \Sigma^+}^{(c)} = C_{K^- n, \Sigma^-}^{(c)} 
  = C_{\pi^+ \Xi^-, \Xi^0}^{(c)} = \sqrt{2} C_{\pi^0 \Xi^-, \Xi^-}$ \\
$ \qquad \qquad = - \sqrt{2} C_{\pi^0 \Xi^0, \Xi^0} = \sqrt{6}\, (D - F) \, ,$ \\
$C_{\pi^0 \Sigma^0, \Lambda}^{(c)} = C_{\pi^+ \Sigma^-, \Lambda}^{(c)}
  = C_{\pi^- \Sigma^+, \Lambda}^{(c)} = C_{\eta \Sigma^+, \Sigma^+}^{(c)} 
  = C_{\eta \Sigma^-, \Sigma^-}^{(c)} = C_{\eta \Sigma^0, \Sigma^0}^{(c)}
  = - C_{\eta \Lambda, \Lambda}^{(c)} = 2 D \, ,$ \\
$C_{\pi^+ \Sigma^-, \Sigma^0}^{(c)} = - C_{\pi^- \Sigma^+, \Sigma^0}^{(c)}
  = - C_{\pi^0 \Sigma^-, \Sigma^-}^{(c)} = C_{\pi^0 \Sigma^+, \Sigma^+}^{(c)}
  = 2 \sqrt{3} F \, ,$ \\
$C_{K^+ \Xi^- , \Lambda}^{(c)} = C_{K^0 \Xi^0 , \Lambda}^{(c)} 
  = C_{\eta p, p}^{(c)} = C_{\eta n, n}^{(c)} = - D + 3 F \, ,$ \\
$\sqrt{2}\,  C_{K^+ \Xi^- , \Sigma^0}^{(c)} = - \sqrt{2}\,  C_{K^0 \Xi^0 , \Sigma^0}^{(c)} 
  = C_{\pi^- p, n}^{(c)} = \sqrt{2}\,  C_{\pi^0 p, p}^{(c)} 
  = - \sqrt{2}\,  C_{\pi^0 n, n}^{(c)} = C_{\bar{K}^0 \Sigma^-, \Xi^-}^{(c)}$ \\
$ \qquad \qquad = C_{K^- \Sigma^+, \Xi^0}^{(c)} = \sqrt{6}\, (D + F) \, .$ 
\end{tabular}}
\eeq
%

The interaction kernels utilized in the various approaches under consideration,
``WT'', ``c'', ``$s$'' and ``$u$'', are obtained by projecting out the $s$-wave part of the 
amplitudes according to Eq.~(\ref{eq:prj_swave}) with
{\setlength{\arraycolsep}{3.0pt} 
\beqa 
V_{\textrm{``WT''}} & = & V^{(a)} \ , \\
V_{\textrm{``c''}}  & = & V^{(a)} + V^{(b)} \ , \\
V_{\textrm{``$s$''}}  & = & V^{(a)} + V^{(b)} + V^{(c)} \ , \\ 
V_{\textrm{``$u$''}}  & = & V^{(a)} + V^{(b)} + V^{(c)} + V^{(d)} \ .
\eeqa}
%


\begin{table}
\centering
{\renewcommand{\arraystretch}{1.5} 
\begin{tabular}{|c|cccccccccc|}
\hline
& $K^- p$ & $\bar{K}^0 n$ & $\pi^0 \Lambda$ & $\pi^0 \Sigma^0$ & $\pi^+ \Sigma^-$
  & $\pi^- \Sigma^+$ & $\eta \Lambda$ & $\eta \Sigma^0$ & $K^- \Xi^+$
  & $K^0 \Xi^0$ \\
\hline
$K^- p$          & 4 & 2 & $\sqrt{3}$ & 1 & 0 & 2 & 3 & $\sqrt{3}$ & 0 & 0 \\
$\bar{K}^0 n$    & & 4 & $-\sqrt{3}$ & 1 & 2 & 0 & 3 & $-\sqrt{3}$ & 0 & 0 \\
$\pi^0 \Lambda$  & & & 0 & 0 & 0 & 0 & 0 & 0 & $\sqrt{3}$ & $-\sqrt{3}$ \\
$\pi^0 \Sigma^0$ & & & & 0 & 4 & 4 & 0 & 0 & 1 & 1 \\
$\pi^+ \Sigma^-$ & & & & & 4 & 0 & 0 & 0 & 2 & 0 \\
$\pi^- \Sigma^+$ & & & & & & 4 & 0 & 0 & 0 & 2 \\
$\eta \Lambda$   & & & & & & & 0 & 0 & 3 & 3 \\
$\eta \Sigma^0$  & & & & & & & & 0 & $\sqrt{3}$ & $-\sqrt{3}$ \\
$K^- \Xi^+$      & & & & & & & & & 4 & 2 \\
$K^0 \Xi^0$      & & & & & & & & & & 4 \\
\hline
\end{tabular}}
\caption{Coefficients $C_{jb,ia}^{(a)} = C_{ia,jb}^{(a)}$ of the leading-order contact 
         interaction.}
\label{clebschWT}
\end{table}

\begin{table}
\centering
\rotatebox{90}{
{\setlength{\tabcolsep}{2.0pt} \renewcommand{\arraystretch}{1.5}
\begin{tabular}{|c|cccccccccc|}
\hline
& $K^- p$ & $\bar{K}^0 n$ & $\pi^0 \Lambda$ & $\pi^0 \Sigma^0$ & $\pi^+ \Sigma^-$ 
  & $\pi^- \Sigma^+$ & $\eta \Lambda$ & $\eta \Sigma^0$ & $K^- \Xi^+$ & $K^0 \Xi^0$ \\
\hline
$K^- p$          & $\sst 4 (b_0 + b_D) m_{K}^2$ & $\sst 2 (b_D + b_F) m_{K}^2$  
  & $\frac{-(b_D + 3 b_F) \mu_{1}^2}{2 \sqrt{3}}$ & $\frac{(b_D - b_F) \mu_{1}^2}{2}$ 
  & 0 & $\sst (b_D - b_F) \mu_{1}^2$ & $\frac{(b_D + 3 b_F) \mu_{2}^2}{6}$
  & $\frac{-(b_D - b_F) \mu_{2}^2}{2 \sqrt{3}}$ & 0 & 0 \\
$\bar{K}^0 n$    & & $\sst 4 (b_0 + b_D) m_{K}^2$ 
  & $\frac{(b_D + 3 b_F) \mu_{1}^2}{2 \sqrt{3}}$ & $\frac{(b_D - b_F) \mu_{1}^2}{2}$
  & $\sst (b_D - b_F) \mu_{1}^2$ & 0 & $\frac{(b_D + 3 b_F) \mu_{2}^2}{6}$
  & $\frac{(b_D - b_F) \mu_{2}^2}{2 \sqrt{3}}$ & 0 & 0 \\
$\pi^0 \Lambda$  & & & $\frac{4 (3 b_0 + b_D) m_{\pi}^2}{3}$ & 0 & 0 & 0 & 0 
  & $\frac{4 b_D m_{\pi}^2}{3}$ & $\frac{-(b_D - 3 b_F) \mu_{1}^2}{2 \sqrt{3}}$
  & $\frac{(b_D - 3 b_F) \mu_{1}^2}{2 \sqrt{3}}$ \\
$\pi^0 \Sigma^0$ & & & & $\sst 4 (b_0 + b_D) m_{\pi}^2$ & 0 & 0 
  & $\frac{4 b_D m_{\pi}^2}{3}$ & 0 & $\frac{(b_D + b_F) \mu_{1}^2}{2}$ 
  & $\frac{(b_D + b_F) \mu_{1}^2}{2}$ \\
$\pi^+ \Sigma^-$ & & & & & $\sst 4 (b_0 + b_D) m_{\pi}^2$ & 0 
  & $\frac{4 b_D m_{\pi}^2}{3}$ & $\frac{4 b_F m_{\pi}^2}{\sqrt{3}}$
  & $\sst (b_D + b_F) \mu_{1}^2$ & 0 \\
$\pi^- \Sigma^+$ & & & & & & $\sst 4 (b_0 + b_D) m_{\pi}^2$
  & $\frac{4 b_D m_{\pi}^2}{3}$ & $\frac{-4 b_F m_{\pi}^2}{\sqrt{3}}$ & 0
  & $\sst (b_D + b_F) \mu_{1}^2$ \\ 
$\eta \Lambda$  & & & & & & & $\frac{4(3 b_0 \mu_{3}^2 + b_D \mu_{4}^2)}{9}$ & 0 
  & $\frac{(b_D - 3 b_F) \mu_{2}^2}{6}$ & $\frac{(b_D - 3 b_F) \mu_{2}^2}{6}$ \\
$\eta \Sigma^0$ & & & & & & & & $\frac{4(b_0 \mu_{3}^2 + b_D m_{\pi}^2)}{3}$ 
  & $\frac{-(b_D + b_F) \mu_{2}^2}{2 \sqrt{3}}$ & $\frac{(b_D + b_F) \mu_{2}^2}{2 \sqrt{3}}$ \\
$K^- \Xi^+$    & & & & & & & & & $\sst 4 (b_0 + b_D) m_{K}^2$ 
  & $\sst 2 (b_D - b_F) m_{K}^2$ \\
$K^0 \Xi^0$    & & & & & & & & & & $\sst 4 (b_0 + b_D) m_{K}^2$ \\
\hline
\end{tabular}}}
\caption{Coefficients $C_{jb,ia}^{(b_1)} = C_{ia,jb}^{(b_1)}$ of the next-to-leading
         order contact interaction, where we have made use of the abbreviations 
         $\mu_{1}^2 = m_{K}^2 + m_{\pi}^2$, $\mu_{2}^2 = 5 m_{K}^2 - 3 m_{\pi}^2$, 
         $\mu_{3}^2 = 4 m_{K}^2 - m_{\pi}^2$ and $\mu_{4}^2 = 16 m_{K}^2 - 7 m_{\pi}^2$.}
\label{clebschC1}
\end{table}

\begin{table}
\centering
\rotatebox{90}{
{\setlength{\tabcolsep}{2.0pt} \renewcommand{\arraystretch}{1.5}
\begin{tabular}{|c|cccccccccc|}
\hline
& $K^- p$ & $\bar{K}^0 n$ & $\pi^0 \Lambda$ & $\pi^0 \Sigma^0$ & $\pi^+ \Sigma^-$ 
  & $\pi^- \Sigma^+$ & $\eta \Lambda$ & $\eta \Sigma^0$ & $K^- \Xi^+$ & $K^0 \Xi^0$ \\
\hline
$K^- p$          & $\sst 2 d_2 + d_3 + 2 d_4$ & $\sst d_1 + d_2 + d_3$  
  & $\frac{-\sqrt{3}(d_1 + d_2)}{2}$ & $\frac{-d_1 - d_2 + 2 d_3}{2}$ 
  & $\sst -2 d_2 + d_3$ & $\sst -d_1 + d_2 + d_3$ & $\frac{d_1 - 3 d_2 + 2 d_3}{2}$
  & $\frac{d_1 - 3 d_2}{2 \sqrt{3}}$ & $\sst -4 d_2 + 2 d_3$ & $\sst -2 d_2 + d_3$ \\
$\bar{K}^0 n$    & & $\sst 2 d_2 + d_3 + 2 d_4$ 
  & $\frac{\sqrt{3}(d_1 + d_2)}{2}$ & $\frac{-d_1 - d_2 + 2 d_3}{2}$
  & $\sst -d_1 + d_2 + d_3$ & $\sst -2 d_2 + d_3$ & $\frac{d_1 - 3 d_2 + 2 d_3}{2}$
  & $\frac{-(d_1 - 3 d_2)}{2 \sqrt{3}}$ & $\sst -2 d_2 + d_3$ & $\sst -4 d_2 + 2 d_3$ \\
$\pi^0 \Lambda$  & & & $\sst 2 d_4$ & 0 & 0 & 0 & 0 & $\sst d_3$ 
  & $\frac{\sqrt{3}(d_1 - d_2)}{2}$ & $\frac{-\sqrt{3}(d_1 - d_2)}{2}$ \\
$\pi^0 \Sigma^0$ & & & & $\sst 2 (d_3 + d_4)$ & $\sst -2 d_2 + d_3$ & $\sst -2 d_2 + d_3$ 
  & $\sst d_3$ & 0 & $\frac{d_1 - d_2 + 2 d_3}{2}$ & $\frac{d_1 - d_2 + 2 d_3}{2}$ \\
$\pi^+ \Sigma^-$ & & & & & $\sst 2 d_2 + d_3 + 2 d_4$ & $\sst -4 d_2 + 2 d_3$ & $\sst d_3$ 
  & $\frac{2 d_1}{\sqrt{3}}$ & $\sst d_1 + d_2 + d_3$ & $\sst -2 d_2 + d_3$ \\
$\pi^- \Sigma^+$ & & & & & & $\sst 2 d_2 + d_3 + 2 d_4$ & $\sst d_3$ 
  & $\frac{-2 d_1}{\sqrt{3}}$ & $\sst -2 d_2 + d_3$ & $\sst d_1 + d_2 + d_3$ \\
$\eta \Lambda$   & & & & & & & $\sst 2 (d_3 + d_4)$ & 0 & $\frac{-d_1 - 3 d_2 + 2 d_3}{2}$ 
  & $\frac{-d_1 - 3 d_2 + 2 d_3}{2}$ \\
$\eta \Sigma^0$  & & & & & & & & $\sst 2 d_4$ & $\frac{-(d_1 + 3 d_2)}{2 \sqrt{3}}$ 
  & $\frac{d_1 + 3 d_2}{2 \sqrt{3}}$ \\
$K^- \Xi^+$      & & & & & & & & & $\sst 2 d_2 + d_3 + 2 d_4$ & $\sst -d_1 + d_2 + d_3$ \\
$K^0 \Xi^0$      & & & & & & & & & & $\sst 2 d_2 + d_3 + 2 d_4$ \\
\hline
\end{tabular}}}
\caption{Coefficients $C_{jb,ia}^{(b_2)} = C_{ia,jb}^{(b_2)}$ of the next-to-leading
         order contact interaction.}
\label{clebschC2}
\end{table}


\end{document}